# Real-time shape approximation and 5-D fingerprinting of single proteins

Erik C. Yusko[1,†,‡], Brandon R. Bruhn[1,†], Olivia Eggenberger[1], Jared Houghtaling[1], Ryan C. Rollings[2], Nathan C. Walsh[2], Santoshi Nandivada[2], Mariya Pindrus[3], Adam R. Hall[4], David Sept[1,5], Jiali Li[2], Devendra S. Kalonia[3], Michael Mayer[1,6,*]

**Affiliations:**
[1]Department of Biomedical Engineering, University of Michigan, Ann Arbor, MI 48109, USA
[2]Department of Physics, University of Arkansas, Fayetteville, Arkansas 72701, USA
[3]Department of Pharmaceutical Sciences, University of Connecticut, Storrs, CT 06269, USA
[4]Department of Biomedical Engineering and Comprehensive Cancer Center, Wake Forest University School of Medicine, Winston Salem, NC 27157, USA
[5]Center for Computational Medicine and Biology, University of Michigan, Ann Arbor, MI 48109, USA
[6]Biophysics Program, University of Michigan, Ann Arbor, MI 48109, USA

[†]These authors contributed equally to this work.
[‡]Current address: Department of Physiology and Biophysics, University of Washington, Seattle, WA 98195, USA
[*]Correspondence should be addressed to M.M. (mimayer@umich.edu)

This work exploits the zeptoliter sensing volume of electrolyte-filled nanopores to determine, simultaneously and in real time, the approximate shape, volume, charge, rotational diffusion coefficient, and dipole moment of individual proteins. We have developed the theory for a quantitative understanding and analysis of modulations in ionic current that arise from rotational dynamics of single proteins as they move through the electric field inside a nanopore. The resulting multi-parametric information raises the possibility to characterize, identify, and quantify individual proteins and protein complexes in a mixture. This approach interrogates single proteins in solution and determines parameters such as the approximate shape and dipole moment, which are excellent protein descriptors and cannot be obtained otherwise from single protein molecules in solution. Taken together, this five-dimensional characterization of biomolecules at the single particle level has the potential for instantaneous protein identification, quantification, and possibly sorting with implications for structural biology, proteomics, biomarker detection, and routine protein analysis.

Methods to characterize, identify, and quantify unlabeled, folded proteins in solution on a single molecule level do not currently exist[1]. If available, such methods would have a disruptive impact on the life sciences and clinical assays by simplifying routine protein analysis, enabling rapid and ultra-sensitive biomarker detection[2], and allowing the analysis of personal proteomes[3]. Furthermore, if such methods instantly provided low-resolution approximations of shape, they could help to reveal the conformation of transient protein complexes or large assemblies that are not accessible by electron microscopy, NMR spectroscopy, X-ray crystallography, or small-angle X-ray scattering[4]. Here, we demonstrate that interrogation of single proteins or protein-protein complexes during their passage through the electric field inside a nanopore can enable characterization of these particles based on spheroidal approximations of their shape, as well as their volume, charge, rotational diffusion coefficient, and dipole moment.

Dipole moment has mostly been neglected as a protein descriptor. Despite the pioneering work by Debye[5] and Oncley[6], neither the usefulness of this parameter for protein identification nor its importance for concentrated protein solutions has hitherto been widely appreciated, and existing methods for determining protein dipole moments are tedious and limited to ensemble measurements. We propose, however, that dipole moment provides a powerful dimension for label-free protein analysis since its magnitude is widely distributed among different proteins (absolute values typically range from 0 to 4,000 Debye)[7]. Dipole moment may therefore approach the usefulness of protein size for identification and would likely exceed the usefulness of protein charge (whose values are less distributed, typically ranging from -40$e$ to +40$e$)[7]. Moreover, the pharmaceutical industry is increasingly recognizing the importance of dipole moment for antibody formulations[8], in part because subcutaneous injection of highly concentrated solutions of monoclonal antibodies (the fastest growing class of therapeutics) can be impractical due to high viscosity and aggregation resulting from dipole alignment[8-10]. Measurements of antibody dipole moments could therefore provide a criterion to select early candidates in the drug discovery process and reduce development costs[11].

Interrogating single protein particles during their passage through a pore is simple in principle[12-16]. It requires a single electrolyte-filled pore that connects two solutions across a thin insulating membrane and serves as a conduit for ions and proteins (Fig. 1a). Electrodes connect the solutions on both sides of the membrane to a high-gain amplifier that applies a constant electric potential difference while measuring the ionic current through the nanopore. This arrangement ensures that virtually the entire voltage drop occurs within the pore, rendering this zone supremely sensitive to transient changes in its ionic conductivity. Consequently, each protein that is driven electrophoretically through the pore displaces conductive electrolyte, distorts the electric field, and reduces the ionic current through the pore. If the volume of the electrolyte-filled pore is sufficiently small compared to the volume of the particle, then the change in ionic current due to the translocating particle is measurable and characterized by its magnitude, $\Delta I$, and duration, $t_d$[17-23]; this current signature is referred to as a resistive pulse. In addition to its exquisite sensitivity to conductivity changes, this small volume transiently separates single proteins from other macromolecules in solution, enabling, for the first time, that rotational dynamics of one protein can be interrogated without artifacts from other macromolecules. For this reason, time-dependent modulations of ionic current as a single protein passes through a nanopore can, under appropriate conditions, relate uniquely to the time-dependent molecular orientation of that protein as well as its shape, volume, charge, rotational diffusion coefficient, and dipole moment (Fig. 1b-e, Supplementary Notes 1-3, and Supplementary Figs. 1-9). Several groups have recently considered, in qualitative terms, the effect of a protein's[17,18,20,24-28] or nanoparticle's shape when analyzing distributed $\Delta I$ signals[29] as well as the effect of a protein's dipole moment on its translocation through an alpha-hemolysin pore in the presence of an AC field[30]. The work presented here develops the theory for a *quantitative understanding* of the dependence of measured $\Delta I$ values on the shape, dipole, and rotational dynamics of a protein inside a nanopore and makes it possible to estimate the volume, approximate shape, rotational diffusion coefficient, and dipole moment of non-spherical proteins in real time (Supplementary Figure 10). We suggest that this ability to measure five parameters simultaneously on single proteins in real time has fundamental implications. For instance, analyzing individual proteins one-by-one may inherently mean that these proteins do not have to be purified for determining their approximate shape or the other four parameters. This consequence would be a paradigm shift compared to existing methods for determining the shape or structure of proteins, which either require purified, concentrated, or crystallized protein samples or cannot examine protein dynamics.

**Theory of spheroids rotating in an electric field**

The main concept underlying the analysis introduced in this work is that rotation of a single non-spherical object during translocation through a cylindrical nanopore[31] modulates the current reduction through the pore and that these modulations can be used to determine the orientation, approximate shape, and volume of the object in the pore (Fig. 1b-e).

*Particle orientation.* Golibersuch[32,33] and others[34-36] demonstrated both theoretically on ideal spheroids and experimentally on red blood cells that a crosswise orientation of an oblate (lentil-shaped object) or prolate (rugby ball-shaped object) distorts the electric field along a tube more dramatically than a lengthwise orientation (Fig. 1b). In the context of current recordings through a nanopore this means that the particle-induced blockade of current, $\Delta I$, is maximal when the spheroidal particle is in its extreme crosswise orientation and minimal in the extreme lengthwise orientation; orientations between these two extremes induce intermediate current reductions. As shown in Figure 1c-e, Fricke[37,38] and later Velick and Gorin[39] as well as Golibersuch[32,33] quantified these effects with an electrical shape factor γ, which is directly proportional to the current reduction $\Delta I$.

The black curve in Figure 1e shows that for randomly rotating spheroids, all possible electrical shape factors are not equally probable; instead, the two extreme shape factors $\gamma_{min}$ and $\gamma_{max}$, which correspond to extreme lengthwise and crosswise orientations, are most probable because γ is less angle-dependent near the two extreme orientations than in intermediate orientations (see Supplementary Note 2 for details).

This U-shaped probability distribution means that the translocation of randomly rotating spheroidal proteins through a nanopore should result in a distribution of Δ*I* values with two maxima, one corresponding to Δ*I*(γ*min*) and one to Δ*I*(γ*max*). In contrast, spherical proteins, with a γ value of 1.5 that is independent of orientation, should result in Normal distributions of Δ*I* values.

*Particle shape.* In addition to these orientation-dependent effects, the particle's volume and shape also affect the extent of electric field line distortion (Fig. 1c). For example, when comparing two oblates of equal volume in a cross-wise orientation, the particle that deviates most from a perfect sphere (i.e. the flatter oblate) distorts the field lines more dramatically than the rounder object. Conversely, in a lengthwise orientation, the flatter of these two particles distorts the field lines less dramatically than the rounder object. In other words, particles with increasingly non-spherical shapes result in a more extreme ratio between the current blockage in their crosswise *versus* lengthwise orientation.

*Particle volume.* For translocation of two particles with the same shape but different volume, both the orientation-dependent minimal and maximal current reductions have larger magnitudes for the larger particle compared to the smaller one. Therefore, the magnitude of the current reduction depends on particle volume, while the ratio between the minimal and maximal current reduction depends on particle shape.

*General implications for nanopore recordings of non-spherical particles.* The dependence of Δ*I* values on the shape and orientation of translocating particles has generally been neglected in nanopore-based protein characterization, thereby introducing uncertainty in measurements of volume for particles that are not perfect spheres. Considering these shape-dependent effects, as proposed here, will likely increase the accuracy of nanopore-based particle characterization as most particles and proteins are not perfect spheres.

## RESULTS

**Tethering proteins resolves translocation events in time and enables determination of charge**

In order to obtain time-resolved values of Δ$I$ from the translocation of single proteins, we slowed down translocation by tethering proteins to a lipid anchor that was embedded in the fluid lipid bilayer coating of the nanopores (Fig. 1b and Fig. 2; see Supplementary Note 4 for a detailed discussion on the effects of lipid tethering and the point of attachment)[20,27,40]. In this way, the speed of protein translocation was dominated by the approximately 100-fold higher viscosity of the lipid coating compared to that of the aqueous electrolyte. In addition, we maximized the possibility that the proteins could rotate and sample all orientations in the nanopore by employing long and flexible tethers (Fig. 2). Finally, the lipid coating minimized non-specific interactions between proteins and the pore wall[20], thus enabling extraction of quantitative data on Brownian rotational and translational dynamics of proteins while they were in the pore[41]. For instance, we took advantage of the resulting translocation times to determine the net charge of all ten proteins and found a strong correlation between the charge from nanopore experiments and reference values for the charge of each protein (Pearson correlation coefficient $r = 0.95$, see Supplementary Fig. 11).

**Current blockades reveal the approximate shape and volume of proteins**

To determine the approximate shape and volume of proteins, we developed two strategies based on the theory developed by Fricke[37,38], Velick and Gorin[39], and Golibersuch[32,33] (Fig. 1c-e). Both strategies approximate the shape of proteins with a spheroid and have different strengths and weaknesses as demonstrated in Figures 3 and 4.

The first strategy estimates shape and volume from distributions of *maximum* Δ$I$ values from *many* translocation events that were obtained from a pure protein solution. In other words, only the peak value of Δ$I$ from each resistive pulse is used for analysis. Maximum Δ$I$ values have been employed in almost all nanopore-based resistive pulse analyses of protein volume to date combined with the assumption of a perfectly spherical particle shape (i.e. $\gamma = 1.5$), thereby foregoing the opportunity to evaluate protein shape. In contrast, Golibersuch showed by examining red blood cells that maximum Δ$I$ values could also be used to approximate the shape of particles.[32] Here, we adopted this concept for the first time to proteins in nanopores. An advantage of using maximum Δ$I$ values to estimate protein shape and volume is that the ratio between the extreme values of current reduction, Δ$I(\gamma_{max})$ and Δ$I(\gamma_{min})$, is relatively insensitive to deviations in pore geometry from a perfect cylinder. A disadvantage of this approach is that shape and volume cannot be determined from a single translocation event because only the maximum Δ$I$ value from each translocation event is analyzed and thus many translocations are required to sample all possible electrical shape factors (see Supplementary Note 2 for discussion).

Determining the shape and volume of spheroids from distributions of maximum Δ$I$ values proceeds in three steps; Figure 3 shows the results from each step (see Supplementary Note 2 and Supplementary Figs. 5-8 for details). First, an algorithm detects resistive pulses from the translocation of hundreds to a few thousands copies of the same protein and determines the maximum amplitude of the current modulation, Δ$I$, with respect to the baseline current for each pulse (Fig. 3a,b). As predicted theoretically in Fig. 1c-e, the resulting distribution of maximum Δ$I$ values is either Normal for spherical proteins (Fig. 3c) or bimodal for non-spherical proteins (Fig. 3d-f as well as Supplementary Fig. 6 and Supplementary Note 2). Second, in order to circumvent binning effects encountered with probability distributions[20], the experimentally determined distribution of Δ$I$ values is converted to an empirical cumulative density function, CDF (Fig. 3c,d, insets), and fit iteratively with an equation that describes the variation in Δ$I$ due to rotation of proteins with non-spherical shape (Supplementary Note 2, Equation 13a,b). We refer to this equation as the convolution model since it also accounts for broadening of the Δ$I$ distribution due to convolution of the true signal with noise (Supplementary Fig. 5) and for bias towards either the crosswise or lengthwise orientation as a result of the electric-field-induced torque on the protein's dipole moment.[42] The bias in a distribution of *maximum* $\Delta I$ values, however, may also be affected by other factors than the dipole moment (as discussed in Supplementary Note 2), which are all accounted for by the same fitting parameter. The values of $\Delta I(\gamma_{min})$ and $\Delta I(\gamma_{max})$ returned by the fitting procedure reflect the two extreme orientations of the protein (red dashed curves in Fig. 3d-f). Third, based on the direct proportionality between $\Delta I$ and $\gamma$ and the geometrical relationship between $\gamma$ and the length-to-diameter ratio $m$ of a spheroid (Supplementary Note 2, Equations 1 and 4-7), we determine the shape and volume that agree best with the experimental distribution of $\Delta I$ values for the protein.

Figure 3g shows the spheroidal approximation of the shape of ten different proteins compared to the respective crystal structure for each protein, illustrating that this analysis yields excellent estimates of protein shape, particularly for proteins that closely resemble a spheroid. Figure 3h,i, for instance, shows that the volume and $m$ values agree well with the expected reference values; the average deviation of both parameters is less than 20% (Supplementary Tables 1-4 list the results of this analysis as well as reference values). These results also show that two proteins with a similar molecular weight and volume but different shape are clearly distinguishable by this analysis; for instance, compare the ellipsoids determined for the IgG$_1$ antibody and GPI-AChE in Fig. 3g.[29]

Independent from these experimental results, we confirmed the accuracy of this approach for shape and volume determination using simulated data that was generated from the theory of biased one-dimensional Brownian diffusion and convolved with current noise. Fitting the simulated data with the convolution model, just as with the experimental data, returned values of shape and volume that were in excellent agreement with the input parameters (Supplementary Note 5 and Supplementary Fig. 12).

Compared to other methods for determining the shape and volume of proteins in aqueous solution such as solution-state NMR spectroscopy, analytical ultracentrifugation, and dynamic light scattering, the nanopore-based approach is faster (seconds to minutes), requires smaller sample volumes (≤10 µL) and lower protein concentrations (pM to nM), and may perform better as the size of proteins or protein complexes increases due to the concomitant potential increase in signal-to-noise ratio. While the resolution of shape is significantly lower than that of NMR spectroscopy for small proteins (<80 kDa), it is higher than the resolution of analytical ultracentrifugation and dynamic light scattering. In addition, although the limited time-resolution of currently available amplifiers requires tethering proteins to the lipid coating (a reaction that occurs *in situ* on the nanopore chip), the nanopore-based approach does not require extensive modification of pure proteins by isotope labeling as it is the case for protein NMR spectroscopy.

As opposed to this first strategy, which analyzes maximum $\Delta I$ values from many translocation events, the second strategy estimates the shape and volume of proteins from *individual* resistive pulses by analyzing *all* current values from the beginning to the end of single translocation events, $\Delta I(t)$, in a stand-alone manner (Fig. 4a). This analysis relies on a single translocating protein to rotate and sample virtually all orientation-dependent γ values such that the resulting single-event, or intra-event, $\Delta I$ distribution reveals $\Delta I(\gamma_{max})$ and $\Delta I(\gamma_{min})$ and thereby the protein's spheroidal shape approximation and its volume from an individual translocation event. The advantage of this strategy, in addition to estimating shape and volume *from the translocation of a single protein*, is that it can also determine the protein's rotational diffusion coefficient and dipole moment from individual resistive pulses based on orientation-dependent modulations in current over time. In fact, estimates of all four parameters can be determined and updated in real time as a single protein travels through the pore (Supplementary Fig. 10). The disadvantage of this simultaneous multiparameter analysis from single molecules is that the analysis is limited to resistive pulses with durations of at least 400 µs to ensure that each protein resides sufficiently long in the pore to sample the full range of electrical shape factors (under the conditions used in this work, approximately 10% of events exceeded this threshold). We chose this duration based on the mean-square angular displacement equation that predicts a protein will sample all possible orientations in less than 400 µs, on average, as long as its rotational diffusion coefficient exceeds 3,000 rad$^2$ s$^{-1}$, which was clearly the case for all tethered proteins examined here (Supplementary Table 4). Other disadvantages of this analysis include that it is more sensitive to deviations of the pore geometry from a perfect cylinder than the multi-event analysis of maximum Δ$I$ values (see Supplementary Note 6) and that the analysis of individual resistive pulses is associated with relatively high uncertainty as with other single-molecule measurements.

Figure 4 shows estimates of the shape and volume of proteins obtained from fitting distributions of intra-event Δ$I(t)$ values from individual resistive pulses with the convolution model in the same way as the distributions of maximum Δ$I$ values from hundreds of pulses (see Supplementary Note 6 and Supplementary Fig. 15). We find that the intra-event Δ$I$ distributions from translocations of individual proteins retain their key features (e.g. minimal and maximal Δ$I$ values) although the current recordings are smoothed due to filtering (see Supplementary Fig. 21). The median protein shapes obtained from this analysis are in reasonable agreement with their crystal structure (Fig. 4d), although the analysis of maximum Δ$I$ values yielded more accurate shapes (Supplementary Note 6 discusses potential reasons for the discrepancy between the two approaches). With regard to the robustness of each stand-alone single molecule measurement, more than half of all measurements yielded values of the length-to-diameter ratio and of volume that were within ± 35% of the median value (Supplementary Fig. 15). Based on this result and the expectation that further improvements are possible, we propose that intra-event analysis has the potential to yield good estimates of shape and volume of single proteins from individual translocation events. Moreover, this strategy of analyzing intra-event Δ$I$ distributions introduces, to the best of our knowledge, the only existing method for estimating, in real time, the shape and volume of single protein molecules in solution. Shape and volume determination on a single particle level is particularly advantageous for analysis of samples with large heterogeneity in size and shape (such as amyloids); ensemble methods such as dynamic light scattering are not well suited for such samples.[27,43] Other techniques for analyzing the shape and volume of single proteins such as cryo-electron microscopy and atomic force microscopy either require freezing or surface immobilization that fixes the orientation of the proteins; therefore, these methods are not well suited for tracking protein dynamics.

**Current fluctuations reveal the rotational diffusion coefficient of single proteins**

Figure 4 shows that monitoring the time-dependent modulations of $\Delta I$ while a single particle moves through a nanopore makes it possible to measure its rotational diffusion coefficient, $D_R$, by tracking its rotation over short time scales and therefore over small fluctuations in angle (Supplementary Note 6 and 7 and Supplementary Fig. 18-20). We carried out this analysis in three steps by transforming the intra-event current signal into an angle (i.e. orientation) *versus* time curve (Supplementary Note 6), calculating the mean-square angular displacement over various time intervals, τ, and fitting its initial slope with a model for rotational diffusion about a single axis (Fig. 4c). Figure 4f shows that the most probable $D_R$ values for tethered proteins obtained from many intra-event analyses of individual resistive pulses were strongly correlated with the expected values of $D_R$ in bulk solution (Pearson's $r$ = 0.93). As expected, the presence of the lipid tether and close proximity of the proteins to the bilayer coating reduced $D_R$ significantly;[44,45] this tether-induced attenuation of rotation was consistent with an apparent viscosity increase by a factor of 211 compared to the viscosity in bulk solution (Supplementary Fig. 18). This value is in excellent agreement with fluorescence polarization measurements of GPI-anchored AChE by Yuan and Axelrod, which revealed that the rotational diffusion coefficient of tethered AChE is 199 times smaller than its expected value in bulk solution.[46] For analyzing the rotational dynamics of proteins in real time as presented here, this tether-induced reduction of $D_R$ was critical as it enabled changes in protein orientation to be resolved in time (Supplementary Figs. 10 and 21).

With regard to the robustness of these measurements, we found that, on average, the relative standard deviation of the most probable value of $D_R$ from distributions of measured single molecule values was 46% from experiment-to-experiment or day-to-day; however, as is typical for many single molecule measurements, the variation from event-to-event was large with a mean absolute deviation of 403% (see Supplementary Fig. 18).

To the best of our knowledge, this approach is the fastest method (sub-millisecond) for estimating the rotational diffusion coefficient of single proteins in solution, albeit with considerable uncertainty at this initial stage of the technology; it is also the only non-fluorescent method to determine $D_R$.[47] While the requirement for tethering proteins precludes direct determination of the bulk value of $D_R$ by this approach, the good correlation shown in Figure 4f demonstrates that bulk $D_R$ values can be estimated from the measured $D_R$ values of tethered proteins.

**Bias in a protein's orientation in a nanopore reveals its dipole moment**

Monitoring the rotational dynamics of proteins at long time scales and hence over large changes in angle shows theoretically (Fig. 2c) and experimentally (Fig. 4c,e) that proteins with a dipole moment do not rotate randomly when they experience the MV m$^{-1}$ electric field intensity inside the pore; instead, the proteins undergo biased Brownian rotation due to electric-field-induced torque on their dipole moment.[5,6] Quantifying this bias in orientation by fitting the intra-event $\Delta I$ distribution from an individual resistive pulse with the convolution model made it possible to calculate a protein's dipole moment by considering the potential energy landscape of a dipole in an electric field (Fig. 4b; see Supplementary Notes 2 and 6 and Supplementary Figs. 16 and 17). In this analysis, the fitting parameter µ of the convolution model is equivalent to the dipole moment and therefore yields its magnitude. In contrast, in the analysis of maximum $\Delta I$ values, the same parameter encompasses additional factors, as discussed before, and hence precludes estimation of dipole moment (Supplementary Note 2).

Fig. 4e shows that the most probable values of dipole moment from this nanopore-based analysis agree well with expected values; the average deviation is less than 25%. With regard to the robustness of this method from experiment-to-experiment or day-to-day: the relative standard deviation of the most probable value from distributions of measured single molecule values was 12% and compares well with dielectric impedance spectroscopy measurements;[48] however, as is typical for many single molecule measurements, the variation from event-to-event was large with a mean absolute deviation of 227% (see Supplementary Fig. 16).

While the uncertainty in each stand-alone single molecule measurement of dipole moment will have to be reduced in order to realize the full potential of this approach, this technique introduces the first experimental method for determining the dipole moment of individual proteins in solution. To this end it exploits a fundamental advantage of single molecule techniques, namely that statistical fluctuations of one particle are easier to interpret and to compare with theoretical models than it would be of an ensemble of particles. An additional advantage of this single particle analysis is that it can estimate dipole moments in real time (Supplementary Fig. 10) and requires only pico- to nanomolar concentrations of proteins. In contrast, the standard method for measuring dipole moment, dielectric impedance spectroscopy, requires micromolar protein concentrations and significantly larger sample volumes.[48]

**Simulations confirm that the shape, volume, rotational diffusion coefficient, and dipole moment of single proteins can be estimated in real time**

An analysis on simulated intra-event data with the convolution model returned values of the determined shape, volume, dipole moment, and rotational diffusion coefficient that were in excellent agreement with the input parameters for the simulation (Supplementary Figs. 10, 13, and 14). These purely theoretical results provide strong complementary evidence for the effectiveness of the methods developed in this work.

**Multiparameter characterization of individual proteins improves protein classification**

To assess the potential of nanopore-based identification and characterization of different proteins in a mixture, we repeated the characterization of glucose-6-phosphate dehydrogenase described in Fig. 3f and added an anti-G6PDH IgG antibody. Thus, in the same experiment, single proteins of G6PDH and protein-protein complexes of G6PDH-IgG were passing through the nanopore. Analysis of intra-event $\Delta I$ distributions from individual resistive pulses returned an estimate of the volume, shape, charge, rotational diffusion coefficient, and dipole moment for single particles passing through the pore. Figure 5 shows that this multiparameter-fingerprinting approach made it possible to distinguish G6PDH from the G6PDH-IgG complex by using a clustering algorithm to classify each translocation event (Fig. 5b; see Supplementary Note 8 and Supplementary Fig. 22 for details)[27,49]. This analysis returned excellent estimates of the size and shape of G6PDH and the G6PDH-IgG complex (Fig. 5a and 5c). In contrast, employing the current standard practice of distinguishing proteins by the $\Delta I$ values and translocation times of individual resistive pulses[50,51] underestimated the amount of the G6PDH-IgG complex formed by 90% and overestimated its volume by 70% (Supplementary Note 8). Figure 5b also confirms several expectations with regard to the difference between G6PDH and its complex with IgG. For instance, individual resistive pulses assigned to the complex correspond to significantly larger molecular volumes and smaller rotational diffusion coefficients than resistive pulses assigned to G6PDH by itself. In addition, the dipole moment of G6PDH is relatively clustered as expected for a protein with well-defined shape and position of amino acids. In contrast, the dipole moment of the complex between G6PDH and the polyclonal anti-G6PDH IgG antibody varies widely as expected for a complex that involves a protein antigen with multiple binding sites and binding of a relatively floppy IgG molecule. This analysis, therefore, provides proof-of-principle for nanopore-based characterization, identification, and quantification at the single protein level and demonstrates the advantage of simultaneous multiparameter characterization for identifying individual proteins or protein-protein complexes over single-variate or bi-variate characterization.

These first results also raise the fundamental question, what benefit may be gained by determining additional descriptors for distinguishing individual molecules in a mixture of hundreds of different proteins. Figure 6 takes a bioinformatics-based approach to address this question. Every pixel in this plot represents the normalized distance between one protein-protein pair in either two or five dimensions. The normalized distances between most protein pairs shift from less than one standard deviation in the two dimensional analysis (lower left corner of the plot) to more than three standard deviations in the five dimensional analysis (upper right corner). The graph therefore illustrates that additional descriptors of proteins beyond the oft-employed protein size and charge make it significantly easier to distinguish proteins from each other. Another question is which protein descriptors are most useful for distinguishing proteins from each other. Ideal descriptors are not correlated with each other and therefore provide orthogonal distinguishing power. Analysis of 780 randomly sampled proteins from the Protein Data Bank revealed that mass, volume, and rotational diffusion constant of proteins are strongly correlated with each other (see Supplementary Fig. 23 and 24), while protein size (i.e. mass or volume) did not correlate strongly with protein charge, shape factor *m*, or dipole moment. Protein charge spanned a range from -40$e$ to +40$e$ with a majority between -10$e$ and +10$e$ and is therefore a somewhat degenerate descriptor. In contrast, dipole moment and the length-to-diameter ratio, *m* – the descriptors made accessible on a single molecule level by the work introduced here – are both widely distributed. Hence, dipole moment and protein shape are compelling candidates for the identification of single proteins by multidimensional fingerprinting.

**DISCUSSION**

The work presented here extends the potential of nanopore-based DNA sequencing to five-dimensional characterization and fingerprinting of proteins and protein complexes. Unlike standard bulk methods, this technique interrogates individual proteins one-at-a-time by taking advantage of the molecular scale volume of the nanopore. This zeptoliter volume ($10^{-21}$ L) temporarily separates individual proteins from other proteins in the bulk solution and inherently forms a focal point for measuring protein-induced changes in ionic conductance with exquisite sensitivity. Hence, only the protein residing in the nanopore modulates the electrical signal. This arrangement together with the lipid coating, which minimizes non-specific interactions and slows down the translocation and rotation of lipid-anchored proteins, enables examination of the translational and rotational dynamics of single proteins long enough in time to determine their approximate shape, volume, charge, rotational diffusion coefficient, and dipole moment. We showed that this approach has advantages in distinguishing a protein from its complex with another protein in a binary mixture.

Based on the spectacular progress in nanopore-based DNA sequencing in the last 17 years[3,52-54], we predict that improvements to the approach introduced here will increase the potential of nanopore-based protein characterization[55]. For instance, the single event (intra-event) analysis likely suffers from deviations in the pore geometry from a perfect cylinder. These irregularities, which are a consequence of the current state of the art fabrication methods, affect the local resistance along the lumen of the pore and hence affect the precision with which the maximum and minimum $\Delta I$ value can be determined. Novel fabrication methods such as He-ion beam fabrication produce pores that are almost perfectly cylindrical and should therefore minimize possible artifacts from this source of error[56]. In addition, the recent development of integrated CMOS current amplifiers[57], which can be produced in parallel to record from hundreds of nanopores simultaneously while reaching at least ten-times higher bandwidth and three-times higher signal to noise ratio compared to the amplifier used in this work[57], will increase the throughput and improve the precision and accuracy of determining the rotational dynamics of proteins on their journey through the pore. Such fast amplifiers may eliminate the need for tethering proteins to lipid anchors[40] while their improved signal to noise ratio combined with the recent development of low-noise nanopore chips[58] will likely reduce the uncertainty in each determined parameter[57,59]. Furthermore, computational approaches that can model proteins with shapes more complex than simple spheroids may increase the resolution of shape determination, while the capability to monitor current modulations with MHz bandwidths[40,57] may open up the possibility to follow transient changes in protein conformation and folding as well as to determine the shape of short-lived protein complexes whose structure and dynamics are not accessible by existing techniques.

We suggest that the ability to measure five parameters simultaneously on single proteins in real time, including parameters that can otherwise not be obtained on a single molecule level, has transformative potential for the analysis and quantification of proteins as well as for the characterization of nanoparticle assemblies. For instance, fast protein identification and quantification in complex mixtures is an unsolved problem[2]. Despite its tremendous capabilities, mass spectrometry has currently limited throughput and is not broadly applicable to meet demand for routine protein analysis[1,2]. Two-dimensional gel electrophoresis remains one of the most important techniques for analyzing complex protein samples, but its reproducibility is limited, and the method is slow and semi-quantitative[60]. We propose that multi-dimensional analysis and fingerprinting of single proteins in nanoscale volumes may be one alternative. The work presented here is only a first step in this direction; if improvements similar to the ones made in nanopore-based DNA sequencing can be realized, we think it has the potential to replace methods such as 2-D gel electrophoresis by providing additional protein descriptors, improved quantification, increased sensitivity, reduced analysis time, and lower cost. Such a capability may ultimately make it feasible to characterize and monitor an individual's proteome with significant implications for personalized medicine[1]. Multiparameter protein characterization on a single molecule level may also reveal biochemically- or clinically-relevant static or dynamic heterogeneities, such as sub-populations of phosphorylated proteins, that are often hidden in ensemble measurements[61]. Moreover, real-time identification of single proteins might ultimately enable single molecule sorting in a fashion analogous to cell sorting.

Finally, this work focused on one of the most relevant and challenging applications of nanoscale shape approximation, namely the characterization of single proteins. The same approach may, however, apply to particles such as DNA origami[62], synthetic nanoparticles[63,64], and nanoparticle assemblies[65], whose characterization on a single particle level is important since they are typically more heterogeneous than proteins and since their charge, shape, volume, and dipole moment affect their assembly characteristics and function[66-68].

**References:**

1       Picotti, P. & Aebersold, R. Selected reaction monitoring-based proteomics: Workflows, potential, pitfalls and future directions. *Nat. Methods.* **9**, 555-566 (2012).
2       Herr, A. E. Disruptive by design: A perspective on engineering in analytical chemistry. *Anal. Chem.* **85**, 7622-7628 (2013).
3       Rusk, N. Disruptive nanopores. *Nat. Methods.* **10**, 35-35 (2013).
4       Sali, A., Glaeser, R., Earnest, T. & Baumeister, W. From words to literature in structural proteomics. *Nature* **422**, 216-225 (2003).
5       Debye, P. *Polar molecules*. 1 edn,  (Dover Publications Inc., 1929).
6       Oncley, J. L. The investigation of proteins by dielectric measurements. *Chem. Rev.* **30**, 433-450 (1942).
7       Felder, C. E., Prilusky, J., Silman, I. & Sussman, J. L. A server and database for dipole moments of proteins. *Nucleic Acids Research* **35**, W512-W521 (2007).
8       Chari, R., Jerath, K., Badkar, A. V. & Kalonia, D. S. Long- and short-range electrostatic interactions affect the rheology of highly concentrated antibody solutions. *Pharm. Res.* **26**, 2607-2618 (2009).
9       Mehl, J. W., Oncley, J. L. & Simha, R. Viscosity and the shape of protein molecules. *Science* **92**, 132-133 (1940).
10      Bonincontro, A. & Risuleo, G. Dielectric spectroscopy as a probe for the investigation of conformational properties of proteins. *Spectrochim. Acta. A.* **59**, 2677-2684 (2003).
11      Hughes, J. P., Rees, S., Kalindjian, S. B. & Philpott, K. L. Principles of early drug discovery. *Br. J. Pharmacol.* **162**, 1239-1249 (2011).
12      Movileanu, L., Howorka, S., Braha, O. & Bayley, H. Detecting protein analytes that modulate transmembrane movement of a polymer chain within a single protein pore. *Nat. Biotechnol.* **18**, 1091-1095 (2000).
13      Siwy, Z. *et al.* Protein biosensors based on biofunctionalized conical gold nanotubes. *J. Am. Chem. Soc.* **127**, 5000-5001 (2005).
14      Han, A. *et al.* Sensing protein molecules using nanofabricated pores. *Appl. Phys. Lett.* **88**, 093901 (2006).
15      Dekker, C. Solid-state nanopores. *Nat. Nanotechnol.* **2**, 209-215 (2007).
16      Wei, R., Gatterdam, V., Wieneke, R., Tampe, R. & Rant, U. Stochastic sensing of proteins with receptor-modified solid-state nanopores. *Nat. Nanotechnol.* **7**, 257-263 (2012).
17      Qin, Z. P., Zhe, J. A. & Wang, G. X. Effects of particle's off-axis position, shape, orientation and entry position on resistance changes of micro Coulter counting devices. *Meas. Sci. Technol.* **22** (2011).
18      Fologea, D., Ledden, B., David, S. M. & Li, J. Electrical characterization of protein molecules by a solid-state nanopore. *Appl. Phys. Lett.* **91**, 053901 (2007).
19      Sexton, L. T. *et al.* Resistive-pulse studies of proteins and protein/antibody complexes using a conical nanotube sensor. *J. Am. Chem. Soc.* **129**, 13144-13152 (2007).

20	Yusko, E. C. *et al.* Controlling protein translocation through nanopores with bio-inspired fluid walls. *Nat. Nanotechnol.* **6**, 253-260 (2011).
21	Robertson, J. W. F. *et al.* Single-molecule mass spectrometry in solution using a solitary nanopore. *Proc. Natl. Acad. Sci. U. S. A.* **104**, 8207-8211 (2007).
22	Reiner, J. E., Kasianowicz, J. J., Nablo, B. J. & Robertson, J. W. F. Theory for polymer analysis using nanopore-based single-molecule mass spectrometry. *Proc. Natl. Acad. Sci.* **107**, 12080-12085 (2010).
23	Reiner, J. E. *et al.* Disease detection and management via single nanopore-based sensors. *Chem. Rev.* **112**, 6431-6451 (2012).
24	Raillon, C. *et al.* Nanopore detection of single molecule RNAP-DNA transcription complex. *Nano. Lett.* **12**, 1157-1164 (2012).
25	Soni, G. V. & Dekker, C. Detection of nucleosomal substructures using solid-state nanopores. *Nano. Lett.* (2012).
26	Di Fiori, N. *et al.* Optoelectronic control of surface charge and translocation dynamics in solid-state nanopores. *Nat. Nanotechnol.* **8**, 946-951 (2013).
27	Yusko, E. C. *et al.* Single-particle characterization of Aβ oligomers in solution. *ACS Nano* **6**, 5909-5919 (2012).
28	German, S. R., Hurd, T. S., White, H. S. & Mega, T. L. Sizing individual au nanoparticles in solution with sub-nanometer resolution. *ACS Nano* (2015).
29	Nir, I., Huttner, D. & Meller, A. Direct sensing and discrimination among ubiquitin and ubiquitin chains using solid-state nanopores. *Biophys. J.* **108**, 2340-2349 (2015).
30	Stefureac, R. I., Kachayev, A. & Lee, J. S. Modulation of the translocation of peptides through nanopores by the application of an ac electric field. *Chem. Comm.* **48**, 1928-1930 (2012).
31	Li, J. *et al.* Ion-beam sculpting at nanometre length scales. *Nature* **412**, 166-169 (2001).
32	Golibersuch, D. C. Observation of aspherical particle rotation in Poiseuille flow via the resistance pulse technique. Part 1. Application to human erythrocytes. *Biophys. J.* **13**, 265-280 (1973).
33	Golibersuch, D. C. Observation of aspherical particle rotation in Poiseuille flow via the resistance pulse technique. Part 2. Application to fused sphere dumbbells. *J. Appl. Phys.* **44**, 2580-2584 (1973).
34	DeBlois, R. W., Uzgiris, E. E., Cluxton, D. H. & Mazzone, H. M. Comparative measurements of size and polydispersity of several insect viruses. *Anal. Biochem.* **90**, 273-288 (1978).
35	Deblois, R. W. & Wesley, R. K. A. Viral sizes, concentrations, and electrophoretic mobilities by nanopar analyzer. *Biophys. J.* **16**, A178-A178 (1976).
36	Smythe, W. R. Flow around a spheroid in a circular tube. *Phys. Fluids* **7**, 633-638 (1964).
37	Fricke, H. The electric permittivity of a dilute suspension of membrane-covered ellipsoids. *J. Appl. Phys.* **24**, 644-646 (1953).
38	Fricke, H. A mathematical treatment of the electrical conductivity of colloids and cell suspensions. *J. Gen. Physiol.* **6**, 375-384 (1924).
39	Velick, S. & Gorin, M. The electrical conductance of suspensions of ellipsoids and its relation to the study of avian erythrocytes. *J. Gen. Physiol.* **23**, 753-771 (1940).
40	Plesa, C. *et al.* Fast translocation of proteins through solid state nanopores. *Nano. Lett.* **13**, 658-663 (2013).
41	Hernandez-Ainsa, S. *et al.* Lipid-coated nanocapillaries for DNA sensing. *Analyst*, 104-106 (2013).
42	Woodside, M. T. *et al.* Direct measurement of the full, sequence-dependent folding landscape of a nucleic acid. *Science* **314**, 1001-1004 (2006).
43	Berge, L. I., Feder, J. & Jossang, T. in *Particle size analysis* (eds N.G. Stanley-Wood & R. W. Lines) 374-383 (The Royal Society of Chemistry, 1992).
44	Deen, W. M. Hindered transport of large molecules in liquid-filled pores. *AIChE Journal* **33**, 1409-1425 (1987).

45 Happel, J. & Brenner, H. *Low reynolds number hydrodynamics: With special applications to particulate media*. 1 edn, (Springer Netherlands, 1983).
46 Yuan, Y. & Axelrod, D. Subnanosecond polarized fluorescence photobleaching - rotational diffusion of acetylcholine-receptors on developing muscle-cells. *Biophys. J.* **69**, 690-700 (1995).
47 Weiss, S. Fluorescence spectroscopy of single biomolecules. *Science* **12**, 1676-1683 (1999).
48 Singh, S., Yadav, S., Shire, S. & Kalonia, D. Dipole-dipole interaction in antibody solutions: Correlation with viscosity behavior at high concentration. *Pharm. Res.* **31**, 2549-2558 (2014).
49 Rousseeuw, P. J. & Kaufman, L. *Finding groups in data: An introduction to cluster analysis*. (John Wiley & Sons, Inc., 1990).
50 Venkatesan, B. M. & Bashir, R. Nanopore sensors for nucleic acid analysis. *Nat. Nanotechnol.* **6**, 615-624 (2011).
51 Li, W. *et al.* Single protein molecule detection by glass nanopores. *ACS Nano* **7**, 4129-4134 (2013).
52 Kasianowicz, J. J., Brandin, E., Branton, D. & Deamer, D. W. Characterization of individual polynucleotide molecules using a membrane channel. *Proc. Natl. Acad. Sci. U. S. A.* **93**, 13770-13773 (1996).
53 Manrao, E. A. *et al.* Reading DNA at single-nucleotide resolution with a mutant MspA nanopore and phi29 DNA polymerase. *Nat. Biotechnol.* **30**, 349-353 (2012).
54 Branton, D. *et al.* The potential and challenges of nanopore sequencing. *Nat. Biotechnol.* **26**, 1146-1153 (2008).
55 Nivala, J., Marks, D. B. & Akeson, M. Unfoldase-mediated protein translocation through an [alpha]-hemolysin nanopore. *Nat. Biotechnol.* **31**, 247-250 (2013).
56 Jijin, Y. *et al.* Rapid and precise scanning helium ion microscope milling of solid-state nanopores for biomolecule detection. *Nanotechnology* **22**, 285310 (2011).
57 Rosenstein, J. K., Wanunu, M., Merchant, C. A., Drndic, M. & Shepard, K. L. Integrated nanopore sensing platform with sub-microsecond temporal resolution. *Nat. Methods.* **9**, 487-492 (2012).
58 Lee, M.-H. *et al.* A low-noise solid-state nanopore platform based on a highly insulating substrate. *Sci. Rep.* **4** (2014).
59 Uram, J. D., Ke, K. & Mayer, M. Noise and bandwidth of current recordings from submicrometer pores and nanopores. *ACS Nano* **2**, 857-872 (2008).
60 Issaq, H. J. & Veenstra, T. D. Two-dimensional polyacrylamide gel electrophoresis (2d-page): Advances and perspectives. *Biotechniques* **44**, 697 - 700 (2008).
61 De Pascalis, A. R. *et al.* Binding of ferredoxin to ferredoxin: Nadp+ oxidoreductase: The role of carboxyl groups, electrostatic surface potential, and molecular dipole moment. *Protein Sci.* **2**, 1126-1135 (1993).
62 Dietz, H., Douglas, S. M. & Shih, W. M. Folding DNA into twisted and curved nanoscale shapes. *Science* **325**, 725-730 (2009).
63 Jin, R. *et al.* Photoinduced conversion of silver nanospheres to nanoprisms. *Science* **294**, 1901-1903 (2001).
64 Cecchini, M. P. *et al.* Rapid ultrasensitive single particle surface-enhanced raman spectroscopy using metallic nanopores. *Nano. Lett.* **13**, 4602-4609 (2013).
65 Auyeung, E. *et al.* Synthetically programmable nanoparticle superlattices using a hollow three-dimensional spacer approach. *Nat. Nanotechnol.* **7**, 24-28 (2012).
66 Alivisatos, A. P. Semiconductor clusters, nanocrystals, and quantum dots. *Science* **271**, 933-937 (1996).
67 Kalsin, A. M. *et al.* Electrostatic self-assembly of binary nanoparticle crystals with a diamond-like lattice. *Science* **312**, 420-424 (2006).
68 Yoo, J. & Aksimentiev, A. In situ structure and dynamics of DNA origami determined through molecular dynamics simulations. *Proc. Natl. Acad. Sci. U. S. A.* **110**, 20099-20104 (2013).

69   Pedone, D., Firnkes, M. & Rant, U. Data analysis of translocation events in nanopore experiments. *Anal. Chem.* **81**, 9689-9694 (2009).
70   Talaga, D. S. & Li, J. L. Single-molecule protein unfolding in solid state nanopores. *J. Am. Chem. Soc.* **131**, 9287-9297 (2009).
71   Maxwell, J. C. *A treatise on electricity and magnetism*. 3rd edn, 435-441 (Clarendon Press, 1904).

**METHODS**

*Materials*. All phospholipids were obtained from Avanti Polar Lipids. Bis(succinimidyl) penta(ethylene glycol) (21581) was purchased from Thermo Scientific. Monoclonal anti-biotin IgG$_1$ (B7653), GPI-anchored acetylcholinesterase (C0663), glucose-6-phosphate dehydrogenase (G5885), L-lactate dehydrogenase (59747), bovine serum albumin (A7638), α-amylase (A4551), and streptavidin were purchased from Sigma Aldrich, Inc. Polyclonal anti-biotin IgG-Fab fragments (800-101-098) were purchased from Rockland and β-phycoerythrin (P-800) was purchased from Life Technologies.

*Methods of Nanopore-Based Sensing Experiments*. To sense proteins, we first formed a supported lipid bilayer containing either 0.15 mol% 1,2-dipalmitoyl-*sn*-glycero-3-phosphoethanolamine-N-capbiotinyl (biotin-PE) lipids or 1 mol% 1-palmitoyl-2-oleoyl-*sn*-glycero-3-phosphoethanolamine (POPE) lipid in a background of 1-palmitoyl-2-oleoyl-*sn*-glycero-3-phosphocholine (POPC) lipids (Avanti Polar Lipids, Inc.). We described details of the bilayer formation in Yusko et al.[20] The dimensions of all nanopores are shown in Supplementary Fig. 25. When biotin-PE lipids were present in the bilayer, we added a solution containing anti-biotin IgG$_1$, Fab, or GPI-anchored acetylcholinesterase to the top solution compartment of the fluidic setup such that the final concentration of protein ranged from 5 pM to 10 nM. When sensing GPI-anchored acetylcholinesterase, we started recording resistive pulses after incubating the bilayer-coated nanopore for 1 h with GPI-anchored acetylcholinesterase (where the solution was 150 mM KCl, 10 mM HEPES, pH = 7.4) to allow time for the GPI-lipid anchor of the protein to insert into the fluid lipid bilayer coating. When POPE lipids were present in the bilayer, we first dissolved bis(succinimidyl) penta(ethylene glycol), a bifunctional crosslinker, in a buffer containing 2 M KCl and 100 mM KHCO$_3$ (pH = 8.4) and immediately added this solution to the top compartment of the fluidic setup such that the final concentration of crosslinker was 10 mg/mL. After 10 min, we rinsed away excess crosslinker and subsequently added β-phycoerythrin, glucose-6-phosphate dehydrogenase, L-lactate dehydrogenase, bovine serum albumin, α-amylase, or butyrylcholinesterase dissolved in the same buffer as the preceding step to the top compartment such that final protein concentration ranged from 1 to 3 µM. After at least 30 min, we rinsed away excess protein and began recording. We recorded resistive pulses at an applied potential difference of -0.04 to -0.115 V with the polarity referring to the top fluid compartment relative to the bottom fluid compartment, which was connected to ground. The electrolyte contained 2 M KCl with either 10 mM HEPES at pH 6.5 for experiments with GPI-anchored acetylcholinesterase; 10 mM HEPES at pH 7.4 for experiments with IgG, Fab, α-amylase, butyrylcholinesterase, and streptavidin; 10 mM $C_6H_7KO_7$ at pH 5.1 for experiments with β-phycoerythrin; 10 mM $C_6H_7KO_7$ at pH 5.2 for experiments with bovine serum albumin; or 10 mM $C_6H_7KO_7$ at pH 6.1 for experiments with glucose-6-phosphate dehydrogenase and L-lactate dehydrogenase. We used Ag/AgCl pellet electrodes (Warner Instruments) to monitor ionic currents through electrolyte-filled nanopores with a patch-clamp amplifier (Axopatch 200B, Molecular Devices Inc.) in voltage-clamp mode (i.e. at constant applied voltage). We set the analog low-pass filter of the amplifier to a cutoff frequency of 100 kHz. We used a digitizer (Digidata 1322) with a sampling frequency of 500 kHz in combination with a program written in LabView to acquire and store data[59]. To distinguish resistive pulses reliably from the electrical noise, we first filtered the data digitally with a Gaussian low-pass filter ($f_c$ =15 kHz) in MATLAB and then used a modified form of the custom written MATLAB routine described in Pedone et al.[69,70]. We calculated the translocation time, $t_d$, as the width of individual resistive-pulse at half of their peak amplitude, also known as the full-width-half-maximum value[20,70]. From this analysis we obtained the $\Delta I$ and $t_d$ values for each resistive pulse, and we only analyzed $\Delta I$ values for resistive-pulses with $t_d$ values greater than 50 µs, since resistive pulses with translocation times faster than 50 µs have attenuated $\Delta I$ values due to the low-pass filter[20,69].

With regard to the success rate of the experiments reported here, we used a total of 68 different nanopores for this work and 21 of these nanopores (31%) yielded measurements. Experiments generally failed due to one of three reasons: First, the baseline current was lower than expected based on pore geometry and electrolyte conductivity prior to coating the nanopore with a lipid bilayer ($I_{baseline} < 0.9 * I_{expected}$), second, the nanopore did not coat, or third, the baseline current after coating was too noisy to detect translocation events. For the 68 nanopores, we obtained the expected baseline current in 73% of attempts, successfully coated the pore in 37% of attempts (cumulative success rate = 27%), and achieved sufficiently low noise for recording after successfully coating the pore in 46% of attempts (cumulative success rate = 12%). These statistics indicate that approximately 1 in 10 experiments yielded a measurement, on average. The success rate was, however, highly dependent on the nanopore chip being used: A subset of approximately 10 nanopores coated successfully in ~80% of attempts until they abruptly failed irreversibly at the first stage described above; on average this failure occurred after 16 experiments.

Supplementary Text and Figures are available in the online version of the paper.

**ACKNOWLEDGEMENTS**
We thank C.L. Asbury for several helpful discussions and review of the manuscript. This work was supported by a Miller Faculty Scholar Award (M.M), the Air Force Office of Scientific Research (M.M. and D.S., grant number 11161568), Oxford Nanopore Technologies (M.M., grant number 350509-N016133), the National Institutes of Health (M.M., grant number 1R01GM081705), the National Human Genome Research Institute (J.L., grant numbers HG003290 and HG004776), Professor J.Golovchenko's Harvard nanopore group for FIB pore preparation (J.L.), a Rackham Pre-Doctoral Fellowship from the University of Michigan (E.C.Y), and the Microfluidics in Biomedical Sciences Training Program from the NIH and BIBIB (B.R.B).

**AUTHOR CONTRIBUTIONS**
E.C.Y., B.R.B, and M.M conceived and designed experiments, analyzed data, and co-wrote the manuscript. E.C.Y., B.R.B., and O.M.E. performed nanopore experiments. R.R, N.W., N.S., A.R.H., and J.L fabricated nanopores. M.P, D.S.K., and B.R.B. measured the dipole moments of proteins with impedance spectroscopy and provided constructive feedback on the manuscript. D.S. performed computational analyses of several protein crystal structures and provided guidance on statistical methods used in the manuscript.

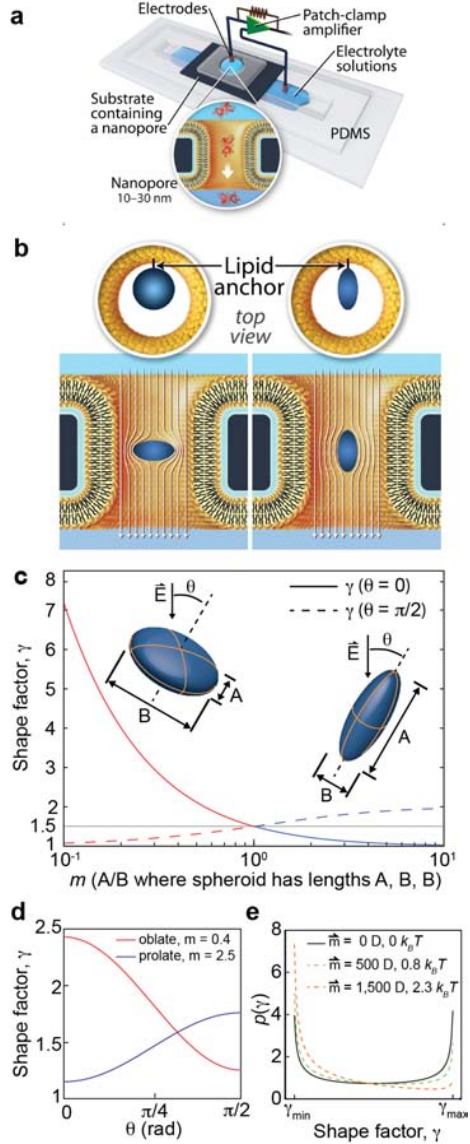

**Figure 1.** Rotational dynamics of individual proteins inside a nanopore reveal a spheroidal approximation of the protein's shape. (**a**) Experimental setup to measure resistive pulses from the translocation of individual proteins. (**b**) Top and side views of a nanopore illustrating the two extreme orientations of a spheroidal protein that is anchored to a fluid lipid coating on the pore wall. A crosswise orientation disturbs the field lines inside the pore more than a lengthwise orientation due to the angle-dependent electrical shape factor $\gamma$.[36] (**c**) Electrical shape factor $\gamma$ of spheroids (prolates in blue curves and oblates in red curves) as a function of their aspect ratio, $m$, for two extreme orientations: when the angle, $\theta$, between the axis of rotation of the ellipsoid relative to the electric field is 0, i.e. $\theta = 0$ (solid curves), and when $\theta = \pi/2$ (dashed curves). For reference, a sphere has a $m$ value equal to 1, and an electrical shape factor $\gamma$ of 1.5 that is independent of its angle $\theta$ (grey line).[32-34,36-38,71] (**d**) Shape factor as a function of $\theta$ for prolates with a defined $m$ value of 2.5 and oblates with an $m$ value of 0.4. (**e**) Bimodal probability distribution of shape factors, $p(\gamma)$, for spheroids without a dipole moment as predicted by Golibersuch (black curve)[32,33] and for spheroidal proteins with a dipole moment of 500 and 1500 Debye pointed parallel to the longest axis of the protein (dashed curves). For the different magnitudes of the dipole moment, the energy difference between $\theta = 0$ and $\theta = \pi/2$ is listed in units of $k_BT$ for a typical electric field of $2\times10^6$ V m$^{-1}$. See Supplementary Notes 2 and 9 for details.

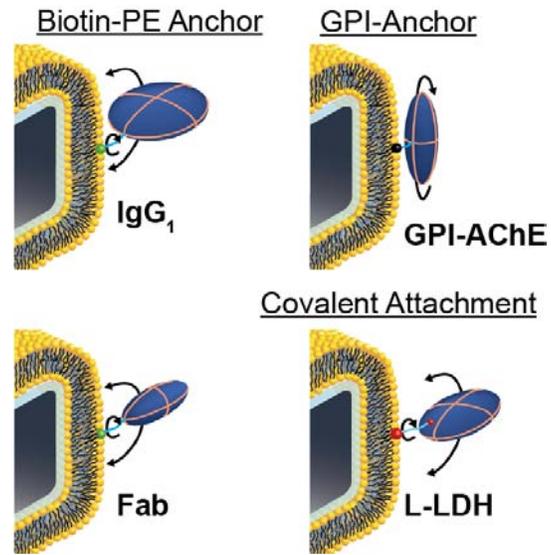

**Figure 2.** Three different strategies of anchoring proteins to the lipid coating used in this work to slow down translocation such that rotational diffusion of the proteins could be resolved in time. A lipid anchor with a biotin group selectively captured anti-biotin antibodies and Fab fragments, an intrinsic GPI anchor captured acetylcholinesterase, and a bi-functional, amine-reactive crosslinker provided a general strategy to attach proteins of interest covalently to ethanolamine lipids in the bilayer coating. All proteins analyzed in this work were tethered with a phospholipid anchor to the bilayer by one of these three strategies. These tethers were sufficiently long ($\geq$ 1.5 nm in their extended conformation) and flexible ($\geq$ 12 σ-bonds) and nanopore diameters were at least twice the volume-equivalent spherical diameter of the examined proteins, such that the proteins were able to rotate and sample all possible orientations.

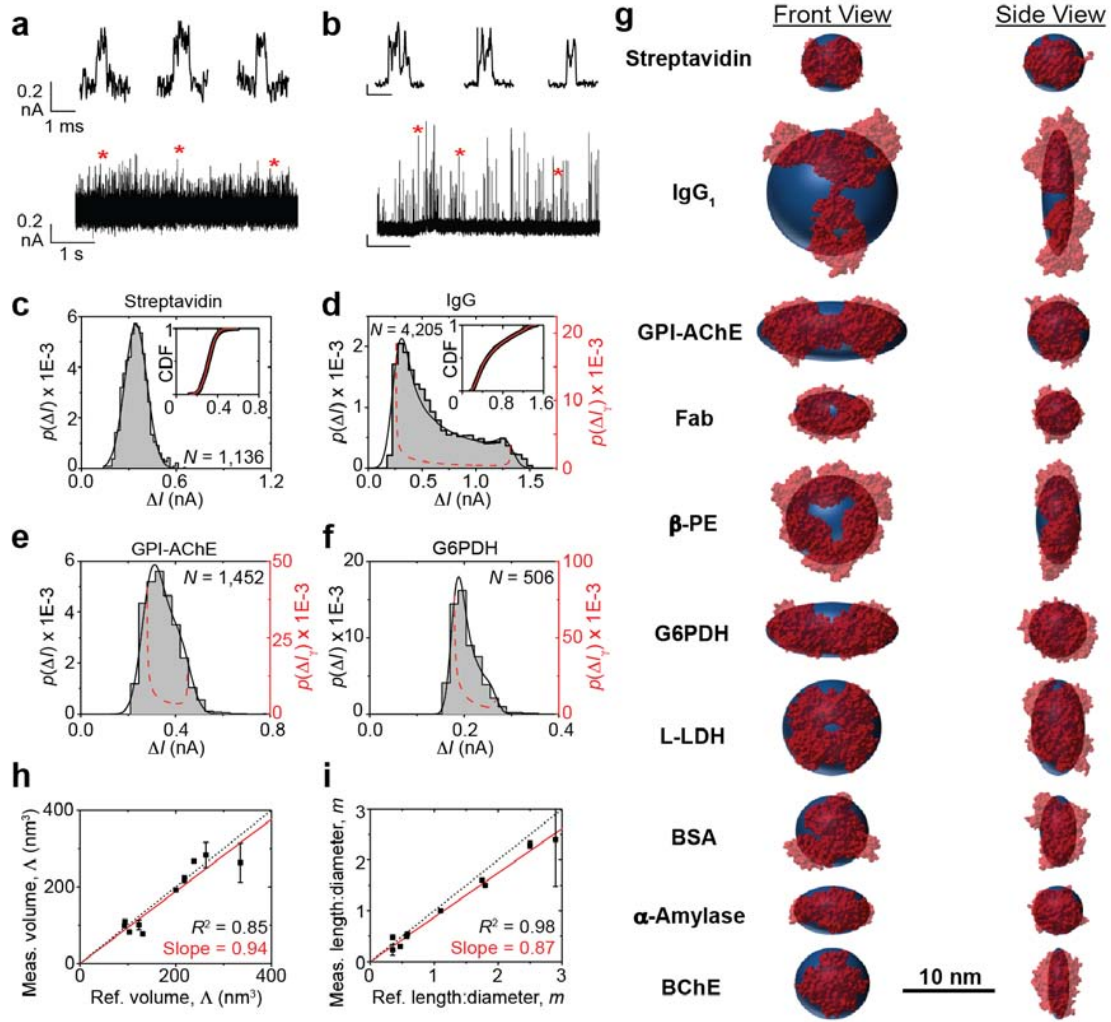

**Figure 3.** Determination of approximate protein shape and volume from histograms of maximum $\Delta I$ values from resistive pulse recordings. (**a, b**) Examples of original current traces as a function of time: upward spikes indicate individual resistive current pulses towards zero current due to the translocation of single streptavidin (**a**) or IgG (**b**) proteins. Resistive pulses marked by an asterisk are shown in detail above. (**c-f**) Histograms of maximum $\Delta I$ values from resistive pulse recordings with streptavidin (**c**), IgG$_1$ (**d**), GPI-AChE (**e**), and G6PDH (**f**) proteins. Black curves show the solution of the convolution model, $p(\Delta I)$, after a non-linear least squares fitting procedure, and red dashed curves show the estimated distribution of $\Delta I$ values due to the distribution of shape factors, $p(\Delta I_\gamma)$. Supplementary Table 1 lists the values of all fitting parameters and the electric field strength used in each experiment. Supplementary Note 2 and Supplementary Fig. 5-7 explain the convolution model and fitting procedure in detail and extend the analyses to all proteins characterized in this work. (**g**) Comparison of the approximate shape of ten proteins as determined by analysis of resistive pulses (blue spheroids) with crystal structures from the Protein Data Bank in red (streptavidin: 3RY1, anti-biotin immunoglobulin G$_1$: 1HZH, GPI-anchored acetylcholinesterase: 3LII, anti-biotin Fab fragment: 1F8T, β-phycoerythrin: 3V57, glucose-6-phosphate dehydrogenase: 4EM5, L-lactate dehydrogenase: 2ZQY, bovine serum albumin: 3V03, α-amylase: 1BLI, and butyrylcholinesterase: 1P0I). (**h**) Comparison of the measured volume by nanopore-based analysis with the expected reference volume. (**i**) Comparison of the measured length-to-diameter ratios $m$ of all proteins with the expected reference values of $m$. Error bars in **h**,**i** represent the standard deviation in most probable values from experiment-to-experiment or day-to-day.

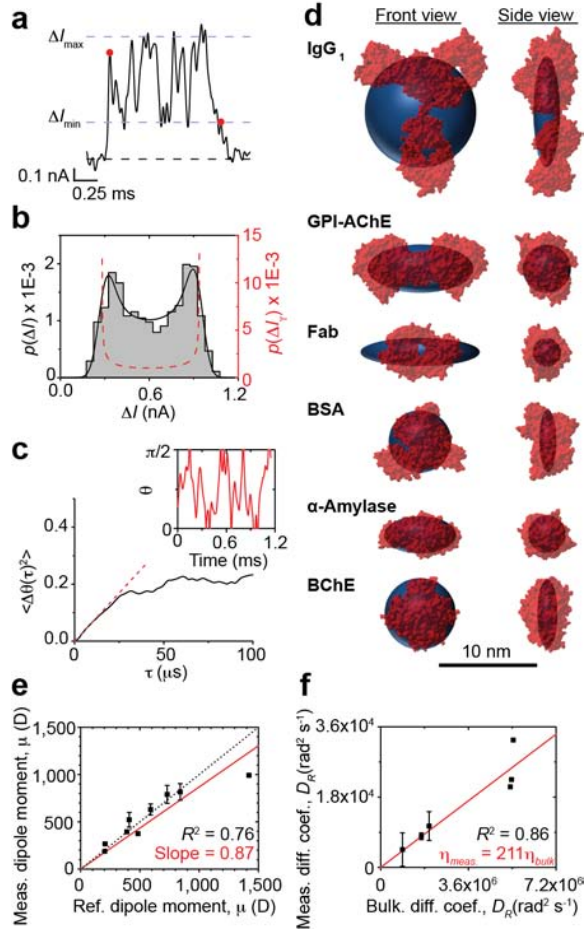

**Figure 4.** Approximate shape, dipole moment, and rotational diffusion coefficient obtained from current modulations *within individual* resistive pulses from the translocation of a single protein. (**a**) Resistive pulse from the translocation of a single IgG$_1$ molecule. Red dots mark the beginning and end of the resistive pulse as identified by an automated algorithm. (**b**) Distribution of all current values within this one resistive pulse. The black curve shows the solution of the convolution model, $p(\Delta I)$, after a non-linear least squares fitting procedure, and the red dashed curve shows the estimated distribution of $\Delta I$ values due to the distribution of shape factors, $p(\Delta I_\gamma)$. (**c**) Mean-square angular displacement curve (black trace) and the initial slope (dashed red line). The inset shows the transformation of intra-event $\Delta I(t)$ to $\theta(t)$. (**d**) Comparison of the approximate shape of proteins as determined by analysis of individual resistive pulses (blue) with crystal structures in red (blue spheroids show the median values of *m* and volume from single event analyses of each protein; see Supplementary Fig. 15 for complete distributions from the single event analyses). (**e**) The most frequently observed dipole moments (in ascending order) of G6PDH, L-LDH, α-amylase, β-phycoerythrin, BSA, Fab, GPI-AChE, IgG$_1$, and BChE agree well with expected reference values of their dipole moments. (**f**) The most frequently observed rotational diffusion coefficients (in ascending order) of IgG$_1$, β-phycoerythrin, GPI-AChE, BChE, Fab, and α-amylase agree well with the expected reference values. The signal-to-noise ratio for G6PDH, L-LDH, and BSA was too small to determine accurate values of $D_R$. Error bars in **e**,**f** represent the standard deviation in most probable values from experiment-to-experiment.

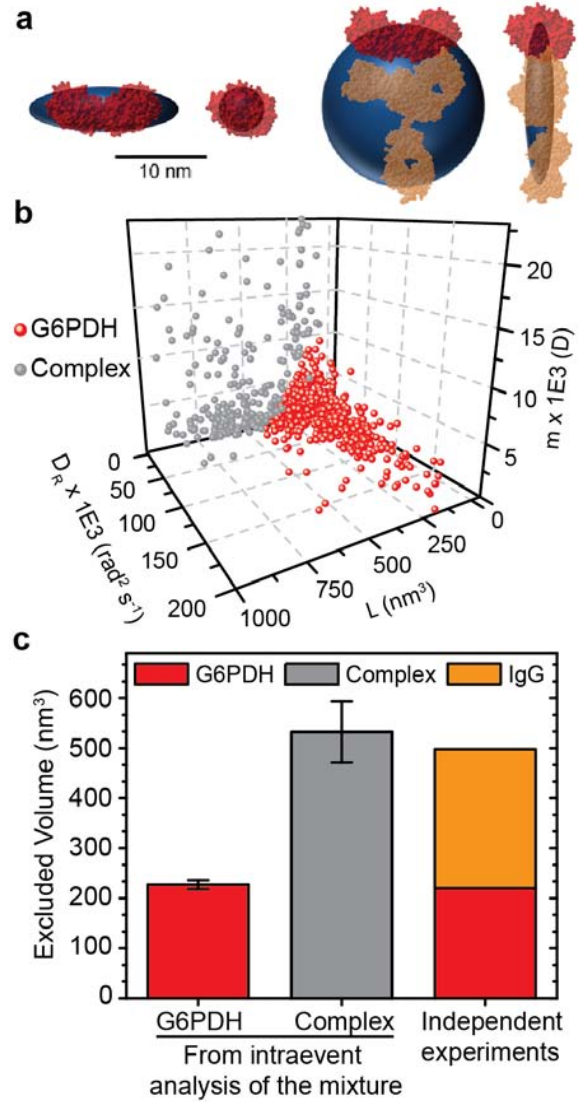

**Figure 5.** Fingerprinting of individual translocation events permits identification and characterization of G6PDH and a G6PDH-IgG complex from a mixture. (**a**) The volume, Λ, and approximate shape of G6PDH (left side) and G6PDH-IgG complex (right side) as determined by analysis of individual resistive pulses is similar to the crystal structures in red. Blue spheroids show the median values of $m$ and Λ determined from single event analyses and classification of each event. (**b**) Values for the volume, rotational diffusion coefficient, and dipole moment determined from individual events. The *kmeans* clustering algorithm in MATLAB classified single events as corresponding to a single G6PDH (red points) or to the G6PDH-IgG complex (grey points) (see Supplementary Note 8 and Supplementary Fig. 22). This single event classification estimated that 28% of events were due to the complex, which is nearly the same proportion of events estimated to be in the complex based on analysis of maximum Δ$I$ values from distributions of hundreds of resistive pulses (Supplementary Fig. 22). (**c**) The volume of G6PDH and the G6PDH-IgG complex determined by single-event analysis and classification of events from the mixture are nearly identical (< 10% deviation) to the sum of volumes obtained for G6PDH in an experiment without anti-G6PDH IgG and the volume of an individual IgG. Error bars represent the standard error of the median volume value (Supplementary Note 8).

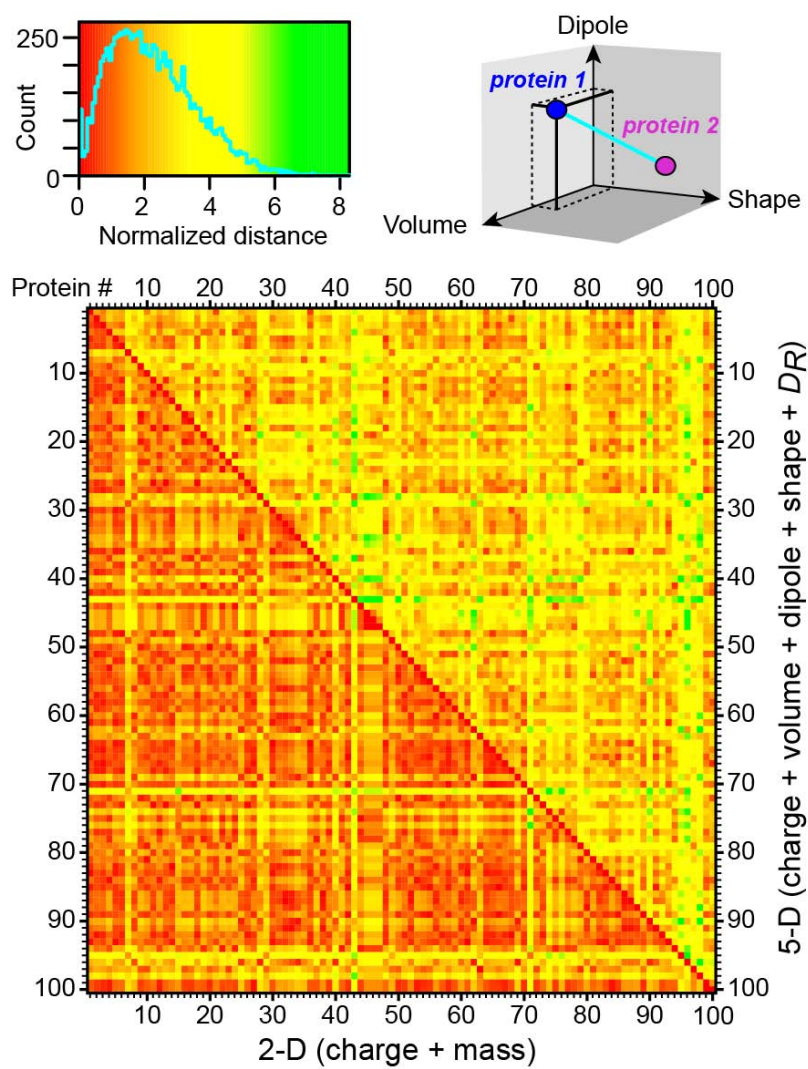

**Figure 6.** Advantage of 5-D fingerprinting over the standard 2-D characterization for protein identification. Using structural and sequence data from the Protein Data Bank, we randomly selected a group of proteins and determined their mass, volume, rotational diffusion constant, shape factor, dipole moment, and charge. Each parameter can be thought of as a dimension, and the heat map shows the separation between each pair of 100 randomly sampled proteins for two dimensions (lower left corner) or five dimensions (upper right corner) calculated using standard normal distributions for each descriptor. This separation is calculated as $\sqrt{\sum_{i=1}^{n} d_i^2}$ where $n$ is the number of dimensions and $d_i$ is the difference between the values of two different proteins in one parameter. Red squares mark protein-protein pairs that are similar in all descriptors (i.e. closely spaced), while yellow and green squares indicate increasing separation. Physical descriptors beyond protein charge and mass such as shape and dipole moment create additional dimensions and facilitate protein identification by increasing the separation between each protein-protein pair.

# Supplementary Text and Figures:

# Real-time shape approximation and 5-D fingerprinting of single proteins

Erik C. Yusko[1,†,‡], Brandon R. Bruhn[1,†], Olivia Eggenberger[1], Jared Houghtaling[1], Ryan C. Rollings[2], Nathan C. Walsh[2], Santoshi Nandivada[2], Mariya Pindrus[3], Adam R. Hall[4], David Sept[1,5], Jiali Li[2], Devendra S. Kalonia[3], Michael Mayer[1,6,*]

**Affiliations:**
[1]Department of Biomedical Engineering, University of Michigan, Ann Arbor, MI 48109, USA
[2]Department of Physics, University of Arkansas, Fayetteville, Arkansas 72701, USA
[3]Department of Pharmaceutical Sciences, University of Connecticut, Storrs, CT 06269, USA
[4]Department of Biomedical Engineering and Comprehensive Cancer Center, Wake Forest University School of Medicine, Winston Salem, NC 27157, USA
[5]Center for Computational Medicine and Biology, University of Michigan, Ann Arbor, MI 48109, USA
[6]Biophysics Program, University of Michigan, Ann Arbor, MI 48109, USA

[†]These authors contributed equally to this work.
[‡]Current address: Department of Physiology and Biophysics, University of Washington, Seattle, WA 98195, USA
[*]Correspondence should be addressed to M.M. (mimayer@umich.edu)

Table of Contents

**Supplementary Note 1. Control experiments indicate that broad distributions of Δ*I* values were not due to impurities or simultaneous translocations**

To confirm that the distributions of Δ*I* values during experiments with monoclonal anti-biotin IgG$_1$ antibodies were not affected by potential impurities in the solution, we performed three control experiments. In one control experiment, we competitively inhibited the binding of IgG$_1$ antibodies to the biotin-PE lipids on the surface by adding an excess concentration of soluble biotin to the aqueous solution of an ongoing experiment (Supplementary Fig. 2a and 2b). Fifteen minutes after the addition of the soluble biotin we observed the frequency of resistive pulses decrease from 34 s$^{-1}$ to 1.3 s$^{-1}$. In the second control experiment, we generated a lipid bilayer coated nanopore that did not contain biotin-PE lipids in the coating and therefore was not specific for the translocation of IgG$_1$ antibodies (Supplementary Fig. 2c). In this experiment, the concentration of the IgG$_1$ antibody was even higher (25 nM compared to 20 nM) than in the original experiment (Supplementary Fig. 2a), and the frequency of translocation events was 2 s$^{-1}$. Since the frequency of events is proportional to concentration, we estimated that if the concentration of IgG$_1$ in this control experiment was 20 nM, we would expect to observe an event frequency of approximately 1.6 s$^{-1}$. From these two control experiments, we estimated that during experiments with biotin-PE lipids in the bilayer coating only 3.8 to 4.7 % of translocation events were due to proteins that were not bound to biotin-PE lipids. Furthermore, almost all (~90%) of the translocation times calculated from resistive-pulses observed in control experiments (where binding to biotin-PE was not possible) were less than 50 μs, and we did not include resistive-pulses with translocation times less than 50 μs in the analysis of Δ*I* distributions because the amplitude would be attenuated due to electronic filtering[1,2]. Consequently, we concluded that the protein we detected in the solution of anti-biotin IgG$_1$ antibodies was bound to biotin-PE lipids specifically. In the final control experiment, we removed any fragments of IgG proteins (i.e. Fab fragments) or other proteins in the IgG stock solution by purifying the solution with a Protein A spin column (Thermo Scientific 89952). Using this purified solution in a nanopore-based sensing experiment, we observed distributions of Δ*I* values similar to those seen with the stock solution, and we determined the same volume and shape of anti-biotin IgG$_1$, within error (Supplementary Table 1). Results from these three control experiments indicate that the resistive pulses in experiments with IgG$_1$ were due to anti-biotin IgG$_1$ proteins and not due to the presence of other proteins or Fab fragments.

Since IgG antibodies can occasionally form dimers[3], we performed dynamic light scattering (DLS) experiments to characterize the hydrodynamic diameter of the IgG$_1$ antibodies. If dimers of IgG$_1$ antibodies were present in solution and contributing to the bimodal distribution of Δ*I* values in Fig. 3 of the main text, we would expect them to be reflected in DLS experiments in a significant fraction because approximately ½ of the resistive pulses had Δ*I* values that can be attributed to either of the bimodal peaks in the distribution of Δ*I* values. Consequently, if dimers were present, we would expect to observe two peaks in the distributions of estimated hydrodynamic diameters of the particles (proteins in this case) in DLS experiments[3]. Supplementary Fig. 2d shows that we only observed one peak corresponding to a hydrodynamic diameter of 10.5 ± 2.0 nm. This value is in good agreement with previously published hydrodynamic diameters of IgG antibodies of 10.9 to 11.0 nm, which were determined in physiologic buffers[3,4]. As additional evidence, we added urea to a concentration of 8 M to denature the IgG protein and disassociate potential aggregates. Again we only observed one peak corresponding to a hydrodynamic diameter of 12.9 ± 2.7 nm (Supplementary Fig. 2d). This hydrodynamic diameter is larger because of the random-coil and ball-like structure of denatured

IgG$_1$ antibodies compared to their native, oblate-shaped structures[3]. Thus, the results presented in Supplementary Fig. 2 confirm that dimers of IgG$_1$ antibodies were not responsible for the bimodal distribution of $\Delta I$ values and that the IgG$_1$ antibodies were stable and functional in the buffered solutions used here.

      GPI-anchored acetylcholinesterase purified from human erythrocyte membranes naturally occurs in a dimeric, prolate-shaped form that is held together by disulfide bonds near the C-terminal tail of the protein[5-9]. To confirm that the GPI-AChE used in this work remained in dimeric form and to detect impurities in solution, we performed a SDS-PAGE experiment. We ran three lanes on the SDS-PAGE gel corresponding to three different treatments of the protein: incubation with SDS, incubation with SDS and β-mercaptoethanol to dissociate the disulfide bond, and incubation with only β-mercaptoethanol to assess whether the disulfide bond in the folded protein was accessible to β-mercaptoethanol as reported in literature[9]. After staining, we observed only one protein band in each lane. When the protein was denatured with only SDS, we observed the dimeric form that ran with an apparent molecular weight of ~140 kDa. In both lanes where the protein was treated with β-mercaptoethanol, we observed only one protein band, running at an apparent molecular weight of ~60 kDa. These apparent molecular weights are lower than the values in reported literature of 160 kDa for the dimer and 80 kDa for the monomer because GPI-AChE is an amphiphilic protein and likely has a higher binding capacity for SDS than more commonly run soluble proteins[5-11]. This increased binding of SDS yields a greater charge to mass ratio and therefore greater migration speed of the protein compared to most soluble proteins, causing GPI-AChE to migrate in the gel as if it had a lower molecular weight. This phenomenon is well known for amphiphilic proteins[12].

      The fact that we only observe one band in each lane of the gel indicates that our samples contained high concentrations of GPI-AChE relative to other contaminants. In the lanes treated with β-mercaptoethanol the absence of a band at ~140 kDa coincident with the appearance of a band at ~60 kDa is consistent with breakage of the disulfide bond holding the dimer together. Moreover, as reported in literature, the disulfide bond was accessible in the native structure of the protein, as indicated by the appearance of a single band at the monomer molecular weight when the protein was treated only with β-mercaptoethanol (and no SDS) prior to running the gel[9]. Consequently, this gel confirms that the GPI-AChE protein in our sample was in its natural dimeric, prolate-shaped form. Moreover, the control experiments in Supplementary Fig. 2a-c indicate that if there were soluble (i.e. not lipid-anchored) contaminants in the solution, they would not be detected, since soluble proteins would not be concentrated on the lipid surface or slowed during translocation through the nanopore.

      To rule out the possibility that the widely distributed $\Delta I$ values were due to two proteins passing through the nanopore simultaneously, we compared the frequency of translocation events with the translocation times for each protein[13]. In the case of streptavidin translocations, we observed approximately 45 translocation events per second and a most-probable translocation time of about 115 μs. Consequently, on average there was a 0.52% probability of a molecule occupying the nanopore at any time, and the probability of two streptavidin proteins occupying the nanopore at the same time would be 0.003%. In the case of the IgG$_1$ translocation events, the maximum frequency we observed was approximately 30 events per second and a most probable translocation time of about 55 μs. Consequently, on average there was a 0.16% probability of an IgG$_1$ protein occupying the nanopore at any time, and the probability of two IgG$_1$ proteins occupying the nanopore at the same time would then be 0.0027%. Even if the first translocation event of an IgG antibody would be exceptionally long lived (e.g. 1000 μs), the probability of a

second antibody to enter the pore during that time would still only be around 3% at an average translocation frequency of 30 Hz. This analysis neglects steric effects, which we expect would be significant given the size of an IgG$_1$ antibody and the dimensions of the nanopores. For GPI-anchored acetylcholinesterase the estimated probability of a two proteins being in the nanopore at the same time was 0.000036%. For Fab fragments, β-phycoerythrin, glucose-6-phosphate dehydrogenase, L-lactate dehydrogenase, BSA, α-amylase, and butyrylcholinesterase, the estimated probability was less than 0.00001%.

Even during the resistive-pulse sensing experiments with streptavidin in which we estimated the highest probability of observing a protein in the nanopore, we did not observe resistive-pulses with multiple current levels that might suggest the translocation of two proteins simultaneously. Consequently, we conclude that the resistive pulses due to each protein detected in this work resulted from the translocation of one protein at a time.

**Supplementary Note 2. Determining the volume and shape of proteins from fitting distributions of maximum Δ$I$ values**

*Equation relating the amplitude of resistive pulses to the volume and electrical shape factor of particles*

The relationship between the magnitude of Δ$I$ and the volume of a particle stems from Maxwell's derivation[14], and it is shown in equation (1)[15-18].

$$\frac{\Delta I}{I} = -\frac{4 \Lambda \gamma}{\pi d_P^2 (l_P + 0.8 d_P)} S\left(\frac{d_M}{d_P}\right) \Rightarrow \Delta I = -\frac{\Lambda V_A \gamma}{\rho (l_P + 0.8 d_P)^2} S\left(\frac{d_M}{d_P}\right), \quad (1)$$

where $\gamma$ is the electrical shape factor[16,19-23], $\Lambda$ (m$^3$) is the excluded volume of the particle, $l_P$ (m) is the length of the pore, $d_P$ (m) is the diameter of the pore, Δ$I$ (A) is the magnitude of the change in the current during translocation of a particle, $I$ (A) is the baseline current, $V_A$ (V) is the applied voltage, and $\rho$ (Ω m) is the resistivity of the electrolyte. $S\left(\frac{d_M}{d_P}\right)$ is a correction factor applied when the diameter of the particle, $d_M$, approaches the diameter of the pore, $d_P$, (i.e. $d_M > 0.5\, d_P$)[15,16]. Under these conditions the electric field in the pore is additionally distorted between the particle and the pore walls resulting in a non-linear increase in the resistance with increasing particle volume[15,16]. Qin *et al.* recently reviewed these correction factors and showed that the most accurate correction factor for all $d_M/d_P$ ratios was developed by Smythe[24] and Deblois et al.[15], equation (2)[25]:

$$S\left(\frac{d_M}{d_P}\right) = \frac{1}{1 - 0.8\left(\frac{d_M}{d_P}\right)^3}. \quad (2)$$

Note that in the majority of resistive-pulse sensing literature, particles and proteins have been considered spherical and consequently $\gamma$ was set to a value of 1.5 and $\Lambda$ was constrained to equal $\frac{1}{6}\pi d_M{}^3$. Substituting these values into equation (1) simplifies it to the more commonly seen form in equation (3)[14,15,17,18,23,25]:

$$\frac{\Delta I}{I} = -\frac{d_M{}^3}{d_P^2(l_P + 0.8 d_P)} S\left(\frac{d_M}{d_P}\right) \Rightarrow \Delta I = -\frac{\pi V_A d_M{}^3}{4\rho(l_P + 0.8 d_P)^2} S\left(\frac{d_M}{d_P}\right), \quad (3)$$

Since in this work we analyzed resistive-pulses due to the translocation of non-spherical proteins and we expected $d_M$ to be less than ½ $d_P$, we set the correction factor to a value of 1[2,17,18]. We used equation (1) and expressed the impeded flow of ions through the nanopore during protein translocation events as reductions in current, $\Delta I$.

The volume exclusion model shown in equation (1) has yielded accurate estimates of volume in a number of prior publications[17,22,26-30]; however, it has also been inadequate under a variety of different experimental conditions[31-35]. The model fails to describe certain current pulses because it does not account for heterogeneity in the distribution of ions, and thus the conductivity of the solution, in the nanopore. Heterogeneity in the distribution of ions results from electrostatic interactions with the surface of the pore and translocating particle. For instance, Lan *et al.* observed biphasic current pulses resulting in part from the accumulation of chloride ions on one side of the particle[31]. In this case, the flow of chloride ions around the particle was inhibited as the particle and pore were both negatively charged. To determine whether such effects are likely to occur under the experimental conditions used here, we performed finite-element simulations nearly identical to those described by Lan *et al.* Supplementary Fig. 9a shows similar local variations in the conductivity of the solution to those reported by Lan *et al.* at a low ionic strength of 10 mM KCl due to the accumulation and depletion of chloride ions on opposite sides of the protein. In contrast, the conductivity of the solution is nearly constant at the high ionic strength of 2 M KCl that we used in the experiments presented here (Supplementary Fig. 9b). In this case, the $\Delta I$ signature (see Supplementary Fig. 9c) is well described by the volume exclusion model shown in equation (1). Consequently, the volume exclusion model is appropriate under the experimental conditions used in this work.

*Electrical shape factor and distributions of shape factors*

The electrical shape factor has been reported in literature since Maxwell derived equations to describe the conductance of solutions that contain insulating (i.e. non-conducting) spheres[14]. Maxwell considered both the volume fraction of the spheres in solution and the deformation of the electric field around these spheres. To account for the distortion of the electric field, Maxwell derived a scaling factor that is dependent on the shape of the insulating particles (i.e. electrical shape factor) and equal to 3/2 or 1.5 for spheres. Several years later, Fricke derived the electrical shape factor for spheroids, and Velick and Gorin developed analytical equations to describe the shape factor for ellipsoids of a general shape[36-38]. In 1954, Smythe numerically tested Maxwell's theory for the specific case of a particle residing in a pore; this work verified the electrical shape factor of 1.5 for spheres as well as the methods described by Fricke, Velick, and Gorin[24]. Around the same time, many groups experimentally proved these theories during resistive-pulse sensing experiments with holes that were micrometers in diameter while sensing various micrometer-sized particles[16,19,23,39,40]. In 1973, Golibersuch observed the rotation of red blood cells within the pore of a resistive-pulse sensor and derived the distribution for electrical shape factors to explain the periodic variations in $\Delta I$ that occurred during the rotation of the blood cell.

The mathematical descriptions for shape factors are analogous among many systems and can be used to describe how electric and magnetic fields deform around insulating particles as well as how ideal fluids flow around obstacles in wind tunnels or in aqueous solutions with laminar flow[19,24,39]. Spheres alter flow and electric fields to the same extent regardless of their orientation; however, spheroid particles alter these fields to a different extent depending on their orientation relative to the direction of the field. Thus, the electrical shape factor is a function of a particle's shape and orientation.

To relate the value of $\Delta I$ to the volume and shape of non-spherical proteins, we considered the possible values of the electrical shape factor, $\gamma$, with the condition that a protein may have an oblate, prolate, or spherical shape. Oblates and prolates have an axis of revolution (shown as the dashed blue line in Fig. 1 of the main text) with length A and secondary axes with length B. Golibersuch elegantly pointed out that equation (4) describes the electrical shape factor, $\gamma$, for these ellipsoids as a function of the angle between the axis of symmetry and the electric field, $\theta$, (Fig. 1)[19,39]:

$$\gamma(\theta) = \gamma_\perp + (\gamma_\parallel - \gamma_\perp)\cos^2(\theta) \tag{4}$$

where $\gamma_\parallel$ and $\gamma_\perp$ are the electrical shape factors when the axis of symmetry is parallel to the electric field (i.e. $\theta = 0, \pi, ...$) and perpendicular to the electric field (i.e. $\theta = \pi/2, 3\pi/2, ...$), respectively. Equation (4) implies that the shape factor for any orientation will range between the values of $\gamma_\parallel$ and $\gamma_\perp$. These factors, $\gamma_\parallel$ and $\gamma_\perp$, are related to the well-described depolarization factors for ellipsoids, $n_\parallel$ and $n_\perp$, by equation (5) and are a function of the length to diameter ratio, $m = A/B$, of an ellipsoid[19,20,40,41].

$$\gamma_\parallel = \frac{1}{1-n_\parallel} \text{ and } \gamma_\perp = \frac{1}{1-n_\perp} \tag{5}$$

where $n_\parallel$ for a prolate spheroid with $m = A/B > 1$ is described by equation (6):

$$n_\parallel = \frac{1}{m^2-1}\left[\frac{m}{\sqrt{m^2-1}}\ln\left(m+\sqrt{m^2-1}\right)-1\right] \tag{6}$$

and $n_\parallel$ for an oblate spheroid with $m = A/B < 1$ is described by equation (7):

$$n_\parallel = \frac{1}{1-m^2}\left[1-\frac{m}{\sqrt{1-m^2}}\cos^{-1}(m)\right] \tag{7}$$

and $n_\perp = (1 - n_\parallel)/2$[19,23,41].

To derive the distribution of shape factors, we assume that ellipsoidal proteins rotate freely such that all angles of $\theta$ are equally likely when $\Delta I$ is measured. By symmetry, values of $\theta$ range between 0 and $\pi/2$. According to Golibersuch, these assumptions enable using substitution of variables to write a probability distribution function for electrical shape factors $P(\gamma)$ based on the probability of observing a certain orientation $P(\theta(\gamma))$, where $\theta$ is a function of $\gamma$ (equation (8))[19]:

$$P(\gamma)d\gamma = P[\theta(\gamma)]\frac{d\theta}{d\gamma}d\gamma \tag{8}$$

Since, by symmetry, values of $\theta$ range between 0 and $\pi/2$ and we assumed that all angles of $\theta$ are equally likely, we solved for $P(\theta)$ by noting that the integral of a probability distribution function equals 1:

$$\int_0^{\pi/2} P(\theta)d\theta = 1 = \int_0^{\pi/2}\frac{2}{\pi}d\theta \Rightarrow P(\theta)d\theta = \frac{2}{\pi}d\theta \tag{9}$$

Combining equation (8) with (9), we obtained:

$$P(\gamma)d\gamma = \frac{2}{\pi}\left(\frac{d\gamma}{d\theta}\right)^{-1}d\gamma. \tag{10}$$

Differentiating equation (4) with respect to $\theta$ (i.e. $\frac{d\gamma}{d\theta}$) and combining the result with equation (10), we obtained a probability density function for the possible shape factors[19].

$$P(\gamma)d\gamma = \frac{1}{\pi\left[(\gamma - \gamma_\perp)(\gamma_P - \gamma)\right]^{1/2}}d\gamma \tag{11}$$

Fig. 1e of the main text (black line) shows that this probability density function (equation (11)) is bimodal and symmetric with peaks at $\gamma_\parallel$ and $\gamma_\perp$. The bimodal character of this distribution reflects the fact that for small deviations in $\theta$ near 0 and near $\pi/2$, there is little change in the value of the shape factor compared to deviations in $\theta$ around $\pi/4$ (Fig. 1d).

    Before attempting to describe the non-Normal distributions of $\Delta I$ values as a consequence of $p(\gamma)$, we considered whether the non-spherical proteins could sample various orientations, and therefore shape factors, in these experiments as well as whether the time-scale of rotation would bias the measurement of maximum $\Delta I$ values. We first considered potential steric limitations on the orientations of the proteins in the nanopore. Figure 2 in the main text shows the expected lipid anchoring locations for the anti-biotin IgG antibody, anti-biotin Fab fragment, GPI-AChE[5]. Since the chemical linker between the lipid head group and the ligand for the IgG$_1$ and Fab fragments was approximately 1.5 nm in length, we expect the anchoring positions shown in Fig. 2 to permit rotation of the proteins in orientations that could generate the minimum and maximum shape factors. We attached the remaining non-spherical proteins characterized in this work to the bilayer *via* a homobifunctional crosslinker with a flexible, 2.2-nm-long polyethylene glycol spacer arm. Since the crosslinker reacted with primary amines (e.g. lysines and glutamines), the anchoring locations on these proteins were randomly distributed across their surface. Consequently, we also expect these proteins to sample the full range of electrical shape factors while passing through the nanopore.

We next examined whether the dipole moment of a protein may align completely in the large electric field in the nanopore (~$10^6$ V m$^{-1}$). Combining the potential energy, $\Delta U$, of a dipole moment in an electric field and the Boltzmann distribution of energies while assuming that the dipole moment was pointed parallel to one of the principal axes of the protein, we expanded on Golibersuch's probability distribution of shape factors to develop a $p(\gamma)$ for a protein with a dipole moment (Fig. 1e in the main text and Supplementary Note 9). To expand on the theories developed by Golibersuch, we considered the possible probability distribution of shape factors if the orientation of the protein were biased by the electric field in the nanopore. The electric field in the nanopore is on the order of $10^6$ V m$^{-1}$, and consequently, we expect the orientation of a protein to be biased by alignment of its dipole moment, $\vec{\mu}$ (Debye ≈ 3.33564×10$^{-30}$ C m), in the electric field, $\vec{E}$ (V m$^{-1}$). Taking into account the potential energy of a dipole in an electric field, $\Delta U = \vec{E} \cdot \vec{\mu} = -E\mu\cos(\phi)$, using the Boltzmann distribution of energies, and assuming the dipole was aligned along the symmetry or equatorial axis, we derived equations (12a) and (12b), respectively (Supplementary Note 9). Equations (12a) and (12b) describe probability distribution functions of shape factors for spheroid proteins when their orientation is biased by the dipole energy in an electric field.

$$P(\gamma)d\gamma = \frac{1}{A}\cosh\left[\frac{E\mu\left(\frac{\gamma - \gamma_\perp}{\gamma_P - \gamma_\perp}\right)^{1/2}}{k_BT}\right]\left[\frac{1}{\pi\left[(\gamma - \gamma_\perp)(\gamma_P - \gamma)\right]^{1/2}}\right]d\gamma \quad (12a)$$

$$P(\gamma)d\gamma = \frac{1}{A}\cosh\left[\frac{E\mu\left(\frac{\gamma - \gamma_P}{\gamma_\perp - \gamma_P}\right)^{1/2}}{k_BT}\right]\left[\frac{1}{\pi\left[(\gamma - \gamma_\perp)(\gamma_P - \gamma)\right]^{1/2}}\right]d\gamma \quad (12b)$$

In equations (12a) and (12b), $A$ is a normalization constant described in Supplementary Note 9. Fig. 1e of the main text demonstrates that for spheroid proteins with dipoles of several thousand Debyes, it is theoretically possible to observe a bimodal distribution of shape factors. The average dipole moments of proteins is approximately 550 Debye (http://bioinfo.weizmann.ac.il/dipol/indexj.html), suggesting that many non-spherical proteins should generate a skewed bimodal distribution of shape factors. Additional factors may bias the orientation of proteins in the nanopore including steric effects, interactions with the pore wall, and alignment of slender proteins prior to entering the nanopore. All of these factors could affect the estimated value of $\Delta U$ or µ in this model. Therefore, an alternative interpretation of these parameters is that they describe the overall bias of the protein's orientation toward $\theta = 0$ or $\pi/2$. Equations (12a) and (12b) cannot describe distributions of $\Delta I$ accurately for proteins that are significantly biased (i.e. $\Delta U$ > ~4 k$_B$T or µ > ~3000 D for a typical pore at 100 mV applied potential) toward intermediate orientations relative to the electric field (i.e. $\theta = \pi/4$). Under these circumstances, the model would not resolve $\Delta I_{min}$ and $\Delta I_{max}$ accurately, underestimating the shape of the protein (i.e. $m$ would approach 1) and overestimating the volume of the protein. Consequently, equations (12a) and (12b) are an approximation of how the orientation, and therefore distribution of shape factors, of a protein with a dipole moment may be biased, and they allow the theoretical distribution of shape factors to become asymmetric.

We also considered whether the orientation of the protein would be significantly biased due to the hydrodynamic drag force, which is orientation dependent for non-spherical particles. To this end, we calculated the drag for an oblate with a relatively extreme shape (4 x 16 x 16 nm) when its axis of symmetry is aligned perpendicular and parallel to the direction of fluid flow (i.e. the direction of translational motion). Assuming the pore is 25 nm long and the protein transits this distance in 100 μs, the average speed of the proteins is 2.5 x $10^{-4}$ m $s^{-1}$, and the corresponding orientation-dependent drag force would range between 26 and 33 fN[42]. Based on these forces, the difference in energy required to move the protein through the entire length of the pore varies by a maximum of roughly 0.04 $k_B T$. As a result, we do not expect hydrodynamic drag to significantly bias the orientation of the protein.

Finally, we considered whether the proteins would rotate in the pore too quickly to be time resolved or whether their rotation would bias the measurement of Δ$I$ values such that we would only observe Δ$I$ values corresponding to $γ_{max}$, and therefore, not resolve Δ$I$ values corresponding to $γ_{min}$. Axelrod observed that GPI-AChE has rotational diffusion coefficients, $D_R$, of 10,000 ± 4,000 $rad^2$ $s^{-1}$ and Timbs *et al.* have observed dramatically reduced mobility (i.e. $D_R$ ≈ 0.003 $rad^2$ $s^{-1}$) of IgG antibodies binding to lipids in a substrate-supported monolayer[43-45]. Consequently, we estimate that the average time for a protein to rotate π/2 radians to be at least 125 μs. Since the majority of the translocation times in these experiments were between 50 and 100 μs (Supplementary Fig. 2), we expect the majority of Δ$I$ values to reflect a single orientation or a very limited range of orientations of the protein in the nanopore. Consequently, we expect the bimodal distributions of Δ$I$ values observed here to reflect accurately the underlying distribution of shape factors with modes at $γ_{min}$ and $γ_{max}$[19]. This prediction is supported by our recent discovery of bimodal distributions of Δ$I$ values from translocation of a single, pure protein[2] and subsequent observations made by Raillon *et al.*[22].

Since the value of Δ$I$ is directly proportional to the electrical shape factor, $γ$, according to equation (1), we expressed equations (12a) and (12b) in terms of Δ$I$. For an oblate this procedure results in equations (13a) and (13b), where the parameters Δ$I_{min}$ and Δ$I_{max}$ correspond to $γ_{min}$ and $γ_{max}$.

$$P(\Delta I_\gamma)d\Delta I_\gamma = \frac{1}{A}\cosh\left[\frac{E\mu\left(\frac{\Delta I - \Delta I_{max}}{\Delta I_{min} - \Delta I_{max}}\right)^{1/2}}{k_B T}\right]\left[\frac{1}{\pi\left[(\Delta I - \Delta I_{max})(\Delta I_{min} - \Delta I)\right]^{1/2}}\right]d\Delta I_\gamma \quad \textbf{(13a)}$$

and

$$P(\Delta I_\gamma)d\Delta I_\gamma = \frac{1}{A}\cosh\left[\frac{E\mu\left(\frac{\Delta I - \Delta I_{min}}{\Delta I_{max} - \Delta I_{min}}\right)^{1/2}}{k_B T}\right]\left[\frac{1}{\pi\left[(\Delta I - \Delta I_{max})(\Delta I_{min} - \Delta I)\right]^{1/2}}\right]d\Delta I_\gamma. \quad \textbf{(13b)}$$

For a prolate, equations (13a) and (13b) are interchanged. These probability distributions are the expected distributions of Δ$I$ values due only to the possible values of the shape factor – they do not include effects that broaden the measured values of Δ$I$ such as current noise as well as other experimental or analytical errors in determining Δ$I$ values.

*Fitting the convolution model to distributions of ΔI values*

To account for experimental and analytical errors in determining $\Delta I$ values, we convolved the expected distribution of $\Delta I$ values due to variation in the electrical shape factor, $p(\Delta I_\gamma)$ (equations (13a) and (13b)), with a Normal distribution, $p(\Delta I_\sigma)$, to generate the a distribution of $\Delta I$ values that one expects to observe experimentally, $p(\Delta I)$. We used this theoretical distribution (herein referred to as the "convolution model") to fit all of the empirical distributions of $\Delta I$ values presented in this work. Supplementary Figure 5 illustrates this method.

When constructing empirical distributions of $\Delta I$ values from many translocation events, we represented each event by its maximum $\Delta I$ value as opposed to its average (e.g. Fig. 3 in the main text). We followed this strategy because representing events by their average value causes bias towards intermediate $\Delta I$ values and may introduce an additional mode besides the two expected modes at $\Delta I_{min}$ and $\Delta I_{max}$, which would result in an improper fit with the convolution model (Supplementary Fig. 5). On the other hand, representing events by their maximum value likely biases the distribution of $\Delta I$ values toward $\Delta I_{max}$ such that the amplitude of the peak corresponding to $\Delta I_{max}$ increases. In this instance, however, the location of $\Delta I_{max}$ and $\Delta I_{min}$ should be preserved such that the shape of the protein can still be determined accurately.

Since the distribution of $\Delta I$ values resulting from the distribution of shape factors, $p(\Delta I_\gamma)$, is different depending whether the dipole moment is assumed to be parallel to the symmetry or equatorial axis of the protein (equations (13a) and (13b), respectively), we fit each empirical distribution of $\Delta I$ values, $P(\Delta I)$, with both of the resulting solutions to the convolution model. Subsequently, we selected the fit that yielded the larger adjusted $R^2$ value as the correct solution. Since the orientation of the dipole moment dictates the preferred orientation of the protein, this procedure effectively determined whether the distribution of $\Delta I$ values was skewed towards $\Delta I_{min}$ or $\Delta I_{max}$.

When fitting the distributions of $\Delta I$ values for Fab, α-amylase, and BChE, we excluded outliers from the upper end of the distributions to determine their shape correctly. For each distribution, we excluded $\Delta I$ values that were greater than a threshold value, which we chose such that the $R^2$ value of the fit with the convolution model was maximized. Conversely, $\Delta I$ values were not excluded for any of the other proteins detected in this work. Finally, we low-pass filtered the data for BChE at 10 kHz as opposed to 15 kHz in order to improve the signal-to-noise ratio.

*Using ΔI$_{min}$ and ΔI$_{max}$ to solve for the volume and shape of proteins*

Given that the probability distribution of shape factors has modes at $\gamma_\parallel$ and $\gamma_\perp$ corresponding to either $\Delta I_{min}$ or $\Delta I_{max}$ values according to equation (1), we expected that if the value of $\Delta I_{min}$ and $\Delta I_{max}$ could be determined quantitatively from the empirical distribution of $\Delta I$ values then the volume and shape of a protein could also be determined. For example, the minimum shape factor for an oblate spheroid occurs at $\theta = \pi/2$ and has a value of $\gamma_\perp(m)$ (equation (4)). Thus, according to equation (1), the minimum mode in the bimodal $\Delta I$ distribution, $\Delta I_{min}$, is a function of $\Lambda$ and $\gamma_\perp(m)$, and the maximum mode in the bimodal $\Delta I$ distribution, $\Delta I_{max}$, is a function of $\Lambda$ and $\gamma_\parallel(m)$. Since both $\gamma_\parallel$ and $\gamma_\perp$ are solely a function of $m$, we developed the system of equations (14) and (15) in which the values of $m$ and $\Lambda$ are the only two unknowns and the values of $\Delta I_{min}$ and $\Delta I_{max}$ are determined from fitting the empirical distributions of $\Delta I$ with the convolution model. By rearranging equation (3), we can write for oblate spheroids with $m < 1$:

$$\Lambda(m) = \begin{cases} \Lambda(\gamma_\perp(m), \Delta I_{min}) \\ \Lambda(\gamma_\parallel(m), \Delta I_{max}) \end{cases} \text{ if } m < 1, \tag{14}$$

and for prolate spheroids with $m > 1$:

$$\Lambda(m) = \begin{cases} \Lambda(\gamma_\parallel(m), \Delta I_{min}) \\ \Lambda(\gamma_\perp(m), \Delta I_{max}) \end{cases} \text{ if } m > 1. \tag{15}$$

Since this system of equations has a piecewise dependence on the value of $m$, we substituted the determined values of $\Delta I_{min}$ and $\Delta I_{max}$ into equations (14) and (15) and used MATLAB to solve the system for the excluded volume of the protein, $\Lambda$, and the value of $m$. For all prolates and relatively spherical oblates, two solutions to this system of equations exist as shown in Supplementary Fig. 7. The solutions for all experiments are summarized in Supplementary Table 1.

For many of the fits, the value of $\sigma$ is reasonable given the standard deviation of the baseline noise, which was typically between 20 and 60 pA. On the other hand, several of the fits returned relatively low estimates of $\sigma$ (e.g. α-Amylase using Pore 10), which may be a result of using maximum $\Delta I$ values to represent long events or due to partial truncation of the $\Delta I$ distributions since only values larger than a certain threshold were detected. Nevertheless, the excellent agreement between the estimated volume of the proteins and their respective shapes (Supplementary Table 4) provide strong evidence that this procedure enables one to approximate the shape and determine the volume of non-spherical proteins by analyzing the distributions of maximum $\Delta I$ values. This method does not assume any information about the protein to extract the parameters shown in Supplementary Table 1.

While results for $m$ and $\Lambda$ from different pores are in good agreement for G6PDH and BSA (<10% difference in $m$ and <20% difference in $\Lambda$), we observed significant pore-to-pore variability for GPI-AChE and IgG$_1$ (Supplementary Table 4). Using all of the 9 possible pore-to-pore comparisons from the results presented in Supplementary Table 1, we found that pore-to-pore variability in $m$ and $\Lambda$ is weakly correlated with differences in pore diameter and length (i.e. -0.3 ≤ Pearson's $r$ ≤ 0.3). In fact, we observed the lowest pore-to-pore variability in $m$ for G6PDH despite the fact that the pores used to characterize this protein have the largest difference in radii of any of the possible pore-to-pore comparisons. Based on these results, it appears that pore-to-pore variability of determined $m$- and $\Lambda$-values does not depend on pore diameter or length. This variability is likely due to variations in the pore geometry that are not accounted for in the model. The model assumes that the pore is perfectly cylindrical (i.e. constant diameter); however, nanopores prepared by ion-beam sculpting generally have an hourglass shape[46]. Even if the maximum $\Delta I$ value for each event is obtained when the protein is centered about the narrowest constriction of the pore, $\Delta I_{min}$ and $\Delta I_{max}$ will vary with the degree of pore tapering. Moreover, sterics may also introduce pore-to-pore variability by restricting certain protein orientations or conformations, particularly for IgG$_1$ since it is a relatively large protein and composed of three domains that move relative to one another. Based on these arguments, it is perhaps unsurprising that we find the largest pore-to-pore variability for the determination of the shape factor, *m*, for IgG. However, even in this most challenging case with a large protein whose shape deviates significantly from an ellipsoid of rotation, the standard deviation of *m*-values is smaller than ±50%, while it is smaller than ±40% for AChE and ±6% for G6PDH and BSA.

*Estimating the volume of spheroidal proteins via dynamic light scattering*

For comparison to the nanopore-based method, we used the technique of dynamic light scattering (DLS) to estimate the volume of each protein detected in this work. We assumed that the proteins were either spherical or spheroidal in shape in order to calculate their volume from the hydrodynamic radius, $r_H$, returned from the DLS measurements (Supplementary Table 3). For spheroidal proteins, we used the length-to-diameter ratio, $m = a/b$, of the particle (listed in Supplementary Table 1) with the corresponding Perrin shape factor, *S*, to calculate the volume based on the following equation[47]:

$$D = \frac{k_B T}{6\pi\eta r_H} = \frac{k_B T}{f_{sphere} \, 2m^{2/3}/S} = \frac{k_B T}{\left[6\pi\eta(ab^2)^{1/3}\right] 2m^{2/3}/S}$$

where $f_{sphere}$ is the friction coefficient of a sphere with the same volume as a spheroid with semi-axes *a*, *b*, and *b*. Furthermore, *S* for an oblate spheroid is equal to:

$$S_{oblate} = \frac{2\tan^{-1}\varepsilon}{\varepsilon}$$

and *S* for a prolate spheroid is equal to:

$$S_{prolate} = \frac{2\tanh^{-1}\varepsilon}{\varepsilon}$$

where $\varepsilon = \sqrt{|a^2 - b^2|}/a$. Combining the above equations yields:

$$r_{H,oblate} = \frac{\sqrt{|a^2 - b^2|}}{\tan^{-1}\left(\frac{\sqrt{|a^2 - b^2|}}{a}\right)}$$

and

$$r_{H,prolate} = \frac{\sqrt{|a^2 - b^2|}}{\tanh^{-1}\left(\frac{\sqrt{|a^2 - b^2|}}{a}\right)}.$$

We solved the preceding two equations numerically in MATLAB to determine the dimensions of each spheroidal protein and calculated the corresponding volume. The resulting spheroidal volumes were in excellent agreement with the volumes that we determined by fitting the convolution model to distributions of Δ*I* values. For reference, we used the crystal structures of these proteins to determine their length-to-diameter ratio, *m*, and subsequently determine their spheroidal volume; these volumes were also in excellent agreement with volumes obtained from analysis of DLS and resistive-pulse sensing experiments (Supplementary Table 3). In contrast, if we assumed the particles were a perfect sphere, the volumes that we determined from the hydrodynamic radius were overestimated for every non-spherical protein. These experiments provide additional evidence that the methods we present in this paper accurately describe the distribution of Δ*I* values for determining the shape and volume of spheroidal proteins.

*Low applied potentials yield consistent estimates of protein shape*

The value of the shape parameter, *m*, determined from fitting distributions of maximum Δ*I* values for IgG$_1$ and GPI-AChE is consistent at relatively low applied potentials but decreases or increases, respectively, as the applied potential is increased (Supplementary Fig. 8). This deviation might result from deformation of the protein due to the electrophoretic force acting on it while in the nanopore as was observed by Freedman *et al.*[48]; however, the amount of deformation that is expected based on theory (see proceeding subsection) is not large enough to account for the change in *m* observed here. Furthermore, Pelta *et al.* have previously shown that proteins do not change shape under similar electric field intensities[49]. Alternatively, this deviation could be due to changes in the size and shape of the hydration shell surrounding the protein or increasing alignment of the protein in the electric field gradient prior to entering the pore with increasing field intensity. In response to this observation, we limited our analyses to distributions of Δ*I* values that were obtained at relatively low potentials where the distributions appeared to be resolved fully. This approach consistently returned accurate estimates of the shape and volume of non-spherical proteins (Supplementary Table 1).

*Forces acting on proteins in a nanopore*

Since the magnitude of the electric field is on the order of $10^6$ V m$^{-1}$ in the nanopore, we considered theoretically whether it was possible for the shape of proteins to be affected by forces in the nanopore. In this work, we expect proteins in the nanopore to be subjected to the following forces:

1) Instantaneous forces due to collisions with water will be on the order of ~500 pN with a net force equal to 0 on time scales of roughly 1 ps[50].
2) Net torque due to the dipole moment in the electric field will have magnitudes similar to thermal energy. Fig. 1e in the main text shows that we expect the torque on a protein to be on the order of 0 to 4 $k_B T$ in this work.
3) Average force due to the net charge of the protein in the electric field, $F_q$, is in the range of 0.1 to 4 pN for the electric field strengths and net charges of proteins used in this work.
4) Average force on the protein due to viscous drag in the aqueous solution, $F_w$, which opposes the electrophoretic force. We approximated this force to range from about 0.026 to 0.033 pN.
5) Average force on the lipid anchor, $F_L$, which also opposes the electrophoretic force, is thus on the order of 0.1 to 4 pN, since the force due to drag in the aqueous solution is negligible (i.e. $F = 0 = F_q - F_w - F_L$).

Since we expect these five forces to be nearly constant through the length of the nanopore, the shape of proteins in the nanopore should also be constant. This expectation is based on the fact that the internal stiffness of a protein and the viscosity of the solution result in highly over-damped motion of the protein. Any external force that affects the global conformation of a protein results in a gradual deformation of the protein toward its equilibrium conformation over a period of nanoseconds and without oscillations. In other words: "the global motions of proteins, especially less rigid ones, are highly overdamped: They creep rather oscillate when subject to applied forces".[50]

The largest constant force listed here is the possible tension within a protein due to the electrophoretic force acting on the net charge of the protein and the opposing drag force exerted by the lipid anchor. To estimate the deformation of the protein acted upon by a net force of 4 pN, we note that the Young's modulus ($E$) of most rigid proteins is on the order of 1 GPa[50]. Considering a protein similar in size and shape to GPI-AChE (e.g. a cross sectional area, $A$, of 5 nm x 5 nm and a length, $L$, of 13 nm), the total deformation (i.e. change in length) of the protein in response to a force of 4 pN is:

$$\Delta L = F * L / (E * A) = 4 \text{ pN} * 13 \text{ nm} / (1 \text{ GPa} * 5 \text{ nm} * 5 \text{ nm}) = 2 \text{ pm} = 0.002 \text{ nm}$$

This estimate for the deformation of a protein due to 4 pN of force illustrates that forces due to the electric field are unlikely to deform the proteins used in this work.

Proteins in the nanopore may also experience transient collisions with the pore wall in which the average force acting on the protein during the collision is equal to the rate of change in momentum of the protein. To estimate this force conservatively, we consider that a 100 kDa protein has an instantaneous velocity of 8.6 m s$^{-1}$ (this velocity is indeterminable on short time scales due to collisions with water molecules, corresponding to about 2 ps or 0.024 nm of distance traveled)[50] and that it collides directly with the nanopore wall, bouncing straight back with the same speed. If this protein collides with the pore wall over a period of 1 ps, the average force acting on the protein during that collision would be roughly 1 nN. This approximation estimates that the protein would deform between 0.076 nm to 0.52 nm depending on which face of the protein struck the wall. Again these deformations are small compared to the sizes of the protein we used in this work.

What kind of deformations might take place if the forces were far larger than we estimate? Suppose the forces acting on the protein in the nanopore did work equal to ~30 k$_B$T (1.23 E-19 Joules); this energy is ten times larger than the energy we estimate for the protein's dipole moments within the electric field of the nanopore. The deformation of the protein can be estimated by considering the stiffness of the protein, $k = E * L$.[50] Using the dimensions of the hypothetical protein that we described in the previous paragraphs, the stiffness of the protein to be $k = 1$ GPa * 13 nm = 13 N m$^{-1}$. Since the energy of a spring is ½$k\Delta x^2$, we can estimate the deformation, $\Delta x$, of the protein to be on the order of 0.14 nm. Consequently, we do not expect rigid proteins to deform significantly due to forces in the nanopore.

Proteins with multiple domains and flexible connecting regions may change shape in the nanopore, however their motion will be overdamped and not subject to oscillatory changes while in the nanopore. As an example, consider IgG$_1$ which has three separate domains that move relative to one another. Similarly, myosin head-groups are linked to the rest of the protein through a flexible domain known to have a stiffness of 4 pN nm$^{-1}$ (0.004 N m$^{-1}$)[50-52]. Using this stiffness, we estimated the maximum distance that the domains of IgG$_1$ might stretch relative to one another by considering the maximum applied force acting on the molecule of 4 pN. Under this force, IgG$_1$ may stretch on average roughly 1 nm. Because we expect these forces to remain constant through the nanopore and since the global motions of proteins are highly overdamped, especially for flexible proteins, this deformation would be nearly constant through the length of the nanopore.

Based on the magnitude of the forces discussed above, we do not expect the proteins used in this work to change shape significantly while in the nanopore. This expectation is supported by the accurate measurements of the size and shape of the ten different proteins detected in this work compared to the size and shape of these proteins as determined from crystal structures (Fig. 3 in the main text and Supplementary Table 4). In further support of this expectation, proteins that bound non-covalently to biotinylated lipids (IgG$_1$, Fab, and streptavidin) translocated through pores in the bound, lipid-anchored state as confirmed by their distributions of translocation times and measured charges (see Supplementary Fig. 11); if the binding pockets were denatured, antigen-binding would likely not occur.

*Description of the assumptions underlying the convolution model*

The following section describes the primary assumptions underlying the convolution model in particular with regard to their validity. To derive this model (i.e. equation), we made four key assumptions:

1) The protein is a spheroid.
2) The dipole moment of the spheroidal protein is aligned with one of the principal axes.
3) While residing in the nanopore, the orientation of the protein is only biased due to its dipole moment.
4) The pore is perfectly cylindrical.

The first assumption states that the protein is a spheroid with principle axes having lengths *A, B,* and *B* (see Fig 2a). We examined approximately 1,000 randomly sampled proteins from the Protein Data Bank and found that the lengths of two of the three principal axes are less than 20% different on average, indicating that most proteins can be approximated as spheroids. Based on our results for IgG$_1$, we have also shown that our approach can be used to characterize proteins with highly irregular shapes. Although the complexity of the shape of IgG$_1$ is not captured in full, our approach still provides low-resolution shape information and yields accurate values for the dipole moment and rotational diffusion coefficient of the protein.

The second assumption is based on the expectation that the dipole moment is most often aligned with a principle axis of a spheroidal protein. For an asymmetrical protein, we expect the dipole moment to be aligned along the longest axis of the protein because the residues that are furthest from the center of the protein contribute most to the magnitude of the dipole moment. For a multimeric protein with rotational symmetry, such as GPI-AChE or β-PE, we expect the dipole moment to lie along the axis of symmetry since the off-axis components of the dipole moment from each subunit cancel each other out. Thus, in both cases it seems reasonable that the dipole moment will be in near alignment with one of the principal axes of the protein. In support of this expectation, we found that the dipole moment was aligned close to one of the principal axes for each of the nine non-spherical proteins examined here by using the Weizmann server to analyze the protein crystal structures.

The third assumption states that the orientation of a protein in the nanopore is only biased by its dipole moment. We expect this to be true since we coated the nanopores with a lipid bilayer to eliminate non-specific interactions[2], anchored the proteins to the coating *via* long ($\geq$ 1.5 nm) and flexible ($\geq$ 12 σ-bonds) tethers so they could sample most orientations, and used nanopore diameters that were at least twice the volume-equivalent spherical diameter of the proteins to minimize steric effects. Under these conditions, we obtained dipole moment measurements for nine different proteins that were in excellent agreement with reference values (see Fig. 4e), supporting our assumption. We expect this assumption to be valid as long as the protein being characterized does not interact with lipids in the nanopore coating. In such cases, however, interactions with the coating could likely be avoided by modifying the bilayer composition (e.g. including a small fraction of PEG-conjugated lipids).

The fourth assumption is that the pore is cylindrical. This assumption does not depend on the protein under investigation and, consequently, does not limit the general applicability of our approach toward other proteins. The good agreement between the measured and expected values of volume for the ten different proteins examined here (see Fig. 3h) supports this assumption. Additionally, the change in the baseline current observed upon coating a nanopore is generally close to the value predicted by theory in which the pore geometry is assumed to be cylindrical[2], further supporting this assumption. Regardless, we discuss how pores that are not perfectly cylindrical may affect the analysis of intra-event $\Delta I$ values in Supplementary Note 6 below.

## Supplementary Note 3. Interpretation of the observed bimodal distributions of $\Delta I$ values from the translocation of non-spherical proteins

To determine whether any explanation might exist for the bimodal distributions of $\Delta I$ values observed here besides the theory presented in Supplementary Note 2, we closely examined the literature to ascertain whether other groups had observed similar signals in nanopore-based sensing experiments resulting from alternative mechanisms. To the best of our knowledge, there is only one such report. In this study, Spiering *et al.* used optical tweezers to characterize the force response of threading a protein bound to a negatively-charged DNA molecule through a nanopore[53]. For a finite range of optical trap positions, the authors found that the potential landscape "exhibits two minima (potential wells), corresponding to two metastable "states"… with the charged protein on either side of the membrane," resulting in bimodal force *versus* time signals. Since the protein is located *outside* of the pore in both of these states (i.e. where the electric field is negligible), the resulting $\Delta I$ values should be close to zero in resistive-pulse sensing experiments. Hence, we do not think that these two states do not correspond to the two modes in the $\Delta I$ distributions that we observe. The potential landscape in our experiments is different than that described by Spiering *et al.* due to the following reasons: (1) the lipid tethers are shorter (length ~1.5 nm) than the pore length (~30 nm) and hence cannot contract and extend to allow a protein to transition from one side of the pore to the other, (2) the charge of the protein-lipid complex is dominated by the charge of the protein rather than the tether, whereas in the case of a DNA tether, the opposite is true, (3) the lipid tethers only extend on one side of the protein instead of both sides as with the DNA tethers, and (4) there is no optical trap potential present in our experiments. Hence, it is extremely unlikely that the two metastable states described by Spiering *et al.* exist under the experimental conditions used in our work.

Skewed and bimodal distributions of $\Delta I$ values have been observed with increasing frequency in the last few years and in each instance the authors suggested that the shape of the $\Delta I$ distributions may have been influenced by the shape and orientation of the macromolecule. For example, early indications that the shape and orientation of a macromolecule can affect the $\Delta I$ signal were reported by Mathé *et al.*, who observed orientation-dependent translocation signals of DNA through α-hemolysin pores[54], and Fologea *et al*. who observed a unimodal but skewed distribution of Δ*I* values due to the translocation of nodular fibrinogen proteins through nanopores[26]. More recently, Raillon *et al*. observed distributions of Δ*I* values that appeared to be bimodal due to the translocation of an untethered, non-spherical RNA polymerase through a nanopore; without additional quantification, the authors attributed this result to different orientations of the RNA polymerase[22]. Finally, Fiori *et al*. observed a bimodal distribution of Δ*I* values due to the translocation of untethered, prolate-shaped protein ubiquitin[55]. Together, these reports indicated that the bimodal distributions presented in our work do not result from the effect of the lipid tether on the potential landscape but rather the shape and orientation of the translocating proteins. Until the work presented here, however, the origin of these biomodal distributions was not understood and it was unknown whether useful information could be obtained from the shape of these distributions of ΔI values.

While we considered a number of other possible explanations for the current signatures that we observe (see Supplementary Note 1 and the subsection titled "Forces acting on proteins in a nanopore" in Supplementary Note 2), eight observations indicate that Δ*I* reflects the rotational dynamics of proteins passing through the nanopore:

1) Streptavidin, which is spherical with a shape factor, *m*, of 1.1, yielded a Normal distribution of Δ*I* values (Fig. 3c in the main text).
2) The values of $\Delta I_{min}$ and $\Delta I_{max}$ that we determined for each protein are consistent with the values predicted by established theory for large particles; Golibersuch originally developed this theory to describe the periodic variations in Δ*I* that occurred during the rotation of a red blood cell within a resistive-pulse sensor.
3) Simulations based on a spheroidal particle undergoing biased random rotation in one dimension yield Δ*I* signals that are comparable to those that we obtained experimentally (Supplementary Note 5).
4) The values of volume (Λ), length-to-diameter ratio (*m*), rotational diffusion coefficient ($D_R$), and dipole moment (μ) that we determined for 9 different proteins are in good agreement with expected values (Fig. 3g-i and Fig. 4e-f); the methods used to determine these parameters critically depend on the assumption that Δ*I* reflects the orientation of non-spherical proteins, as described by the theory in Supplementary Note 2.
5) $D_R$ of bivalently-bound IgG$_1$ is significantly less than $D_R$ of monovalently-bound IgG$_1$ (see Supplementary Note 7), indicating that Δ*I* reflects the rotational dynamics of the protein in the nanopore.
6) Translocation of IgG$_1$ and GPI-AChE through the same nanopore result in markedly different distributions of Δ*I* values despite their similar molecular weights, indicating that Δ*I* is related to the shape of the protein.
7) Kolmogorov Smirnov (KS) tests indicate that the convolution model, which incorporates the effect of protein shape and orientation combined with noise to predict distributions of Δ*I* values, is not significantly different from the empirical distribution in 11 out of 13 cases (Supplementary Fig. 6), indicating that the model that underlies our analysis and employs the effect of protein orientation and shape on Δ*I* describes the data very well.
8) From a fundamental physical chemistry perspective it is also reasonable to assume that proteins rotate while moving through the pore. In the case of non-spherical proteins this rotation will change the electric field lines and hence modulate the current based on Maxwell's and Golibersuch's equations. This expectation is supported by simulations (see Point 3 of this list). We think it is extremely unlikely that proteins translocate through the pores in one constant orientation over several hundreds of microseconds given that we demonstrated before that the fluid bilayer coating circumvents non-specific protein adsorption to the pore walls.

**Supplementary Note 4. Effect of lipid anchoring on the measurement of protein properties**

Since we anchored each protein to a lipid in the bilayer coating of the nanopore to slow down translocation, we considered whether anchoring may have any other effects on the five parameters measured in this work. First, we do not expect protein shape or volume to be affected by anchoring. As discussed in Supplementary Note 2, the force exerted by the lipid anchor that opposes the electrophoretic force is unlikely to deform the protein. In addition, the chemical modifications involved in the crosslinking procedure are unlikely to cause denaturation as such modifications are standard practice in various biochemical assays that rely on retention of protein function. Our expectation that protein shape and volume are unaffected by anchoring is supported by the excellent agreement between the measured size and shape of the ten different proteins detected in this work with reference values (Fig. 3g-i in the main text).

We do expect the distribution of translocation times to reflect the net charge, $z$, of the protein-lipid complex as a whole. Hence, we subtracted 1 from the expected value of $z$ (Supplementary Figure 11j and Table 4) for each protein (except GPI-AChE) to correct for the net charge of the lipid anchor. For each protein that was crosslinked to the bilayer, we also subtracted 0.93 from the expected value of $z$ based on the "charge regulation" model by Menon and Zydney[56] to account for the consumption of a positively charged amine group.

Tethering a protein to a lipid anchor is known to slow rotation significantly, which we exploited in order to resolve in time the rotational dynamics of proteins residing in the nanopore. Proteins in free solution generally have rotational diffusion coefficients, $D_R$, that are on the order of $10^6$ to $10^7$ rad$^2$ s$^{-1}$ [57,58], while lipid anchored proteins have been shown to rotate over 2 orders of magnitude more slowly[43-45]. For instance, GPI-AChE rotates about 199 times slower when tethered to the bilayer than in bulk (see Supplementary Table 4). Likewise, our measurements indicate that the proteins examined here rotate 211 times slower on average when tethered (see Fig. 4f). There is a strong correlation (Pearson's $r = 0.93$) between the measured and bulk values of $D_R$, indicating that the factor by which tethering slowed rotation was comparable between proteins. This result is likely due to the fact that all proteins were attached to the bilayer by similarly long and flexible tethers.

We do not expect the tether itself to bias protein orientation and thereby affect the measurements of dipole moment, $\mu$; however, the crosslinking reaction consumes a positively charged amine and thus will affect $\mu$. To determine the extent by which crosslinking and removal of the positively charged amine affect $\mu$, we modified the crystal structure for BSA (PDB ID: 3V03) by replacing a single, randomly-chosen lysine residue on the protein surface with a glycine residue and calculated $\mu$ for the modified protein using the Weizmann server (http://bioinfo.weizmann.ac.il/dipol/). We found that the median percent difference between $\mu$ for the native protein and 10 modified versions of the protein was approximately 12 percent and ranged from 1 to 38 percent. In line with this relatively small change, we observed good agreement between the values of $\mu$ determined in nanopore experiments and those measured with impedance spectroscopy (Fig. 4e).

Finally, we examined whether the point of attachment of the lipid anchor may affect the ability of the proteins to sample all possible electrical shape factors (corresponding to $0 \leq \theta \leq \pi/2$ in Fig. 1c). To maximize the possibility that the proteins could sample all orientations ($\theta$) regardless of the method or point of attachment, we used long ($\geq 1.5$ nm in their extended conformation) and flexible ($\geq 12$ σ-bonds) lipid tethers as well as nanopore diameters that were at least twice the volume-equivalent spherical diameter of the examined proteins. Despite these precautions, we do expect points of attachment near the axis of symmetry on oblates to restrict certain orientations and thereby introduce some degree of error into our approach. While such cases are possible when using our crosslinking strategy in which the point of attachment is approximately random, we expect these cases to occur relatively infrequently. For instance, we anticipate that these cases occur in ~20% of the attachments for L-LDH assuming that all points on the surface of the protein are equally likely to serve as the point of attachment. Hence, we expect the point of attachment to have a relatively minor effect on the most probable values of the parameter distributions. The assumption that the proteins are able to sample the full range of electrical shape factors is supported by the good agreement between the measured and expected values for the parameters examined in this work (Fig. 3g-i and Fig. 4e,f). The point of attachment is, however, likely to contribute to the uncertainty in a fraction of the stand-alone single protein measurements.

**Supplementary Note 5. Simulating translocation events due to spheroidal particles**

We numerically simulated translocation events due to spheroidal particles in MATLAB in order to provide support for the analysis methods developed in this work. Input parameters for the simulations included $\Delta I_{min}$, $\Delta I_{max}$, the dipole moment or $\mu$, the rotational diffusion coefficient or $D_R$, pore geometry (i.e. length and diameter), the resistivity of the solution, the standard deviation of the noise, and the duration of each event or $t_d$. To generate an intra-event $\Delta I$ signal, we first simulated a spheroidal particle undergoing a biased random walk in one dimension by adapting the model developed by Gauthier and Slater for translational motion[59]. In our model, bias is introduced solely due to the electric field acting on the dipole moment of the particle, which was assumed to be pointed parallel to one of the principal axes. We simulated discrete 1-ns-long time steps in which the angle of the particle relative to the electric field, $\theta$, changed by a fixed step size, $\Delta\theta = 2D_R\Delta t$. For each time step, the following equation gives the probability that the particle will move in the positive or negative direction:

$$p_\pm = \frac{1}{1+e^{\Delta U/k_B T}} = \frac{1}{1+e^{\pm E\mu\left[\cos(\theta-\Delta\theta)-\cos(\theta+\Delta\theta)\right]/(2k_B T)}} \quad (16)$$

which was implemented in the simulations *via* the random number generator in MATLAB. Note that the change in potential energy, $\Delta U$, is divided by a factor of 2 since the particle is initially located halfway in between the two possible final orientations. After simulating the entire event, we converted $\theta(t)$ to $\Delta I(t)$ based on equation (4) and sampled the signal at a rate of 500 kHz to mimic the sampling conditions of the real electronic recordings. Finally, we added Gaussian noise to the signal (unless indicated otherwise) and proceeded with analyzing these simulated signals in the same manner as the resistive-pulse signals obtained during an experiment.

Supplementary Figure 12 shows results from fitting the convolution model to a cumulative distribution of maximum $\Delta I$ values from simulated translocation events. The convolution model described the experimental data extremely well ($R^2 = 0.999$) and yielded estimates of the length-to-diameter ratio, $m$, and excluded volume, $\Lambda$, that were within 10% of their expected values. As hypothesized in Supplementary Note 2, the distribution was biased toward $\Delta I_{max}$ more than expected (i.e. based on the dipole moment only), which is likely a result of representing each event by its maximum value. These results suggest that fitting distributions of maximum $\Delta I$ values yields accurate estimates of shape and volume but not dipole moment.

Supplementary Figure 13 shows distributions of the length-to-diameter ratio, $m$, and excluded volume, $\Lambda$, determined from fitting the convolution model to simulated intra-event $\Delta I$ signals (analysis of intra-event $\Delta I$ values is presented in Supplementary Note 6). The median values of $m$ and $\Lambda$ exactly match the expected values despite the relatively low signal-to-noise ratio of the data (SNR = $[I_{RMS, Signal} / I_{RMS, Noise}]^2$), which is lower than that observed in any of the experiments summarized in Supplementary Figure 15 wherein the signal-to-noise ratio was at least 1.4 and the noise was also Gaussian. These results suggest that the error in determining $m$ and $\Lambda$ from fitting experimental intra-event $\Delta I$ signals, as described in Supplementary Note 6, is not due to low signal-to-noise ratios. Furthermore, these results highlight the ability of the convolution model to account for the presence of noise.

Supplementary Figure 14 shows the distribution of $\mu$ and $D_R$ that we obtained from analyzing simulated intra-event $\Delta I$ signals. These distributions were described well by a lognormal distribution ($R^2 > 0.98$) similar to our experimental results. The most probable value of $\mu$ determined from fitting each intra-event $\Delta I$ signal with the convolution model was in excellent agreement with the expected (i.e. input) value over the range of values measured in this work (Supplementary Fig. 14c). Similarly, the most probable value of $D_R$ determined from analyzing each intra-event $\Delta I$ signal similarly was in agreement with the input value; however, our analysis methods systematically underestimated $D_R$ by about 10 percent (Supplementary Fig. 14d). This underestimation is likely due to a slight leveling off of the MSAD curve between the first two points (an example MSAD curve is shown in Fig. 4 in the main text), which might be rectified by increasing the sampling frequency of the signal. Regardless, these results suggest that the analysis methods developed in this work yield accurate estimates of the dipole moment and rotational diffusion coefficient of a spheroidal particle as long as its orientation is biased purely by its dipole moment.

We want to emphasize that these simulation results were not acquired by performing a simple backwards calculation. The data used here was simulated based on the probability of the particle rotating in one direction or another (equation (16)), and thus it is accomplished in a manner that is independent from the analysis methods described in equations (8) through (13).

**Supplementary Note 6. Analysis of intra-event $\Delta I$ values**

*Distributions of m and $\Lambda$ determined from fitting intra-event $\Delta I$ values*

Supplementary Figure 15 shows distributions of the length-to-diameter ratio, $m$, and excluded volume, $\Lambda$, determined from fitting the convolution model to all intra-event $\Delta I$ signals longer than 0.4 ms from experiments with IgG$_1$, GPI-AChE, Fab, BSA, α-amylase, and BChE. In general, the median value of $m$ from each experiment corresponds to a shape that is more elongated than we expect (i.e. the median values were less than expected for oblates and greater than expected for prolates), and the median value of Λ is lower than the results we obtained by analyzing distributions of maximum $\Delta I$ values and also lower than what is expected from the crystal structure of each protein (Supplementary Table 1). The discrepancy between the values of these parameters obtained by the two different analysis strategies may result from the shape of the nanopore, which the model assumes is perfectly cylindrical (i.e. constant diameter); however, the pore may have a varying diameter. The intra-event $\Delta I$ signal would be expected to reflect changes in pore diameter[60] and will include $\Delta I$ values from when the protein is in the widest regions of the pore. In contrast, the maximum $\Delta I$ value from each event most likely occurs when the protein is near the narrowest constriction of the pore. One might also expect low $\Delta I$ values as the protein enters and exits the nanopore; however, the electric field is highly non-uniform and dense at the edges of the pore, which is thought to offset this effect or even result in larger than expected $\Delta I$ values[61,62]. In the current model, the effect of pore shape and the non-uniformity of the electric field near the pore entrance and exit are not considered and could result in lower than expected $\Delta I$ values. For the intra-event (i.e. single-event) analysis, these low $\Delta I$ values would yield more elongated shapes and lower volumes than expected. If these hypotheses are true, this single-event analysis could be improved by (1) using pores that more closely match a perfect cylinder, (2) by excluding $\Delta I$ values from the beginning and end of the signals, and (3) by knowing the exact geometry of the nanopore in combination with an improved description of the electric-field in and around the pore. Despite the increased uncertainty in the results from the single-event analysis compared to the analysis of distributions of maximum $\Delta I$ values from many events, the values $m$ and Λ determined from fitting intra-event $\Delta I$ signals still can be used to identify and characterize proteins as evidenced by the repeatability between different experiments for IgG$_1$, Fab, BSA, and α-amylase.

*Determining the dipole moment of a protein from fitting intra-event ΔI values*

In the main text, we plotted the most probable value of the biasing parameter or dipole moment, μ, determined from fitting the convolution model to all intra-event signals longer than 0.4 ms for IgG$_1$, GPI-AChE, Fab, β-PE, G6PDH, L-LDH, BSA, α-amylase, and BChE (Fig. 4e in the main text). Supplementary Figure 16 shows histograms of the values of μ that were returned from fitting each event in all experiments. In every case, the distribution of μ was described well by a lognormal distribution ($R^2 > 0.94$); we expected distributions of this shape based on simulations (see Supplementary Note 5). Moreover, the most probable value of μ in each distribution was indicative of the dipole moment of the protein. Only the permanent dipole moment of a protein biases its orientation inside the nanopore as the dipole moment induced by the electric field is roughly parallel to the field and hence does not affect the torque exerted on the protein[63]. The dipole moment estimates were in good agreement with measurements from dielectric impedance spectroscopy and calculations from crystal structures returned by the software HydroPro and the Weizmann server (Supplementary Table 4). Dielectric impedance spectroscopy was performed as described previously[64] using a buffer of 1 mM KCl and 1 mM HEPES (pH = 7.4) for IgG$_1$ and Fab or 1 mM phosphate (pH = 5.2) for BSA. Moreover, these results were repeatable between different nanopores; the difference in the estimated dipole moment (i.e. most probable values of μ) from experiments with different nanopores was always less than 20 percent, indicating that, with bilayer coated walls, pore-dependent effects did not significantly bias the orientation of the protein.

*Determining the rotational diffusion coefficient of a protein in a nanopore*

To determine the rotational diffusion coefficient, $D_R$, of a protein during a translocation event, we first fit the convolution model to the intra-event $\Delta I$ signal at a bandwidth of 15 kHz to estimate $\Delta I_{min}$ and $\Delta I_{max}$. Using these values, we determined the volume and shape of the protein as described in Supplementary Note 2; this procedure also reveals the maximum and minimum shape factors of the protein based on equations (5) through (7). Using these values we calculated $\theta(t)$ based on equation (4). From this trajectory, we calculated the mean-squared-angular displacement (MSAD) of the protein using overlapping time intervals (i.e. 0 to 4 μs, 2 to 6 μs, 4 to 8 μs, etc.). Since $\theta(t)$ can be "clipped" (i.e. equation (4) yields imaginary values of $\theta(t)$ for $\Delta I$ values that are not between $\Delta I_{min}$ and $\Delta I_{max}$), we only calculated angular displacement between two non-clipped values when computing the MSAD. By symmetry of the spheroid, multiple orientations of the particle are equivalent to $\theta$ in the range of 0 to $\pi/2$ (for example, the orientation of $3\pi/2$ is equivalent in this equation to the orientation of $\pi/2$). This degeneracy in the estimate of $\theta$ means that the trajectory of the MSAD will fail to describe the rotation of the protein accurately for long time scales; rather, the trajectory of $\theta(t)$ should be used only to estimate changes in $\theta$ over short time scales. This degeneracy, combined with the periodicity of rotation, causes the MSAD curve to level off asymptotically (see Fig. 4c in the main text for an example). Hence, we only fit the MSAD curve with a tangent line that passes through the origin to estimate the initial slope of the MSAD curve and reveal the rotational diffusion coefficient, $D_R$. According to the Langevin torque equation, $D_R$ is equal to the initial slope of the MSAD curve divided by 2 for rotation around a single axis[65]. Since filtering attenuates frequency components of the $\Delta I$ signal at which rotation occurs, we calculated $D_R$ at various cut-off frequencies and fit this data with the logistic equation to estimate the value of $D_R$ at infinite bandwidth, which corresponds to the upper horizontal asymptote of the fit (Fig. 18a shows an example). On average, these fits described the experimental data extremely well ($R^2 > 0.96$). We calculated the overall bandwidth of the signal according to the following equation[66]:

$$f_c = \sqrt{1/\left[1/f_{c1}^2 + 1/f_{c2}^2\right]}$$

where $f_{c1}$ is the cutoff frequency of the recording electronics (57 kHz)[1] and $f_{c2}$ is the cutoff frequency of the digital Gaussian filter (ranges from 15 to 57 kHz). Note that the diffusion coefficients determined in this analysis describe rotation about the equatorial axes of the protein since rotation about the axis of symmetry is not reflected in the intra-event $\Delta I$ signal. The diffusion coefficient that describes rotation about the equatorial axes of a spheroid is within 10 percent of the average diffusion coefficient for length-to-diameter ratios ($m$) that are less than 1.5.

Supplementary Figure 18 shows histograms of the values of $D_R$ that were returned from fitting all events longer than 0.4 ms for experiments with IgG$_1$, GPI-AChE, Fab, β-PE, α-amylase, and BChE. We excluded all other experiments (10 of 26) from this analysis due to their relatively low signal-to-noise ratios, which yielded values of $D_R$ that were erroneously high and similar to values obtained from analyzing signals consisting of only Gaussian noise (~50,000 rad$^2$ s$^{-1}$). As with the distributions of μ, each distribution of $D_R$ was described well by a lognormal distribution ($R^2 > 0.96$), wherein the most probable value was strongly correlated (Pearson's $r = 0.93$) with the theoretical rotational diffusion coefficient for each protein in bulk solution (Fig. 4f and Supplementary Table 4). The most probable values measured here were 211 times lower on average than the values of $D_R$ in bulk solution, which is consistent with previous findings for GPI-AChE (see Supplementary Table 4). The rotational diffusion coefficient of the relatively flexible $IgG_1$ antibody was similar in two of the three nanopores; this result suggests that additional pore-dependent effects (e.g. steric effects) not taken into account by this model might impact the rotation of proteins in a nanopore. Supplementary Note 5 shows results from simulated data that support the methods described in this section.

**Supplementary Note 7. Bivalently-bound $IgG_1$ rotates slower than monovalently-bound $IgG_1$**

To provide additional evidence that $\Delta I$ values reflect the orientation of a non-spherical protein residing in the nanopore, we measured resistive-pulses resulting from the translocation of anti-biotin $IgG_1$ bound to one or two biotin-PE lipids in the nanopore coating. Bivalently-bound $IgG_1$ should have reduced translational and rotational diffusion coefficients compared to monovalently-bound $IgG_1$ due to the additional drag associated with the second lipid anchor. To test this hypothesis, we performed an experiment in which the conditions initially favored bivalent binding of $IgG_1$ to the lipid coating, and gradually throughout the experiment, we changed the conditions to favor monovalent binding of $IgG_1$. To favor bivalent binding of $IgG_1$, we used a ratio of lipid-anchored biotin to $IgG_1$ that was 33-fold greater than that used in other experiments involving the same protein (i.e. 2 nM $IgG_1$ and 1 mol% biotin-PE versus 10 nM $IgG_1$ and 0.15 mol% biotin-PE). To shift toward conditions favoring monovalent binding, we introduced soluble biotin at sequentially increasing concentrations (1, 10, and 100 nM for 30 min each) to out-compete the lipid-anchored biotin in binding $IgG_1$, thereby reducing the fraction of bivalently-bound $IgG_1$ and increasing the fraction of monovalently-bound $IgG_1$ throughout the course of the experiment. $IgG_1$ proteins that were not bound to a lipid-anchored ligand were not detected[2].

To determine the *translational* diffusion coefficient of lipid-anchored $IgG_1$ in the presence of 0 and 100 nM of soluble biotin, we fit each distribution of translocation times with Schrödinger's first-passage probability density function. In the absence of soluble biotin wherein bivalent binding is favored, the translational diffusion coefficient was 1.05 nm$^2$ μs$^{-1}$, whereas in the presence of 100 nM of soluble biotin wherein monovalent binding is favored, the diffusion coefficient increased to 1.37 nm$^2$ μs$^{-1}$. This increase by a factor of 1.3 is in agreement with work by van Lengerich *et al.*, who previously estimated that a particle with a single lipid anchor should diffuse laterally about 1.5 times faster than a particle with two lipid anchors[67]. This result supports our expectation that the ratio of bivalently-bound to monovalently-bound $IgG_1$ decreases with the concentration of soluble biotin. We also found that the charge of the protein-lipid complex changed from -3.25 in the absence of soluble biotin (i.e. conditions favoring bivalent binding) to -1.53 in the presence of 100 nM of soluble biotin (i.e. conditions favoring monovalent binding); this change in the value of the charge by -1.7 is slightly larger in magnitude than the theoretically expected value of -1 (the expected charge of one biotin-PE lipid) but this deviation is likely within the error of the measurement. The main aspect for the discussion here is that the negative charge *decreased* in magnitude as expected when fewer IgG molecules are bound bivalently.

We next obtained distributions of *rotational* diffusion coefficients ($D_R$) by analyzing intra-event $\Delta I$ values (see Supplementary Note 6) for $IgG_1$ in the presence of increasing concentrations of soluble biotin (Supplementary Figure 19). The most probable value of $D_R$ increases with the concentration of soluble biotin as expected for conditions that favor monovalent over bivalent binding (Supplementary Figure 19c; Pearson's correlation coefficient $r$ = 1.00). In the absence of soluble biotin, $D_R$ was approximately 100 rad$^2$ s$^{-1}$. In the presence of 100 nM of soluble biotin, $D_R$ increased more than an order of magnitude to 1,744 rad$^2$ s$^{-1}$, approaching the expected value for monovalently-bound IgG$_1$ of 4,500 rad$^2$ s$^{-1}$ (see Supplementary Table 4). As with the results for the translational diffusion coefficient, this trend indicates that the ratio of bivalently-bound to monovalently-bound IgG$_1$ decreases with the concentration of soluble biotin. Together, these results provide strong evidence that $\Delta I$ values reflect the rotational dynamics of the protein since we observed more than 17-times faster rotational diffusion in the same nanopore as we change the experimental conditions from favoring bivalent binding to favoring monovalent binding.

The distribution of maximum $\Delta I$ values is also affected by the ratio of monovalently-bound to bivalently-bound IgG$_1$, as shown in Supplementary Figure 20. The distribution becomes more biased toward low $\Delta I$ values as the fraction of bivalently-bound IgG$_1$ increases, suggesting that bivalently-bound IgG$_1$ is less likely to sample cross-wise orientations during a translocation event than monovalently-bound IgG$_1$. One likely explanation for this bias in orientation is that non-spherical proteins orient length-wise prior to entering the pore due to the strong electric field gradient that they experience once they approach the pore[61,68] and as a consequence of the reduced rotational diffusion coefficient of bivalently-bound IgG$_1$ compared to monovalently-bound IgG$_1$, bivalently-bound IgG$_1$ is less likely to reorient by Brownian motion with dipole-induced bias during an event of a given duration. Alternatively, the bias toward low $\Delta I$ values might result from steric effects that limit crosswise orientations since the second lipid anchor of bivalently-bound IgG$_1$ may restrict the possible range of configurations the protein can assume for a given position of the first lipid anchor. Regardless, the results of increased bias as a consequence of two lipid anchors strongly support the conclusion that $\Delta I$ reflects the orientation and shape of non-spherical protein residing in the nanopore.

**Supplementary Note 8. Distinguishing an antigen and antibody-antigen complex in a single nanopore experiment**

Supplementary Figure 22 illustrates the ability of the methods developed in this work to characterize and identify a single protein, G6PDH, and a protein-protein complex, G6PDH-IgG, in the same solution. Supplementary Fig. 22a-i shows results from analysis of maximum $\Delta I$ values (the procedures for this analysis are described in the figure caption). Supplementary Fig. 22j-l shows results from analysis of all intra-event $\Delta I$ values.

To classify each translocation event as either G6PDH or G6PDH-IgG, we analyzed intra-event $\Delta I$ values as described in Supplementary Note 6 to determine the volume, shape, charge-related $t_d$ value, rotational diffusion coefficient, and dipole moment from each protein or protein complex moving through the nanopore. This procedure identified 787 translocation events that were longer than 400 μs. We normalized the values for each parameter by their standard deviations and classified each event using the clustering algorithm *kmeans* in MATLAB[69,70]. Briefly, the *kmeans* clustering algorithm minimizes, across all clusters, the sum of the distance between all points in the cluster to the centroid of the cluster. To assess the quality of all cluster analyses and provide an error for the values assigned to parameters, we ran a bootstrap method in which 1,000 datasets were created by random resampling with replacement of the original dataset[71]. We then ran the cluster analysis on these 1,000 datasets. The clustering procedure was always robust with approximately 90% of the data (727 events) consistently being classified as either G6PDH or G6PDH-IgG (at least 95% of the time).

We performed the cluster analysis on several combinations of these five parameters and found that a 3D cluster analysis based on the volume, dipole moment, and rotational diffusion coefficient provided the best separation between clusters as well as the most accurate characterization of the volumes for G6PDH (3% difference) and the G6PDH-IgG complex (7% difference). For instance, Fig. 5c in the main text shows that this technique determined a volume for G6PDH of 227 ± 9 nm$^3$ compared to the volume of 220 nm$^3$ determined from distributions of maximum $\Delta I$ values in an independent experiment; similarly, this analysis determined the volume of the complex to be 530 ± 64 nm$^3$, and we expected a volume for the complex of 497 nm$^3$ (the volume of G6PDH plus the volume of an IgG protein). The volume of the complex determined from this intra-event analysis was also in excellent agreement with that determined from analysis of distributions of maximum $\Delta I$ values, which is shown in Supplementary Fig. 22i. Furthermore, both the analysis of maximum $\Delta I$ values (Supplementary Fig. 22f) and analysis of intra-event $\Delta I$ values followed by cluster analysis revealed that after the addition of anti-G6PDH IgG, the proportion of events due to the G6PDH-IgG complex was between 27 to 28 percent. The agreement between these two values provides additional evidence that the classification of events from single-event analysis was accurate. For reference, two-dimensional projections of the 3D scatter plot in Fig. 5b of the main text are shown in Supplementary Fig. 22j-l.

Prior to this work, the standard practice for distinguishing between proteins in a mixture would have been to analyze scatter plots of $t_d$ values *vs.* $\Delta I$ values. To illustrate the benefits of the multi-parameter characterization based on methods developed in this work, we performed a two-dimensional cluster analysis on the same data set used above, using only $t_d$ values and average $\Delta I$ values. This analysis found that the protein complex represented only 2.5 ± 0.5 % of events, which is ~90% lower than the values determined by single-event analysis or analysis of distributions of maximum $\Delta I$ values (Supplementary Fig. 22). Moreover, this analysis failed to determine the volume of the complex accurately as it returned a value of 833 ± 50 nm$^3$, which is 68% greater than the estimated volume of the complex of 497 nm$^3$ determined from independent experiments.

**Supplementary Note 9. Derivation of probability distribution of shape factors for proteins with a dipole moment**

To derive a probability distribution of shape factors that takes into account a bias for a specific orientation based on the dipole moment of a protein and the electric field, we used the Boltzmann distribution of energies:

$$\frac{N_i}{N} = \frac{g_i \exp\left[-U_i/k_B T\right]}{\sum g_j \exp\left[-U_j/k_B T\right]} \quad (17)$$

where $g_i$ is the number of states that have the same energy level, $U_i$ is the energy level of state $i$, $N_i$ is the number of molecules with energy level $i$, $N$ is the total number of molecules in the system, and $k_B T$ is the thermal energy. The denominator of equation (17) is the partition function, and we will label it $Z$. Assuming that all of the energy affecting the orientation of the protein is in the form of the potential energy of a dipole in an electric field, then $g_i$ is constant for all energy states and cancels out of equation (17). The potential energy of a dipole in an electric field is:

$$\Delta U = \vec{E} \cdot \vec{\mu} = -E\mu\cos(\phi) \tag{18}$$

where $E$ is the electric field, $\mu$ is the dipole moment, and $\phi$ is the angle between the moment and the electric field. Combining equations (17) and (18), the proportion of molecules at an angle, $\phi$, is:

$$\frac{N_\phi}{N} = \frac{1}{Z}\exp\left[\frac{E\mu\cos(\phi)}{k_B T}\right] \tag{19}$$

and therefore the probability of observing an angle $\phi$ is:

$$P_\phi = \frac{c}{Z}\exp\left[\frac{E\mu\cos(\phi)}{k_B T}\right] \tag{20}$$

where $c$ is a normalization constant.

Considering a simple scenario in which the dipole moment is parallel with the symmetry or equatorial axis and accounting for the two possible orientations of the dipole moment relative to the electric field for a given orientation (i.e. $\theta$) due to symmetry, we obtained equations (21a) and (21b) for $\phi = \theta$ and $\theta + \pi/2$ from equation (20):

$$P_\theta = \frac{c}{Z}\exp\left[\frac{E\mu\cos(\theta)}{k_B T}\right] + \frac{c}{Z}\exp\left[\frac{E\mu\cos(\pi-\theta)}{k_B T}\right] = \frac{c}{Z}\cosh\left[\frac{E\mu\cos(\theta)}{k_B T}\right] \tag{21a}$$

and

$$P_\theta = \frac{c}{Z}\exp\left[\frac{E\mu\cos\left(\theta+\frac{\pi}{2}\right)}{k_B T}\right] + \frac{c}{Z}\exp\left[\frac{E\mu\cos\left(\pi-\left(\theta+\frac{\pi}{2}\right)\right)}{k_B T}\right] = \frac{c}{Z}\cosh\left[\frac{E\mu\sin(\theta)}{k_B T}\right] \tag{21b}$$

To express $\cos(\phi)$ in terms of the electrical shape factor we first rearranged equation (4), which describes $\gamma$ as a function of $\theta$, to obtain:

$$\cos(\theta) = \left(\frac{\gamma - \gamma_\perp}{\gamma_P - \gamma_\perp}\right)^{1/2} \tag{22}$$

Substituting equation (22) into equations (21a) and (21b), we obtain:

$$P_\theta = \frac{c}{Z}\cosh\left[\frac{E\mu\left(\frac{\gamma-\gamma_\perp}{\gamma_P-\gamma_\perp}\right)^{1/2}}{k_BT}\right] \quad (23a)$$

and

$$P_\theta = \frac{c}{Z}\cosh\left[\frac{E\mu\sin\left(\cos^{-1}\left[\left(\frac{\gamma-\gamma_\perp}{\gamma_P-\gamma_\perp}\right)^{1/2}\right]\right)}{k_BT}\right] = \frac{c}{Z}\cosh\left[\frac{E\mu\left(\frac{\gamma-\gamma_P}{\gamma_\perp-\gamma_P}\right)^{1/2}}{k_BT}\right] \quad (23b)$$

Equations (23a) and (23b) express the probability of observing an angle $\theta$ as a function of the shape factor, $P(\theta(\gamma))$. As in the derivation by Golibersuch, we used substitution of variables to transform $P(\theta(\gamma))$ into $P(\gamma)$:

$$P(\gamma)d\gamma = P_\theta\frac{d\theta}{d\gamma}d\gamma = P_\theta\left(\frac{d\gamma}{d\theta}\right)^{-1}d\gamma \quad (24)$$

and differentiated equation (4) with respect to $\theta$, $\frac{d\gamma}{d\theta}$. Substituting this result into equation (24), we obtained equations (25a) and (25b):

$$P(\gamma)d\gamma = \frac{c}{Z}\cosh\left[\frac{E\mu\left(\frac{\gamma-\gamma_\perp}{\gamma_P-\gamma_\perp}\right)^{1/2}}{k_BT}\right]\left[\frac{1}{\pi\left[(\gamma-\gamma_\perp)(\gamma_P-\gamma)\right]^{1/2}}\right]d\gamma \quad (25a)$$

and

$$P(\gamma)d\gamma = \frac{c}{Z}\cosh\left[\frac{E\mu\left(\frac{\gamma-\gamma_P}{\gamma_\perp-\gamma_P}\right)^{1/2}}{k_BT}\right]\left[\frac{1}{\pi\left[(\gamma-\gamma_\perp)(\gamma_P-\gamma)\right]^{1/2}}\right]d\gamma \quad (25b)$$

To solve for the normalization constants, we integrated equations (25a) and (25b) and set each equation equal to 1 (i.e. $\int_{\gamma_P}^{\gamma_\perp}P(\gamma)d\gamma = 1$). This procedure cancels out the partition function $Z$ and yields:

$$P(\gamma)d\gamma = \frac{1}{A}\cosh\left[\frac{E\mu\left(\frac{\gamma-\gamma_\perp}{\gamma_P-\gamma_\perp}\right)^{1/2}}{k_BT}\right]\left[\frac{1}{\pi\left[(\gamma-\gamma_\perp)(\gamma_P-\gamma)\right]^{1/2}}\right]d\gamma \quad (26a)$$

and

$$P(\gamma)d\gamma = \frac{1}{A}\cosh\left[\frac{E\mu\left(\frac{\gamma - \gamma_P}{\gamma_\perp - \gamma_P}\right)^{1/2}}{k_BT}\right]\left[\frac{1}{\pi\left[(\gamma - \gamma_\perp)(\gamma_P - \gamma)\right]^{1/2}}\right]d\gamma \quad (26b)$$

where A is described by:

$$A = \int_{\gamma_P}^{\gamma_\perp}\cosh\left[\frac{E\mu\left(\frac{\gamma - \gamma_\perp}{\gamma_P - \gamma_\perp}\right)^{1/2}}{k_BT}\right]\left[\frac{1}{\pi\left[(\gamma - \gamma_\perp)(\gamma_P - \gamma)\right]^{1/2}}\right]d\gamma \quad (27a)$$

and

$$A = \int_{\gamma_P}^{\gamma_\perp}\cosh\left[\frac{E\mu\left(\frac{\gamma - \gamma_P}{\gamma_\perp - \gamma_P}\right)^{1/2}}{k_BT}\right]\left[\frac{1}{\pi\left[(\gamma - \gamma_\perp)(\gamma_P - \gamma)\right]^{1/2}}\right]d\gamma \quad (27b)$$

Equations (26a) and (26b) are identical to equations (12a) and (12b) in Supplementary Note 2.

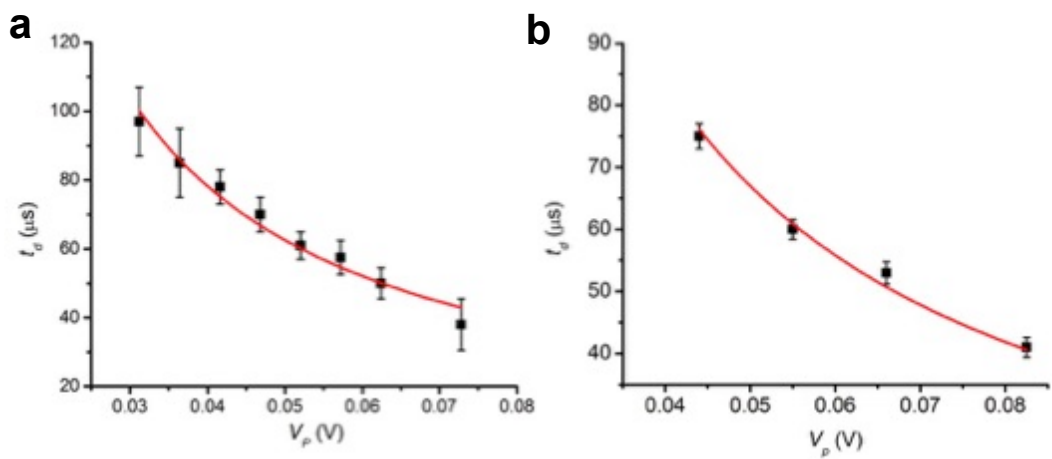

**Supplementary Figure 1. Most probable $t_d$ values for the monoclonal anti-biotin IgG$_1$ antibody (a) and GPI-AChE (b) as a function of the voltage drop, $V_P$, across a bilayer-coated nanopore containing biotin-PE.** The inverse relationship between translocation time and applied voltage as well as the excellent agreement between theory (red curve) and experiment indicate that the lipid-anchored proteins completely passed through the nanopore. The red curve was obtained by a best-fit of equation $t_d = l_P^2 \, k_B T / (|z|eV_p D_L)$ as described in Yusko et al.[2], where the only fitting parameter is the net charge of the protein, $z$. $l_P$ is the length of the nanopore with the bilayer coating, $k_B T$ is the thermal energy (1.38E-23 J K$^{-1}$ × 295 K), $V_p$ is the voltage drop across the nanopore, and $D_L$ is the diffusion coefficient of the lipids in the bilayer as determined from FRAP experiments. For the IgG$_1$ antibody (a), the fit returned a value for $z$ of -3.5 ± 0.1 (in 2 M KCl with pH = 7.4 in 10 mM HEPES) with $R^2$ = 0.98, *p-value* < 0.001 ($N$ = 8), which is the expected value for the charge of this monoclonal antibody based on capillary electrophoresis experiments[2]. The value used for $D_L$ was 1.35E-12 m$^2$ s$^{-1}$ determined from FRAP experiments[2], and the value of $l_P$ was 24 nm. For the GPI-AChE (b), the fit returned a value for $z$ of -2.7 ± 0.1 (in 2 M KCl with pH = 6.1 in 10 mM HEPES) with $R^2$ = 0.99, *p-value* < 0.001 ($N$ = 4). For comparison, the theoretical charge of GPI-AChE at zero ionic strength and pH 7.4 is -12 to -16[7,10]. $D_L$ was 1.6 E-12 m$^2$ s$^{-1}$ and $l_P$ = 24 nm. The bilayer coating in (a) contained 0.15% biotin-PE, 0.8% Rh-PE, and ~99% POPC, and the bilayer coating in (b) contained only 0.8% Rh-PE, and ~99.2% POPC.

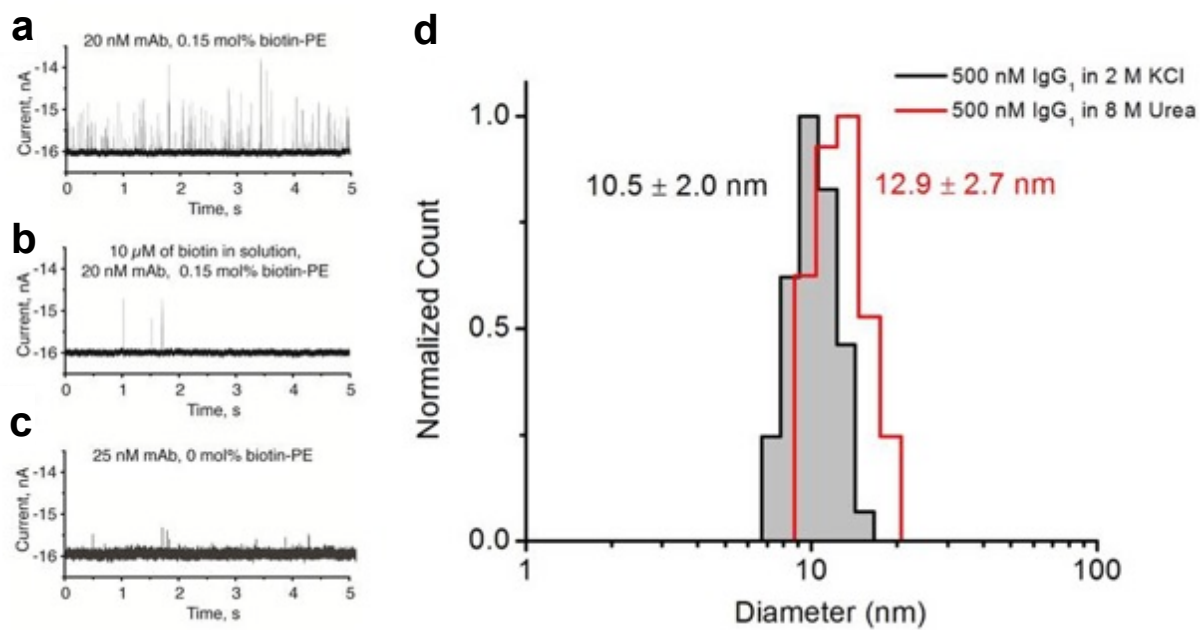

**Supplementary Figure 2. Detection of monoclonal anti-biotin IgG$_1$ antibody with a bilayer-coated nanopore and dynamic light scattering experiments.** a) Current *versus* time trace showing resistive pulses due to translocation of IgG$_1$ antibodies that were bound to biotin-PE lipids in the bilayer coating. Resistive pulses occurred at a frequency of 34 s$^{-1}$. b) Current *versus* time trace recorded after the addition of excess biotin (10 μM) to the solution and containing a reduced frequency of resistive pulses (1.3 s$^{-1}$). c) Current *versus* time trace recorded using the same nanopore as (a) and (b) but with a bilayer coating that did not contain biotin-PE lipids. Resistive-pulses occurred at a frequency of 2 s$^{-1}$. The experiments were performed using pore 2 (Supplementary Fig. 25). d) Hydrodynamic diameter of IgG$_1$ antibodies determined from dynamic light scattering experiments. IgG$_1$ antibodies were at a concentration of 500 nM in aqueous solutions identical to the recording electrolyte (2 M KCl and 10 mM HEPES at pH = 7.4) during the dynamic light scattering experiment. Where indicated, 8 M of urea was added to the solution in order to denature all proteins. The dynamic light scattering results are the combination of 5 runs, each 60 s in duration. Results show the intensity-weighted calculation for the hydrodynamic diameter. The instrument was a Brookhaven 90Plus Particle Sizer and used a 658 nm laser at an angle of 90º to the detector. The absence of a second peak indicates that IgG$_1$ antibodies were not fragmented or present in dimers in 2 M KCl even at concentrations 500 fold greater than in the resistive-pulse sensing experiments.

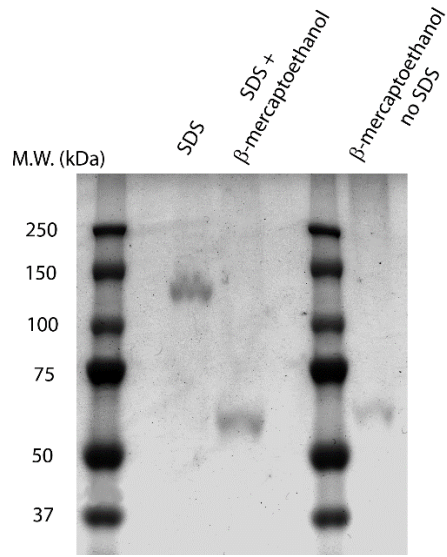

**Supplementary Figure 3. Solutions containing GPI-AChE contained the dimeric, prolate shaped form of GPI-AChE.** 2 μg of protein in Tris-Tricine sample buffer was added to each lane after treatment with 5% w/v SDS, 5% w/v SDS and 7.5 % v/v β-mercaptoethanol, or 7.5% v/v β-mercaptoethanol only. In the samples that contained SDS, the solution was heated to 95°C for 5 min to denature the protein. The gel was a 7.5% Tris-HCl TGX gel from BioRad, and the running buffer was Tris-Glysine buffer (25 mM Tris, 192 mM Glycine, 0.1% SDS). After running the gel, the gel was placed in 100 mL of deionized water and placed in the microwave for 30 s (careful not to boil the solution). The gel was rinsed twice for 3 to 5 min each time. The gel was then immersed in Coomassie staining solution (70 mg of Coomassie brilliant blue in 1 L of water; after 4 h, 3 mL of concentrated HCl was added) and heated in the microwave for 10s (again careful not to boil). The gel was left to stain overnight and destained with pure water[72].

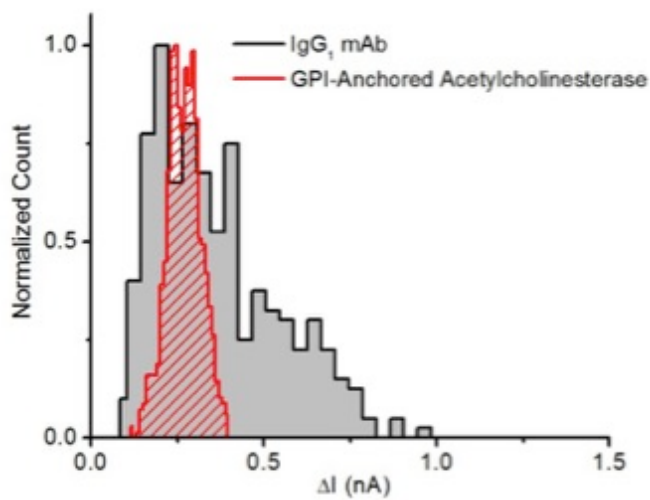

**Supplementary Figure 4. Histograms of the $\Delta I$ values due to the translocation of the IgG$_1$ antibody (150 kDa) and GPI-anchored acetylcholinesterase (160 kDa) through the same nanopore.** The experiments were performed using pore 3 (Supplementary Fig. 25). Though both distributions are bimodal, the relatively narrow distribution of $\Delta I$ values due to GPI-anchored acetylcholinesterase compared to that of the IgG$_1$ antibody confirms that the large molecular weight of the IgG$_1$ antibody was not the reason for broadly distributed $\Delta I$ values. Currents were recorded at an applied potential difference of -100 mV.

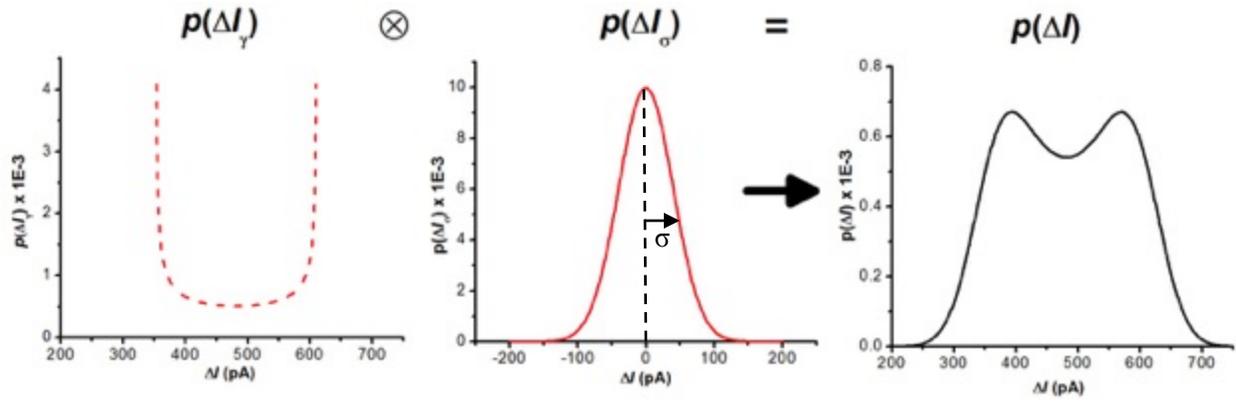

**Supplementary Figure 5. Example convolution of the probability distribution of Δ*I* values one expects due to the distribution of shape factors, *p*(Δ*I*$_\gamma$) (equations (13a) and (13b)), and the error in determining individual Δ*I* values, *p*(Δ*I*$_\sigma$) (a Normal distribution function).** The solution to the convolution is the probability distribution of Δ*I* values one expects to observe, *p*(Δ*I*). During the fitting procedure, the theoretical cumulative distribution, *p*(Δ*I*), is compared to the empirical cumulative distribution of Δ*I* values, *P*(Δ*I*), and the Levenberg-Marquardt non-linear least squares fitting algorithm in MATLAB generates new values for the fitting parameters Δ*I*$_{min}$, Δ*I*$_{max}$, μ, and σ, thereby creating new iterations of *p*(Δ*I*$_\gamma$) and *p*(Δ*I*$_\sigma$). This process repeats until the fit converges, which typically takes around 20 iterations.

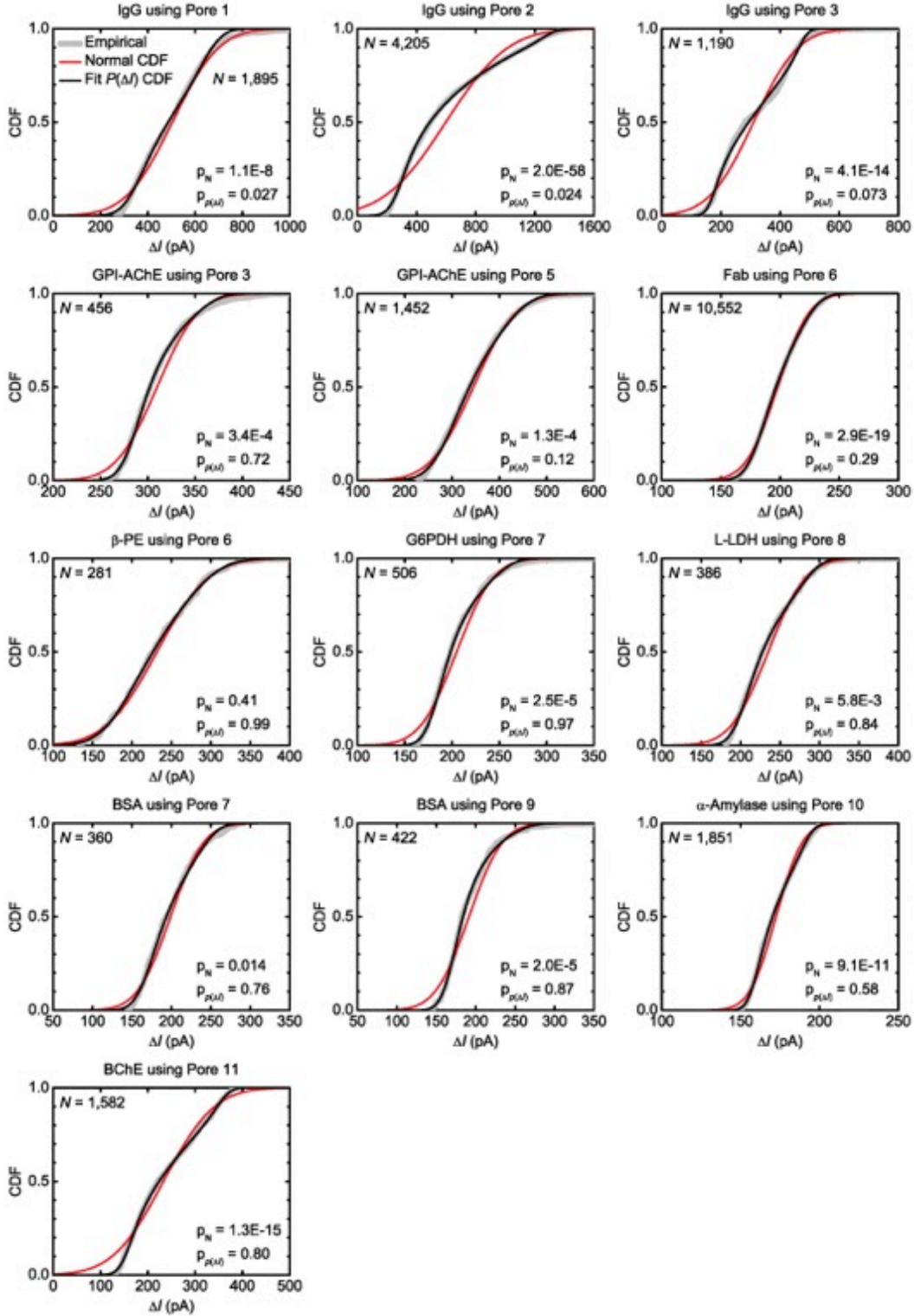

**Supplementary Figure 6. Empirical cumulative distributions (grey curves) of Δ*I* values due to the translocation of non-spherical proteins compared to a best-fit Normal distribution (red curves) and the solution to the convolution model, $p(\Delta I)$ (black curves).** In each case, Kolmogorov Smirnov (KS) tests were used to determine if the empirical distribution was different from the best-fit Normal distribution or $p(\Delta I)$. Resulting *p*-values are shown in the figure panels. In KS-tests, the null hypothesis

is that the two distributions are the same, and therefore, a *p*-value ≤ 0.05 indicates that the difference between two distributions is statistically significant at the α = 0.05 level. For all of the non-spherical proteins except β-phycoerythrin, the distribution of Δ$I$ values was different from a Normal distribution ($p_N$ < 0.05). In contrast, the difference between the empirical distribution and convolution model, $p(\Delta I)$, was not statistically significant in 11 out of 13 cases (*p*-value ≥ 0.05).

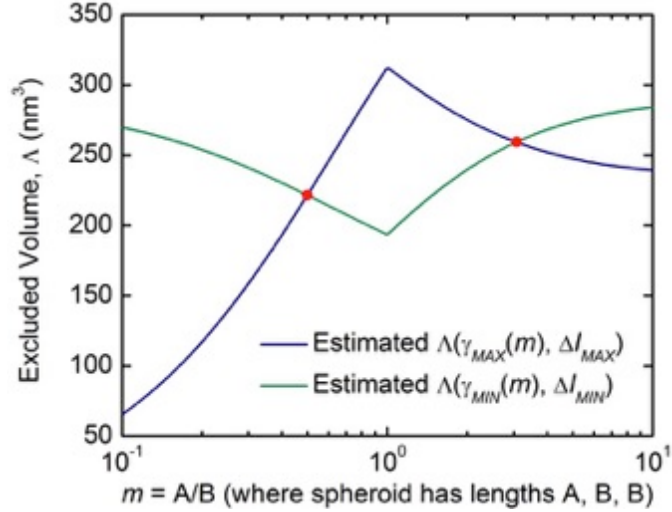

**Supplementary Figure 7. Estimating the excluded volume as a function of *m* using $\Delta I_{min}$ and $\Delta I_{max}$ values illustrates that there are two solutions to equations (14) and (15) for prolate shaped proteins.** This figure shows this result graphically by plotting the estimated volume of GPI-anchored acetylcholinesterase as a function of *m* for Pore 5. The two red dots indicate the two solutions to the system of equations ($m = 0.50$, $\Lambda = 222$ nm$^3$ and $m = 3.1$, $\Lambda = 259$ nm$^3$). In order to simplify the graph, we described the electrical shape factor with the notation $\gamma_{MAX}$ or $\gamma_{MIN}$. We used this notation because for prolates ($m > 1$) $\gamma_{MAX} = \gamma_{\perp}$ and for oblates ($m < 1$) $\gamma_{MAX} = \gamma_{\parallel}$ (see equations (14) and (15)). The opposite is true for $\gamma_{MIN}$.

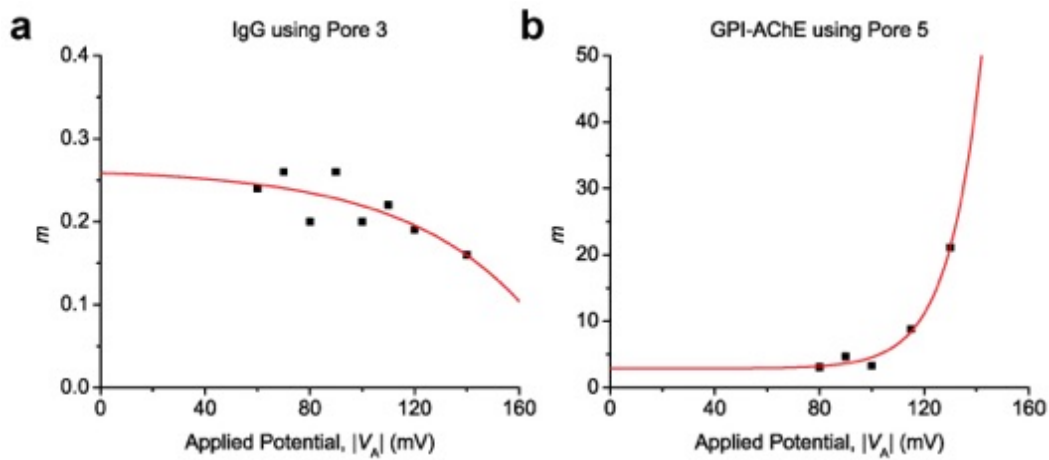

**Supplementary Figure 8. The dependence of a protein's length-to-diameter ratio, *m*, on the applied potential, $V_A$, for IgG$_1$ (a) and GPI-AChE (b).** We determined the value of *m* at different applied potentials by fitting the convolution model to distributions of maximum $\Delta I$ values. Interestingly, *m* is consistent at low potentials, while its value changes to indicate an increasingly elongated protein (i.e. *m* approaches 0 for oblates or approaches ∞ for prolates) with increasing potential. To clearly illustrate this trend, we fit the results with an exponential growth function, $m = m_0 + A \cdot \exp(|V_A|/\tau)$ where $A$ may be positive or negative. Considering that the fits asymptotically approached $m \approx 0.26$ for IgG$_1$ and $m \approx 2.9$ for GPI-AChE and the expected value of *m* is between 0.2 and 0.5 for IgG$_1$ and 2.9 for GPI-AChE (Supplementary Table 4), this result suggests that low potentials yield accurate estimates of the protein shape.

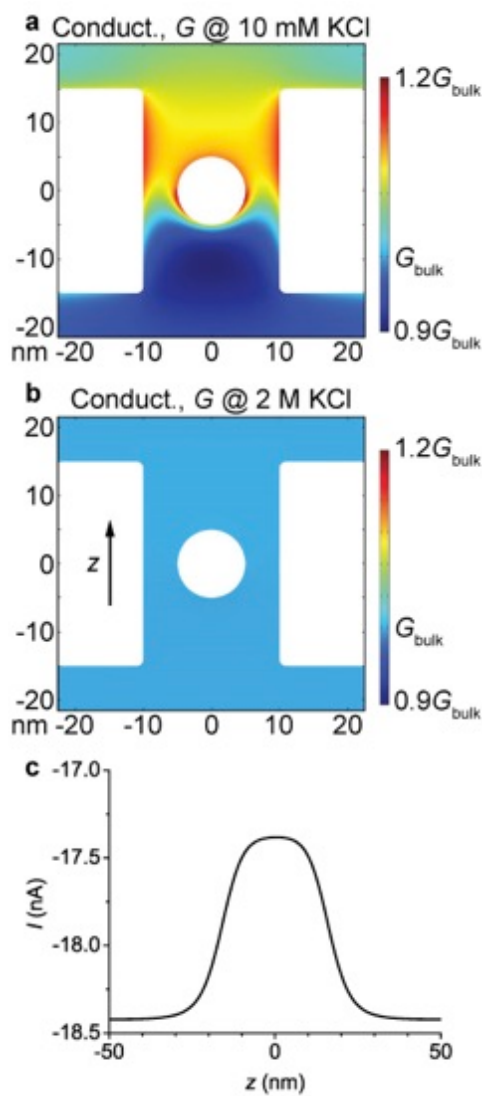

**Supplementary Figure 9. Finite-element simulations indicate that local variations in the conductivity of the solution are negligible under the experimental conditions used in this work.** We performed the simulations in COMSOL Multiphysics 4.4 (COMSOL Inc.). The electric field intensity inside the pore was set to 3 MV m$^{-1}$, the protein charge was set to -10, the charge density of the pore walls was set to 2 mC m$^{-2}$ to account for the non-zwitterionic lipids in the nanopore coating, the protein diameter was set to 10 nm, and the pore diameter and length were set to 20 and 30 nm, respectively. All boundary conditions were identical to those used by Lan et al.[31]. The upper semi-infinite boundary at z = 20 μm had a fixed negative potential relative to the lower boundary. a-b) 2-D heat maps showing the conductivity of the electrolyte solution throughout a vertical cross-section of the nanopore in the presence of (a) 10 mM KCl and (b) 2 M KCl. The color scale of each map was normalized to the conductivity in bulk solution, $G_{bulk}$. At low ionic strength, the conductivity varies significantly due to the accumulation and depletion of chloride ions on opposite sides of the protein; at high ionic strength as used in our experiments, this effect is essentially absent. c) A position-current (I-z) curve obtained by varying the position of the protein in the presence of 2 M KCl. The $\Delta I$ value of this curve is roughly 1.04 nA, which is in excellent agreement with the expected value of 1.00 nA obtained by using the volume exclusion model shown in equation (1). The near perfect symmetry of this curve further indicates that variations in conductivity are negligible at the high ionic strength used in this work.

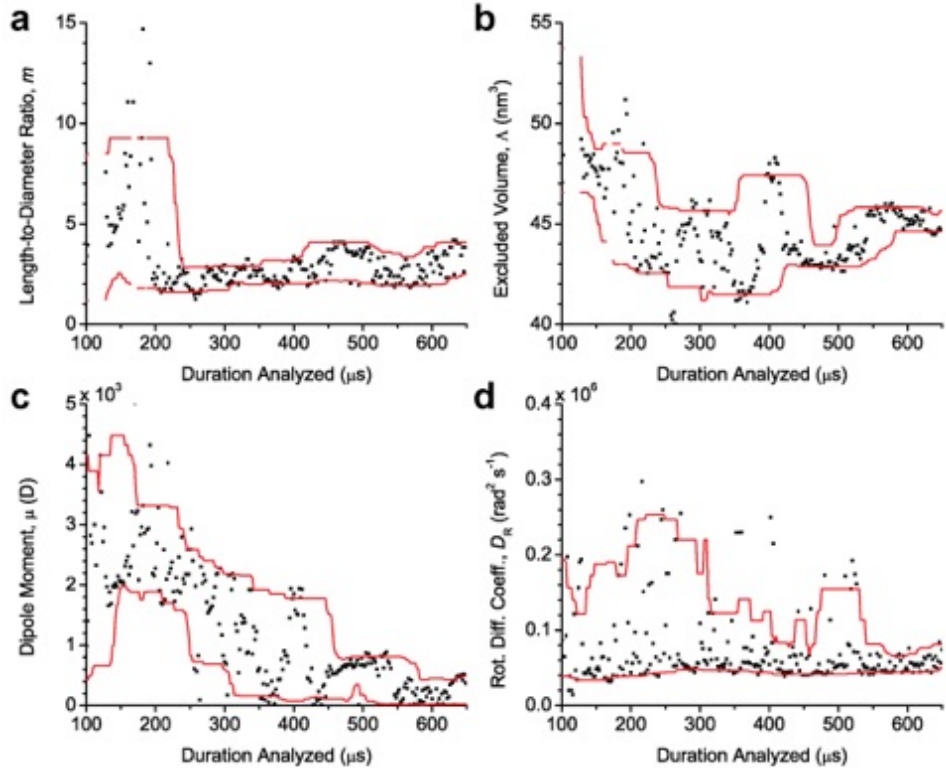

**Supplementary Figure 10. Analysis of intra-event Δ$I$ signals can yield parameter estimates in real-time.** Measurements of the (a) length-to-diameter ratio, (b) excluded volume, (c) dipole moment, and (d) rotational diffusion coefficient obtained by progressively analyzing the current modulations (i.e. intra-event Δ$I$ values) of a single resistive-pulse resulting from the translocation of an individual anti-biotin Fab fragment. The red lines are moving $10^{th}$ and $90^{th}$ percentiles (smoothing window = 50 points). As the protein spends additional time in the pore, more data is acquired and analyzed; consequently, the spread in the determined parameter values narrows and the determined magnitudes of each parameter converge to their final values. The figure also shows that, for this particular event due to the translocation of a single anti-biotin Fab fragment, the variation in each parameter had narrowed to about 20% of its initial spread after approximately 550 μs before the end of the resistive-pulse. These results show that by analyzing a translocation event as it is occurring, it is possible to obtain parameter estimates while the protein still resides in the pore.

**Supplementary Figure 11. Determining the charge of proteins by fitting translocation time distributions with a first-passage-time model.** a-i) Histograms of translocation times from the nanopore experiments summarized in Supplementary Table 1 (bin width = 15 μs). We fit each distribution with Schrödinger's first-passage probability density function

$$P(t_d) = \frac{l_P}{\sqrt{4\pi D_L t_d^3}} e^{-(l_P - vt_d)^2/4D_L t_d}$$ as

described by Ling and Ling[73], where the electrophoretic drift velocity

$$v = |z|eV_P D_L / l_P k_B T$$ as described by

Yusko et al.[2] and the fitting parameters are the protein charge, $z$, and the diffusion coefficient of the lipids in the bilayer coating, $D_L$. We used a bin width of 2 μs when fitting the data, which corresponds to the sampling period of the current recordings. The most probable value of the translocation time is indicated by the dotted black line and corresponds to the maximum of the fit. The error in $z$ is shown in parentheses next to its best-fit value, which we estimated by fitting the data with $D_L$ fixed at its best-fit value ± standard error of the mean. j) Measured versus expected charges. Measured and expected values for anti-biotin IgG$_1$, anti-biotin Fab, and streptavidin were previously determined by Yusko et al.[2] via nanopore and capillary electrophoresis experiments, respectively (black squares). The expected value for BSA was acquired from literature[74] (green circle). The expected values for the remaining proteins were estimated from protein crystal structures via the PROPKA web interface (http://propka.ki.ku.dk/)[75-78] (blue triangles). GPI-AChE and BChE were excluded from this plot due to a lack of a reference value. We subtracted 1 from the expected value for each protein except

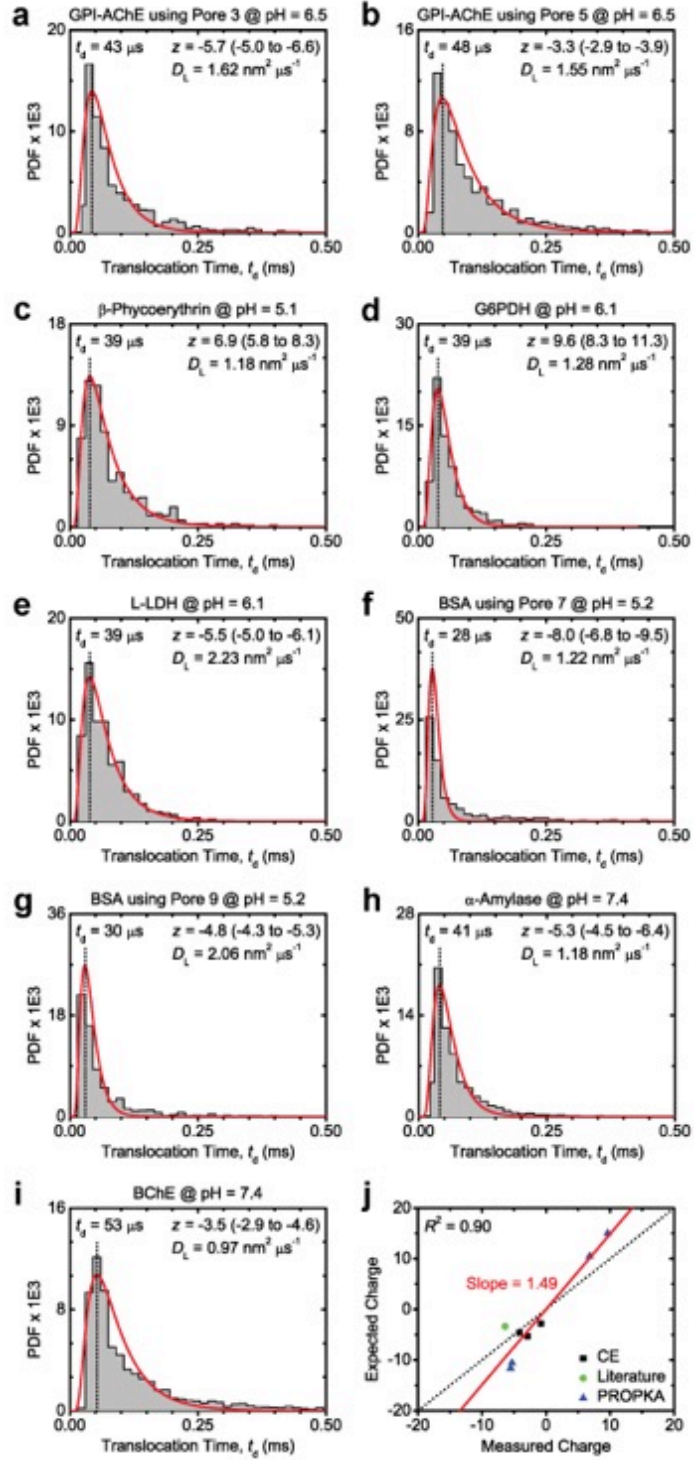

GPI-AChE to account for the net charge of the lipid anchor. For each protein that was covalently attached to the bilayer, we also subtracted 0.93 from the expected value to account for the reaction of a primary amine on the protein surface to form an amide bond[79]. There is a strong positive correlation ($r$ = 0.95) between the measured and expected values; however, the measured values are systematically lower in magnitude than the expected values. This underestimation may be due to inaccuracies in the PROPKA method or the high ionic strength of the recording solution used in nanopore experiments.

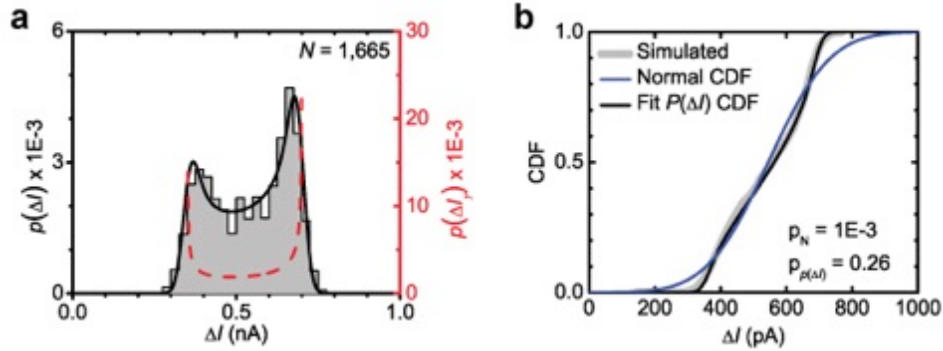

**Supplementary Figure 12. Distributions of maximum Δ*I* values from simulated translocation events.** a) A histogram of maximum Δ*I* values from simulated translocation events. The black curve shows the solution of the convolution model, $p(\Delta I)$, after a non-linear least squares fitting procedure, and the red dashed curve shows the estimated distribution of Δ*I* values due to the distribution of shape factors, $p(\Delta I_\gamma)$. b) The cumulative distribution of the same data shown in (a) (grey curve) compared to a best-fit Normal distribution (blue curve) and the solution to the convolution model (black curve). *p*-values shown in the figure resulted from Kolmogorov Smirnov (KS) tests that compared the simulated, empirical cumulative distribution to the model Normal distribution, $p_N$, or the convolution model, $p_{p(\Delta I)}$. Since, the value $p_N$ was less than 0.05, the KS-test indicated that the distribution was not Normal at the α = 0.05 level. In contrast, the value of $p_{p(\Delta I)}$ was greater than 0.05 and therefore not significantly different from the convolution model; this result indicates that the model describes the empirical distribution well. For the simulations, we used input parameters that were based on the experiment done with IgG$_1$ in pore 1 (e.g. Δ*I*$_{min}$ and Δ*I*$_{max}$ were 329 and 678 pA, corresponding to values of *m* and Λ of 0.37 and 292 nm$^3$). We simulated 2,000 events with translocation times that were sampled from Schrödinger's first-passage probability density function[73]. The signal processing algorithm detected 1,922 events wherein 1,665 of these events were fully time resolved (i.e. $t_d$ > 50 μs). From the fit, we calculated values for *m* and Λ of 0.38 (2.7% greater than the expected value) and 310 nm$^3$ (6.2% greater than the expected value).

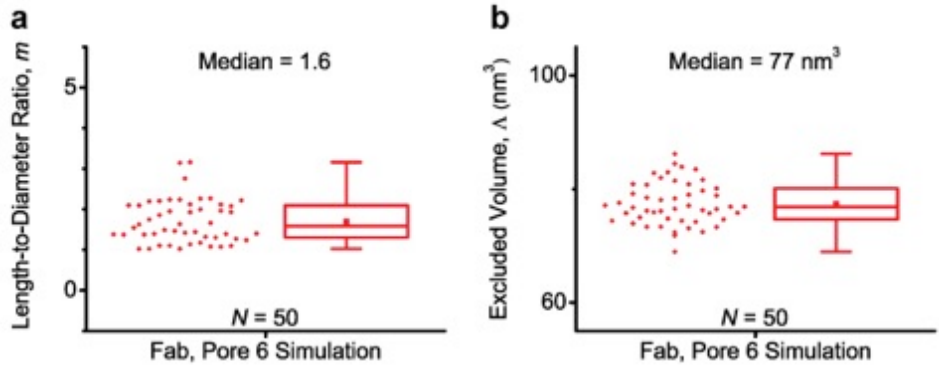

**Supplementary Figure 13. Distributions of the length-to-diameter ratio, *m* (a), and excluded volume, Λ (b), determined from fitting the convolution model to simulated intra-event Δ*I* signals.** The box represents the 1$^{st}$ and 3$^{rd}$ quartiles of the data, the horizontal line is the median value, the point inside the box shows the mean value, and the whiskers extend to data points that are within 1.5 × IQR. For the simulations, we used input parameters that were based on the experiments done with Fab in pore 6 (e.g. $\Delta I_{min}$ and $\Delta I_{max}$ were 178 and 231 pA, corresponding to values of *m* and Λ of 1.6 and 77 nm$^3$). The data was low-pass filtered at 15 kHz. The standard deviation of the noise added to each signal was 26.5 pA, while the standard deviation of the intra-event Δ*I* signals was typically around 26.8 pA, corresponding to a signal-to-noise ratio of roughly 1.02. The duration of each event was 1 ms.

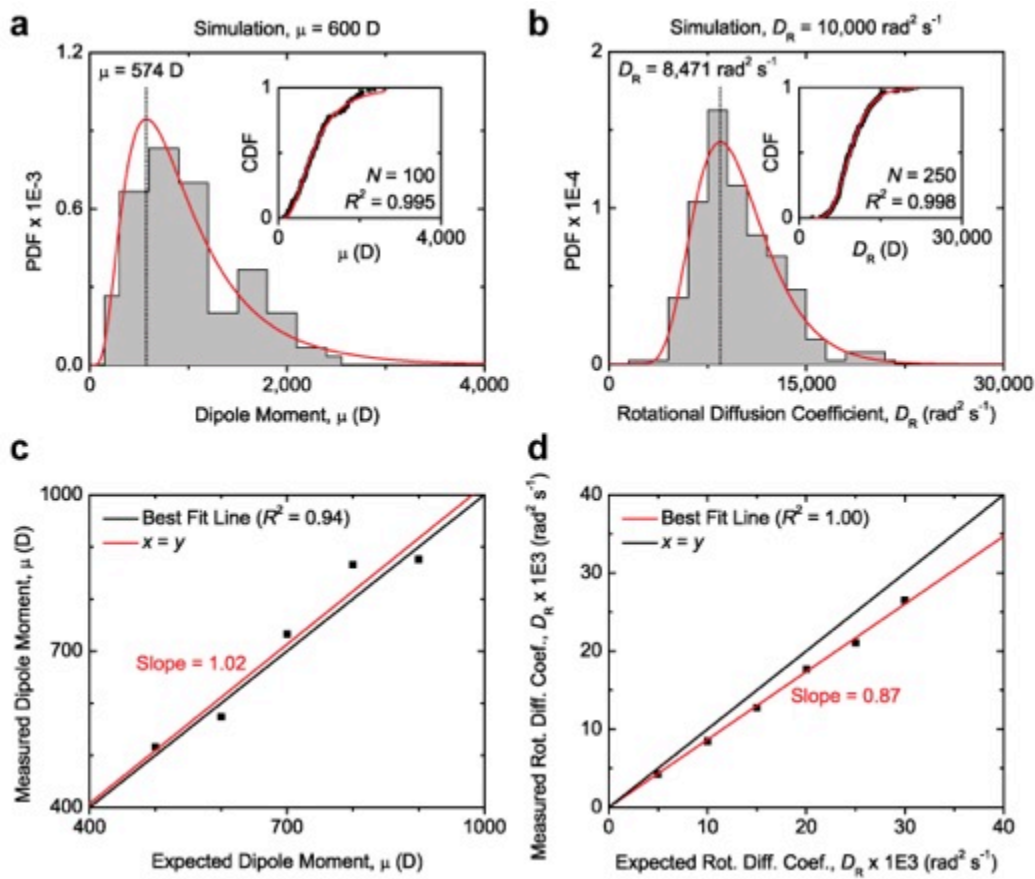

**Supplementary Figure 14. Dipole moments, μ, and rotational diffusion coefficients, $D_R$, determined from analyzing simulated translocation events due to spheroidal particles.** a-b) Distributions of dipole moments and rotational diffusion coefficients determined from analyzing simulated, 1-ms-long translocation events. The inset in each plot shows the empirical cumulative distribution (black squares) fit with a lognormal cumulative distribution function (CDF) (red line). KS-tests indicated the difference between the empirical distribution and best-fit curve was not significant in all cases at a confidence level of α = 0.10. The derivative of the CDF is the probability density function (PDF), which is plotted in red with the histograms of dipole moments and rotational diffusion coefficients. The most probable value is indicated by the dotted black line and corresponds to the maximum of the lognormal fit. c-d) Measured *versus* expected (i.e. input) dipole moments and rotational diffusion coefficients. The ideal outcome wherein the measured values are equal to the input values is shown in black and the best fit line is shown in red. For the simulations where we varied μ, we used input parameters that were based on the experiment done with $IgG_1$ in pore 1. For the simulations where we varied $D_R$, we kept μ fixed at 500 Debyes and did not add noise to the signal; lower signal-to-noise ratios resulted in additional error as expected. Furthermore, we calculated $D_R$ at bandwidths ranging up to 60 kHz for each event to determine the value of $D_R$ at infinite bandwidth, as illustrated in Supplementary Fig. 18a.

**Supplementary Figure 15. Distributions of the length-to-diameter ratio, *m*, and excluded volume, Λ, determined from fitting the convolution model to all intra-event Δ*I* signals longer than 0.4 ms for IgG$_1$ (a-b), GPI-AChE (c-d), Fab (e-f), BSA (g-h), α-Amylase (i-j), and BChE (k-l).** Each box represents the 2$^{nd}$ and 3$^{rd}$ quartiles of the data, the horizontal line is the median value, the square point aligned with the box shows the mean value, and the whiskers extend to data points that are within 1.5 × IQR. Only prolate solutions are shown for GPI-AChE, Fab, and α-amylase. The applied potential was -60 mV for IgG$_1$ using pore 3, -115 mV for GPI-AChE, and -100 mV for all other experiments.

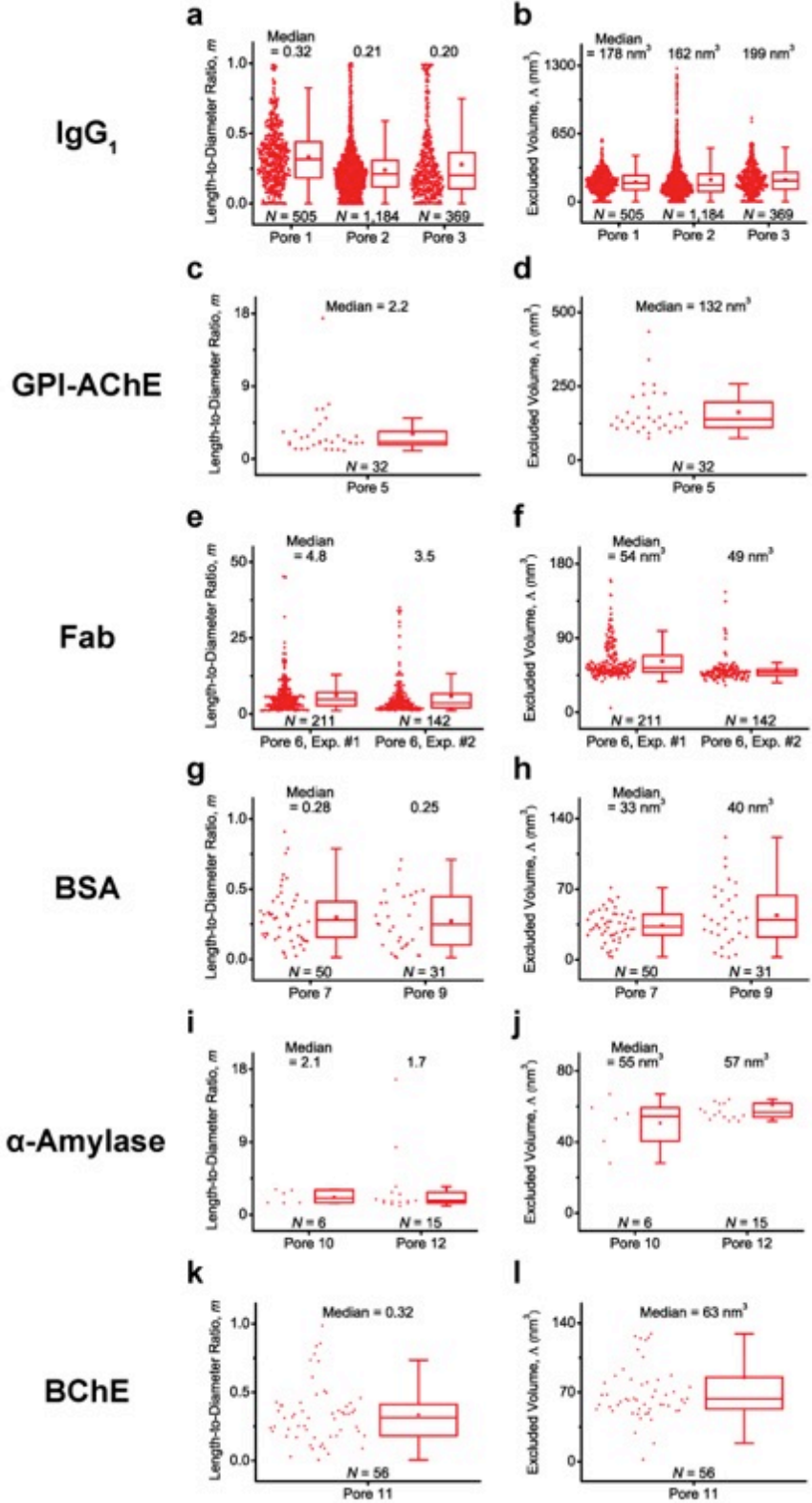

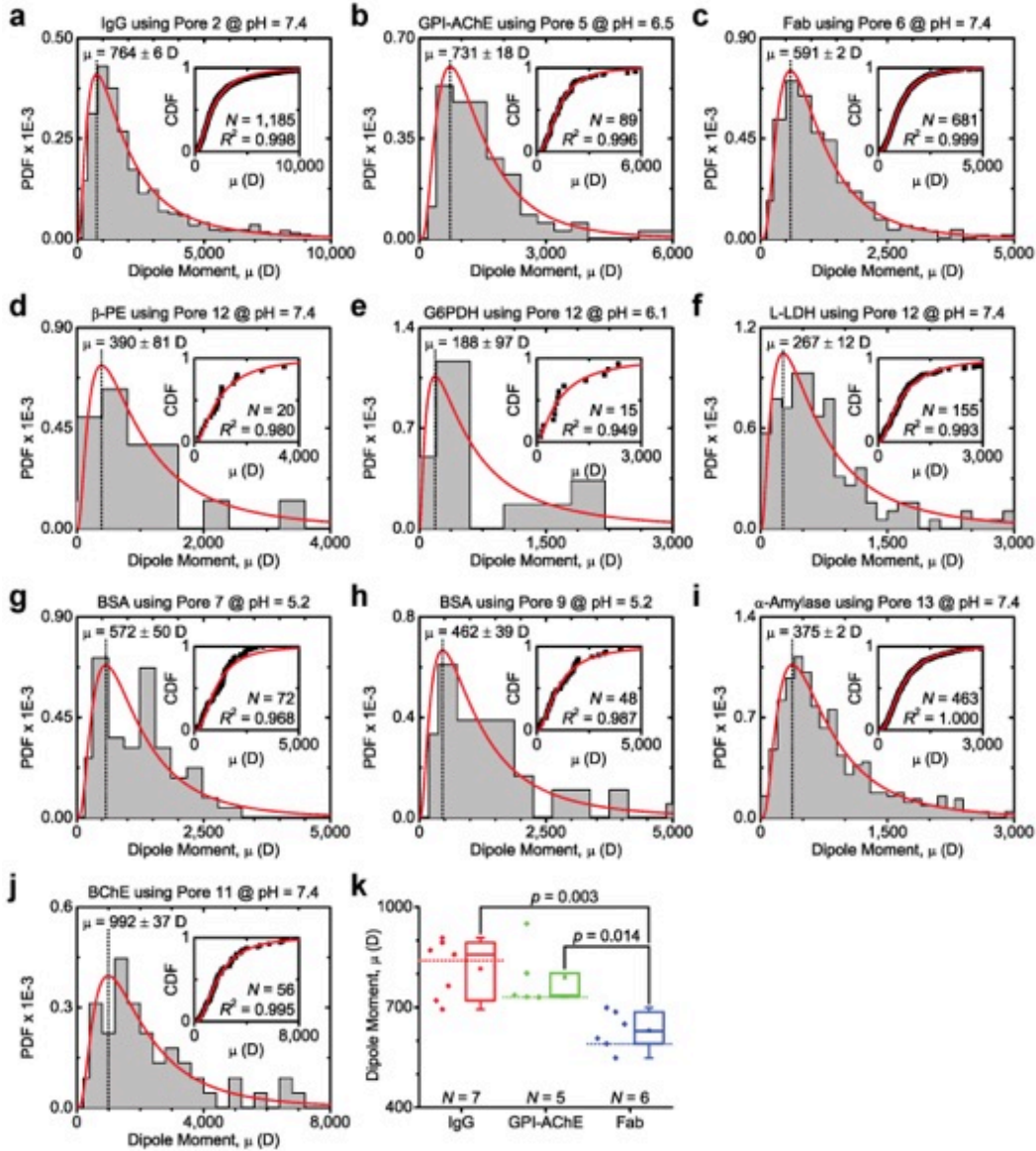

**Supplementary Figure 16. Dipole moments, μ, of IgG$_1$ (a), GPI-AChE (b), Fab (c), β-PE (d), G6PDH (e), L-LDH (f), BSA (g-h), α-amylase (i), and BChE (j) determined from fitting intra-event Δ*I* values with the convolution model.** a-j) The inset in each plot shows the empirical cumulative distribution (black squares) and corresponding fit with a lognormal cumulative distribution function (CDF) (red line). KS-tests indicated the difference between the empirical distribution and best-fit curve was not significant in all cases at a confidence level of α = 0.10. The derivative of the CDF is the probability density function (PDF), which is plotted in red with the histogram of dipole moments. The most probable value of the dipole moment is indicated by the dotted black line and corresponds to the maximum of the lognormal fit. The 95% confidence interval for each most probable value is provided in the plot. During the fitting procedure, only events with durations greater than 0.4 ms were analyzed. The applied potential was -100 mV for all experiments with IgG$_1$, Fab, G6PDH, BSA, and BChE; -115 mV for the experiment with GPI-AChE; -140 mV for the experiment with β-PE; and -200 mV for the experiments with L-LDH and α-amylase. k) A box plot of the most probable value of μ obtained from different experiments for IgG$_1$, GPI-AChE, and Fab. Each box represents the 2$^{nd}$ and 3$^{rd}$ quartiles of the data, the solid horizontal line is the median value, the square point aligned with the box shows the mean value, and the whiskers

extend to data points that are within 1.5 × IQR. The dotted horizontal line indicates the reference value for each protein (see Supplementary Table 4). A Tukey test indicated the distribution for Fab is significantly different than the distributions for the other two proteins ($p$-values are indicated on the plot). For IgG$_1$, we obtained the following values: 908 D in pore 1, 764 D in pore 2, 894 D in pore 3, 871 D in pore 8, 720 D in pore 16, 858 D in pore 17, and 694 in pore 18. For GPI-AChE, we obtained the following values: 731 in pore 5, 736 D in pore 16, 730 D in pore 20, and 950 and 802 D in pore 21. For Fab, we obtained the following values: 548 and 591 D in pore 6, 686 D in pore 11, 699 and 650 D in pore 13, and 607 D in pore 14.

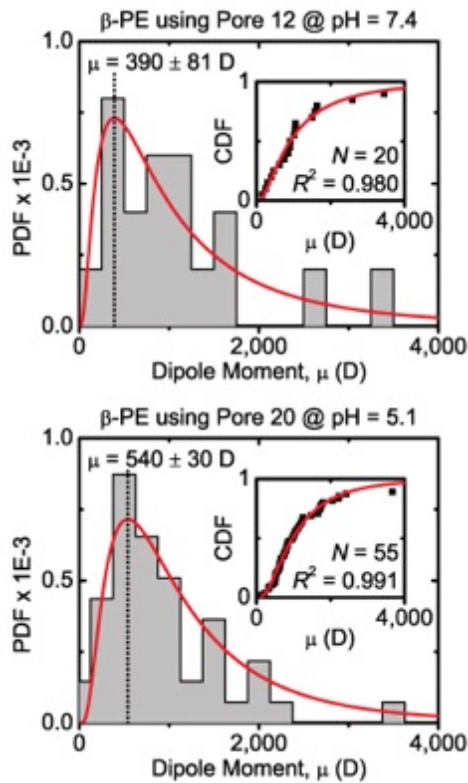

**Supplementary Figure 17. Variation of the dipole moment, μ, of β-phycoerythrin (β-PE) as a function of pH.** The inset in each plot shows the empirical cumulative distribution (black squares) and corresponding fit with a lognormal cumulative distribution function (CDF) (red line). The derivative of the CDF is the probability density function (PDF), which is plotted in red with the histogram of dipole moments. The most probable value of the dipole moment is indicated by the dotted black line and corresponds to the maximum of the lognormal fit. The 95% confidence interval for each most probable value is provided in the plot. Based on theory, β-PE's expected values of μ are 395 D at pH 7.4 and 489 D at pH 5.1, which are in reasonable agreement with the measured values. An additional measurement at pH 5.1 obtained using pore 19 yielded a most probable value with relatively high uncertainty of 774 ± 371 D ($N = 10$). During the fitting procedure, only events with durations greater than 0.4 ms were analyzed. The applied potential was -100 mV in all cases.

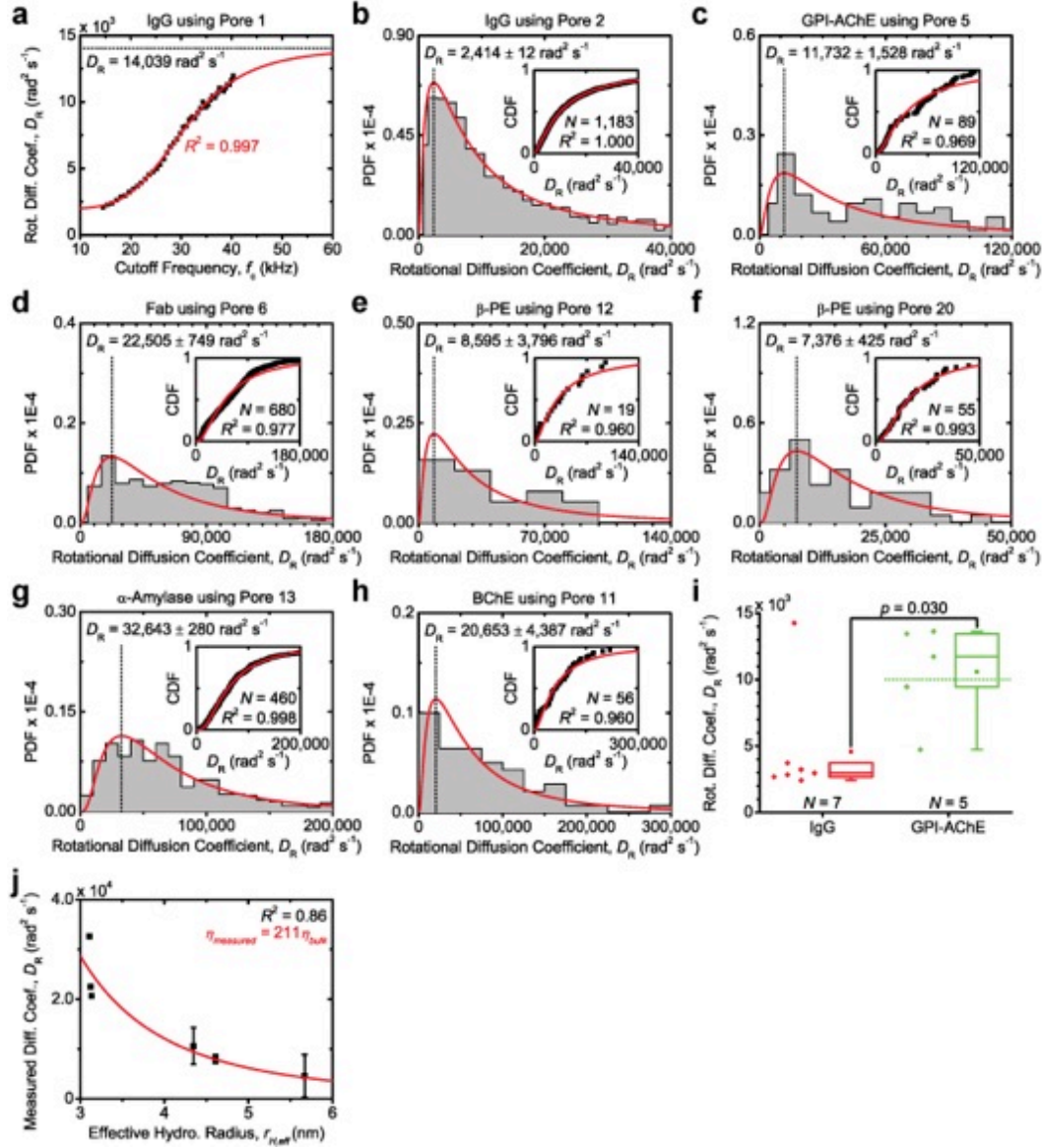

**Supplementary Figure 18. Rotational diffusion coefficients, $D_R$, of IgG$_1$ (a-b), GPI-AChE (c), Fab (d), β-PE (e-f), α-Amylase (g), and BChE (h) determined from analysis of intra-event ΔI values.** a) Rotational diffusion coefficient *versus* the low-pass cutoff frequency for a single event due to IgG$_1$ in pore 1. The curve was fit with the logistic equation to determine $D_R$ at infinite bandwidth, which is denoted by the dotted black line. We used this procedure to determine the values of $D_R$ for all proteins and subsequently generate the histograms in panes (b) through (i). b-h) The inset in each plot shows the empirical cumulative distribution (black squares) fit with a lognormal cumulative distribution function (CDF) (red line). KS-tests indicated the difference between the empirical distribution and best-fit curve was not significant in every case except for the experiment with Fab using Pore 6 (panel (f)) at a confidence level of α = 0.10. The derivative of the CDF is the probability density function (PDF), which is plotted in red with the histogram of rotational diffusion coefficients. The most probable value of the rotational diffusion coefficient is indicated by the dotted black line and corresponds to the maximum of the lognormal fit. The 95% confidence interval for each most probable value is provided in the plot. Only events with durations greater than 0.4 ms were analyzed. The applied potential was -100 mV for all experiments with the IgG$_1$ antibody, Fab, and BChE; -115 mV for the experiment with GPI-AChE; -140

mV for the experiment with β-PE; and -200 mV for the experiment with α-amylase. i) A box plot of the most probable value of $D_R$ obtained from different experiments for IgG$_1$ and GPI-AChE. Each box represents the 2$^{nd}$ and 3$^{rd}$ quartiles of the data, the solid horizontal line is the median value, the square point aligned with the box shows the mean value, and the whiskers extend to data points that are within 1.5 × IQR. The dotted horizontal line for GPI-AChE indicates its reference value. A *t*-test indicated the two distributions are significantly different (the *p*-value is indicated on the plot). For IgG$_1$, we obtained the following values: 14,262 rad$^2$ s$^{-1}$ in pore 1; 2,414 rad$^2$ s$^{-1}$ in pore 2; 3,227 rad$^2$ s$^{-1}$ in pore 3; 2,934 rad$^2$ s$^{-1}$ in pore 8; 2,958 rad$^2$ s$^{-1}$ in pore 16; 2,680 rad$^2$ s$^{-1}$ in pore 17; and 3,727 rad$^2$ s$^{-1}$ in pore 18. For GPI-AChE, we obtained the following values: 11,732 rad$^2$ s$^{-1}$ for pore 5; 4,725 rad$^2$ s$^{-1}$ for pore 16; 9,456 rad$^2$ s$^{-1}$ for pore 20; and 13,630 and 13,446 rad$^2$ s$^{-1}$ for pore 21. j) A fit of the Stokes-Einstein equation for rotational diffusion, $D_R = k_B T / (8\pi\eta r_{H,eff}^3)$, to the measured most probable values of $D_R$. The only fitting parameter was apparent viscosity, η, which had a best-fit value that was 211 times higher than the viscosity of water.

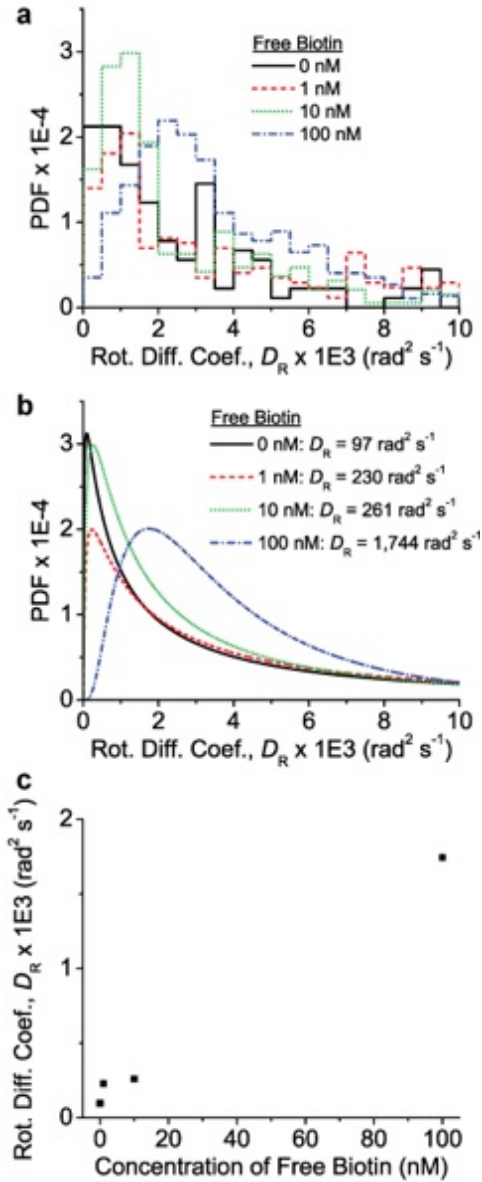

**Supplementary Figure 19. The measured rotational diffusion coefficient, $D_R$, of lipid-anchored IgG$_1$ decreases with the ratio of bivalently-bound to monovalently-bound IgG$_1$.** a) Histograms of $D_R$ values determined from analysis of intra-event $\Delta I$ values. We analyzed 179, 343, 382, and 739 events obtained in the presence of 0, 1, 10, and 100 nM free biotin, respectively. b) Lognormal fits of the $D_R$ distributions. We fit each empirical cumulative distribution with a lognormal cumulative distribution function (not shown). The $R^2$ values of the fits are 0.989, 0.998, 0.987, and 0.996 for 0, 1, 10, and 100 nM free biotin, respectively. The legend displays the most probable value of each fit. c) The most probable value of $D_R$ as a function of the concentration of free biotin. We attribute the increase in $D_R$ with the concentration of free biotin to a decrease in the ratio of bivalently-bound to monovalently-bound IgG$_1$. All recordings were obtained with pore 13 at an applied potential of -100 mV and pH of 7.4.

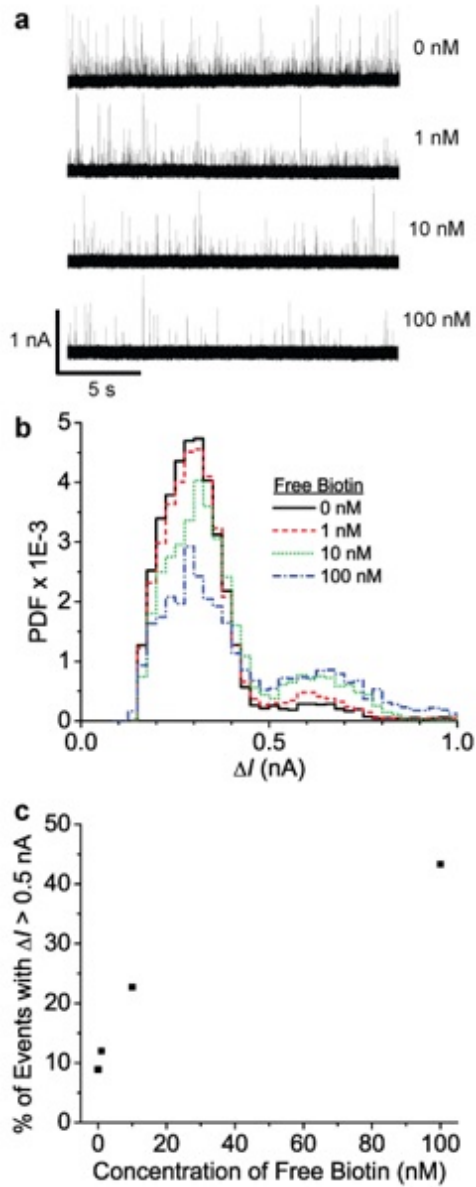

**Supplementary Figure 20. The distribution of maximum Δ*I* values for IgG$_1$ is more biased toward low values when the fraction of bivalently-bound IgG$_1$ is relatively high.** a) Raw current traces low-pass filtered at 15 kHz and obtained in the presence of different concentrations of free biotin. All traces are 20 s long. b) Histograms of maximum Δ*I* values from resistive-pulse recordings. We obtained 7,243, 11,939, 6,279, and 3,598 events in the presence of 0, 1, 10, and 100 nM free biotin, respectively. c) The percentage of events with a maximum Δ*I* value greater than 0.5 nA as a function of the concentration of free biotin. The distribution of maximum Δ*I* values becomes less biased toward low values as the concentration of free biotin increases (i.e. the ratio of monovalently-bound to bivalently-bound IgG$_1$ increases). All recordings were obtained with pore 13 at an applied potential of -100 mV and pH of 7.4.

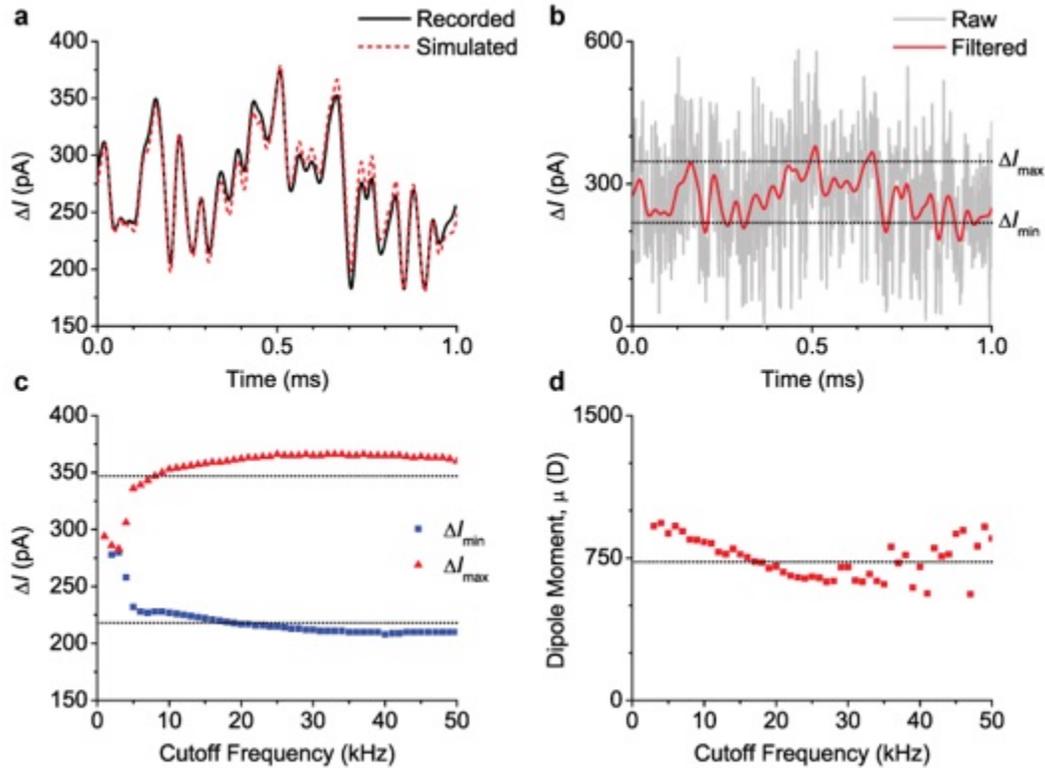

**Supplementary Figure 21. Effect of the recording electronics and low-pass filtering on intra-event $\Delta I$ values.** a) A comparison between a simulated intra-event $\Delta I$ signal that was filtered digitally at 15 kHz (dashed red curve) and a waveform obtained by inputting the unfiltered simulated signal into the experimental setup using a function generator, recording at 500 kHz, and filtering digitally at 15 kHz (black curve). The two curves are nearly identical with an average difference of less than 0.3 pA and a Pearson correlation coefficient of 0.98, indicating the recording electronics do not significantly distort the signal at a bandwidth of 15 kHz. Any deviation between the two curves likely results from additional noise introduced by the recording setup. b) The same intra-event $\Delta I$ signal from (a) before and after filtering digitally at 15 kHz (gray and red curves, respectively). The dotted black lines show the known values for $\Delta I_{min}$ and $\Delta I_{max}$. Although filtering smoothes the signal (and dramatically reduces the noise), the filtered signal still samples $\Delta I_{min}$ and $\Delta I_{max}$ and maintains its bias toward $\Delta I_{min}$. Consequently, fitting the filtered signal with the convolution model still yields accurate values for $\Delta I_{min}$, $\Delta I_{max}$, and μ as shown in (c) and (d). c) Values of $\Delta I_{min}$ and $\Delta I_{max}$ determined from analyzing the same simulated intra-event $\Delta I$ signal at different cutoff frequencies. The dotted black lines show the known values for $\Delta I_{min}$ and $\Delta I_{max}$. The values do not vary considerably with cut-off frequency except at low frequencies (<5 kHz). For instance, there is a 5.6% difference between the values of $\Delta I_{min}$ at 15 and 50 kHz and a 0.6% difference between the values of $\Delta I_{max}$ at 15 and 50 kHz. d) Values of the dipole moment, μ, determined from analyzing the same simulated intra-event $\Delta I$ signal at different cutoff frequencies. The dotted black line shows the known value for μ. Dipole moment has little dependence on cut-off frequency (e.g. there is a 10.2% difference between the values at 15 and 50 kHz), although the results are scattered to a greater degree at high frequencies likely due to a decrease in the signal-to-noise ratio with cut-off frequency. For the simulations, we used input parameters that were based on the expected values for GPI-AChE (see Supplementary Table 4).

**Supplementary Figure 22. Determining the volume and shape of an antibody-antigen complex from individual resistive-pulses.** a) Current trace showing resistive pulses due to the translocation of G6PDH in the absence of antibody. b) Current trace recorded after incubation with 15 μM polyclonal anti-G6PDH IgG for 1 hr. After incubation, we rinsed the chip with recording buffer to remove unbound IgG. c-d) Histograms of maximum Δ$I$ values recorded before and after incubation with anti-G6PDH IgG. Insets show the same data over a reduced y-axis scale. We observed a significant increase in the number of events with large Δ$I$ values after incubation with IgG (e.g. the percentage of events with values larger than 500 pA increased from 0.01 to 9 percent). e) Empirical cumulative distribution (CDF) of Δ$I$ values due to the translocation of G6PDH (grey curve) and the fit of this data to the convolution model (black curve). f) Empirical CDF of Δ$I$ values due to the translocation of *both* G6PDH and the antibody-antigen complex (red dotted curve). To generate a CDF due to the translocation of the complex *only* (i.e. remove Δ$I$ values due to the translocation of unbound G6PDH), we subtracted the CDF due to the translocation of G6PDH *only* (e) after scaling this distribution such that the difference between the two empirical CDFs was minimized at low Δ$I$ values (250 to 350 pA). We expect the majority of Δ$I$ values in this range to result from the translocation of unbound G6PDH. The optimal scaling factor was 0.73, suggesting that roughly 27 percent of translocation events were due to the antibody-antigen complex. g-h) Blue spheroids show the volume and shape of G6PDH and the antibody-antigen complex determined by fitting the

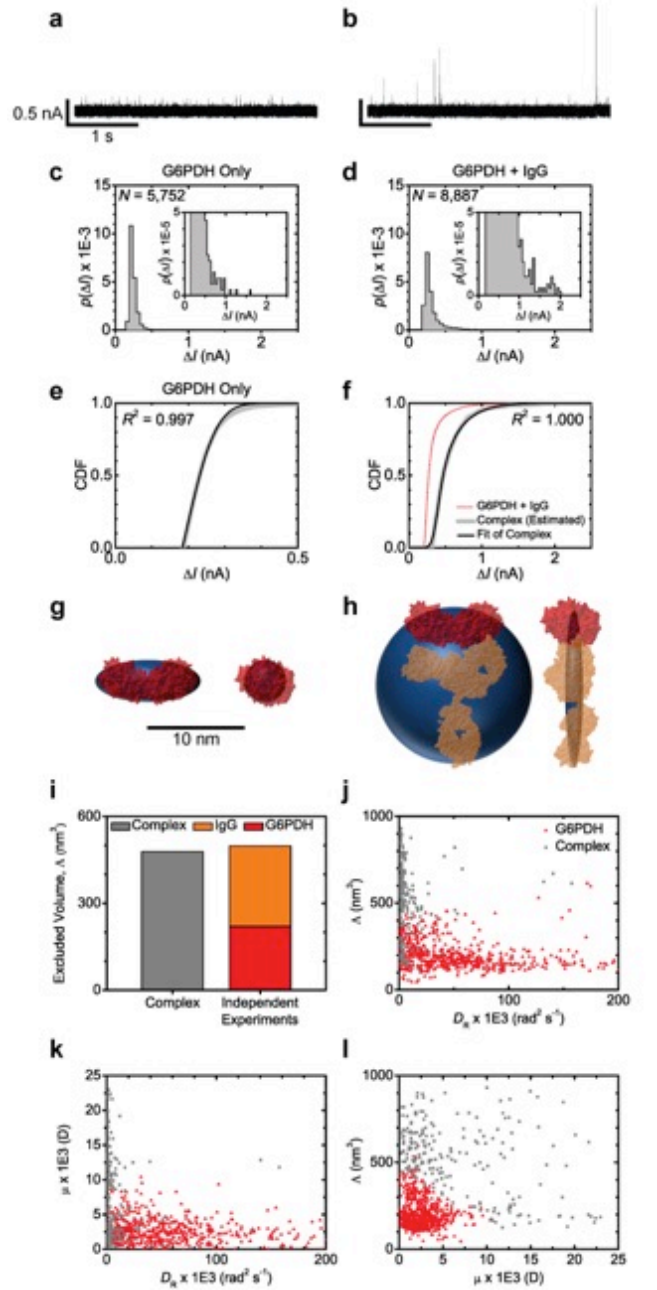

empirical CDFs shown in panes (e) and (f). The crystal structure of G6PDH and IgG are shown in red and orange, respectively. i) Bar plot showing excellent agreement between the volume of the antibody-antigen complex determined from analyzing maximum Δ$I$ values from this experiment and the sum of the volumes of G6PDH and IgG determined that were determined individually in other nanopore experiments (see Supplementary Table 1). j-i) Scatter plots showing the 2D projections of the 3D plot in Fig 5c of the main text. These plots show that resistive pulses assigned to the complex correspond to larger molecular volumes and smaller rotational diffusion coefficients than resistive pulses assigned to G6PDH. The dipole moment of G6PDH is relatively clustered as expected for a protein with well-defined shape and position of amino acids. In contrast, the dipole moment of the complex between G6PDH and the polyclonal anti-G6PDH IgG antibody varies widely as expected since IgG may bind at multiple locations and is a relatively floppy molecule. All recordings were obtained with pore 15 at an applied potential of -100 mV and pH of 6.1. We purchased polyclonal anti-G6PDH IgG (A9521) from Sigma Aldrich, Inc.

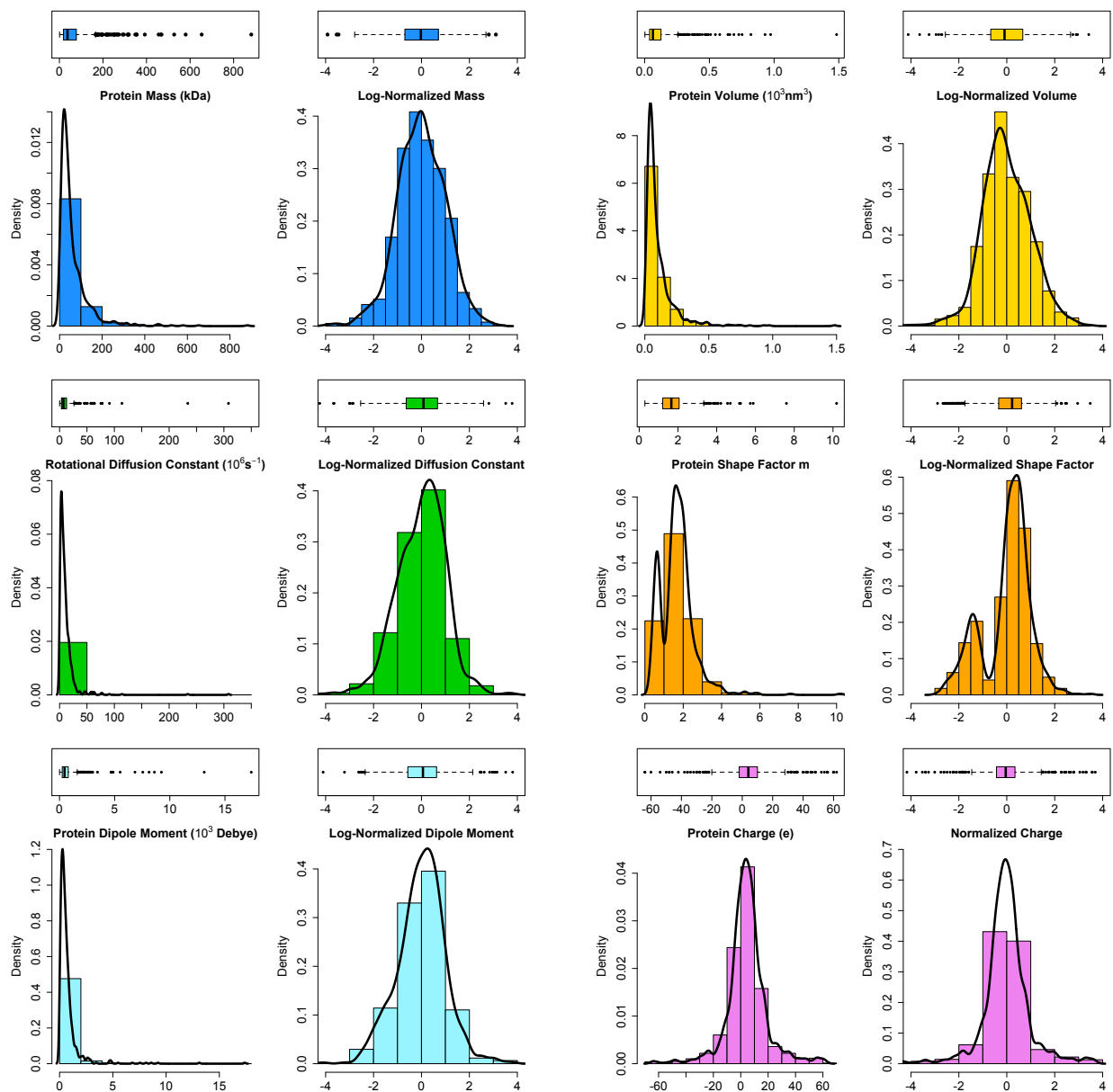

**Supplementary Figure 23. Histograms, boxplots, and density distributions of calculated physical descriptors for 780 proteins.** Using structural and sequence data, we randomly selected a group of proteins and determined their mass, volume, rotational diffusion constant, shape factor, dipole moment, and charge. The distributions on the left show the raw data for each quantity. To properly normalize the data, we first did log-transforms of all quantities, except charge, and then calculated standard normal distributions (shown on the right). As dimensionless, standard normal distributions, we can define a meaningful protein-protein distance in a space that combines multiple descriptors (e.g. charge and mass).

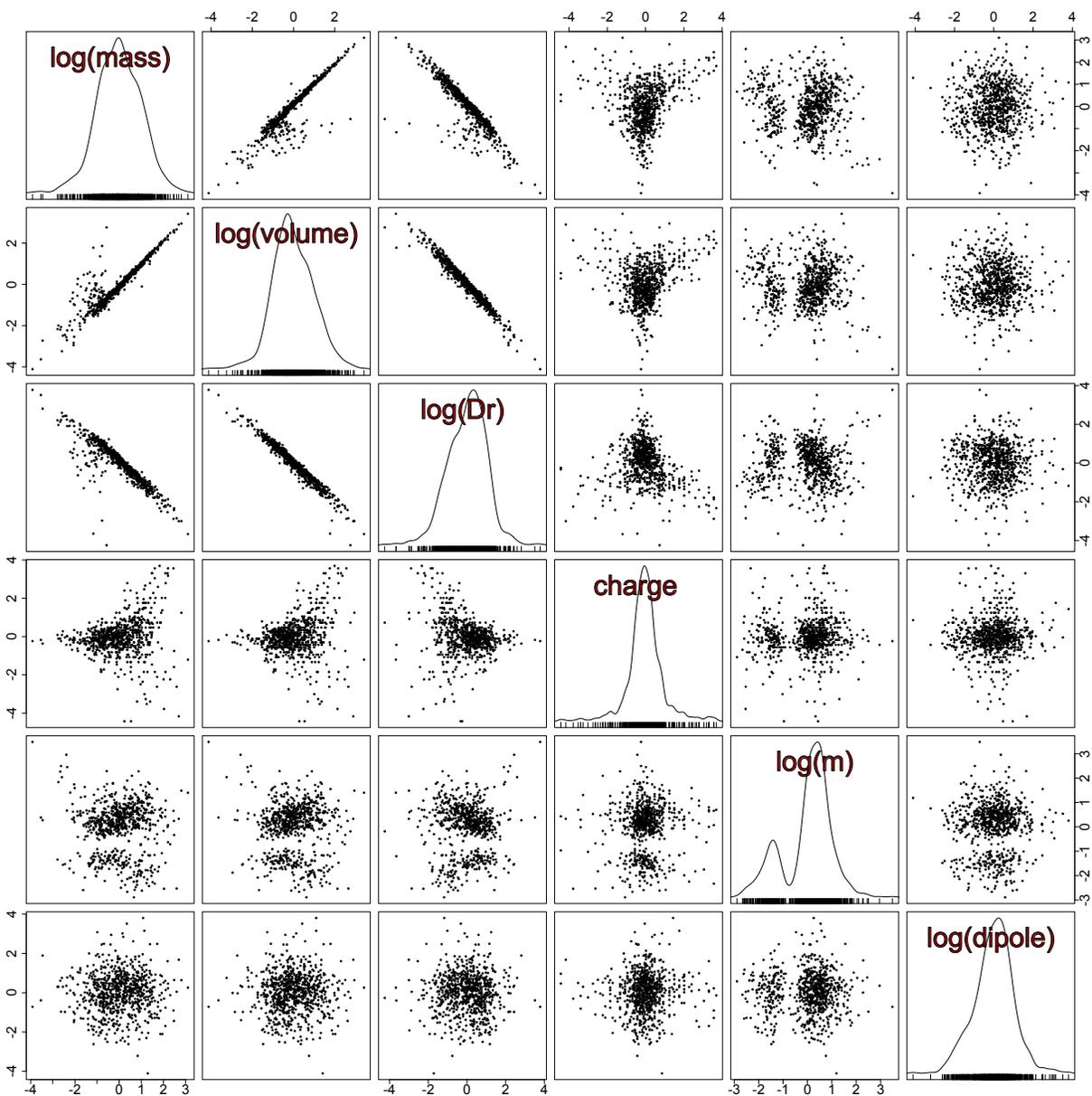

**Supplementary Figure 24. A scatter plot matrix showing the relationships between the log-normalized quantities in Supplementary Figure 23.** Mass, volume, and rotational diffusion constant show a high-degree of correlation; however charge, the length-to-diameter ratio (*m*), and the dipole moment show little correlation with any other descriptor.

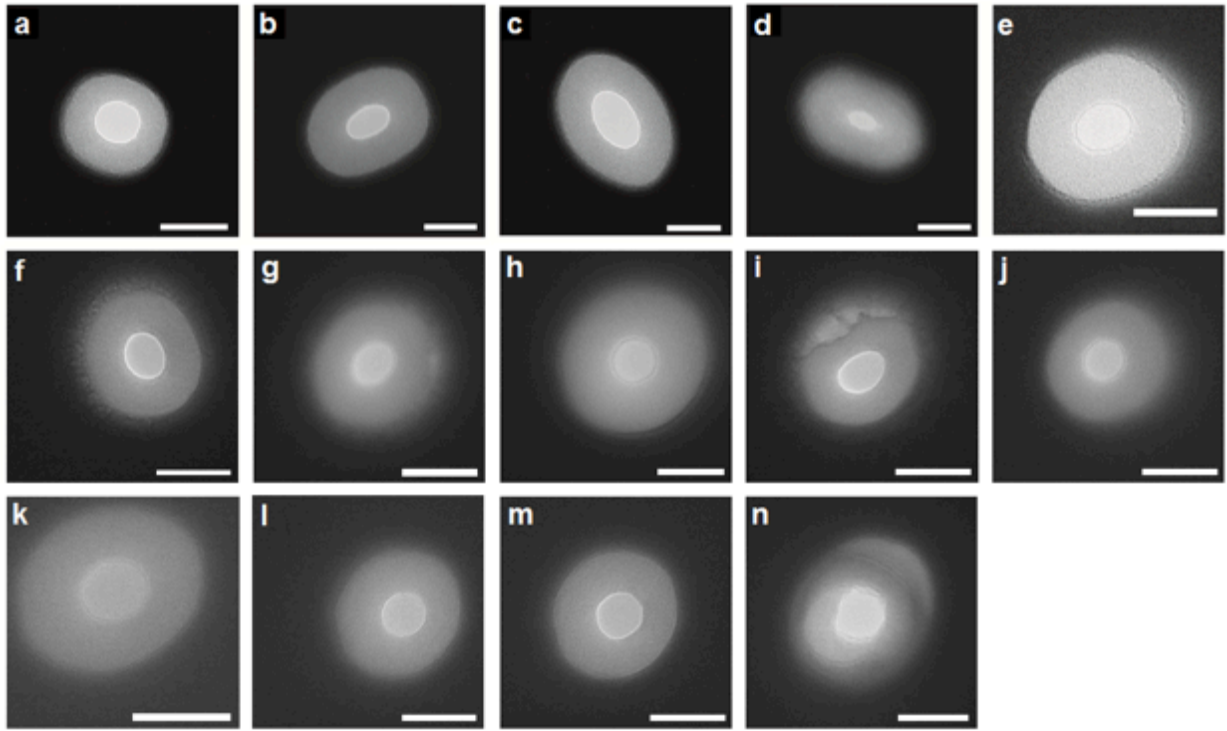

**Supplementary Figure 25. Transmission electron micrographs of the nanopores used in this work.** The brightest part in the center of each image depicts the shape and size of the nanopore and the surrounding circle with reduced brightness reflects the channel leading to the nanopore[2,80]. All scale bars are 50 nm. Nanopores shown are pore 1(a), pore 2 (b), pore 3 (c), pore 4 (d), pore 5 (e), pore 6 (f), pore 7 (g), pore 8 (h), pore 9 (i), pore 10 (j), pore 11 (k), pore 12 (l), pore 13 (m), and pore 14 (n). The dimensions of the nanopores (in units of nm) without the lipid bilayer coating were: for pore 1 $r_P = 16.1$ and $l_P = 21.3$; for pore 2 $r_P = 16.4$ and $l_P = 17.3$; for pore 3 $r_P = 22.7$ and $l_P = 16.2$; for pore 4 $r_P = 9.6$ and $l_P = 18.0$; for pore 5 $r_P = 16.0$ and $l_P = 15.0$; for pore 6 $r_P = 14.2$ and $l_P = 10.0$; for pore 7 $r_P = 14.0$ and $l_P = 15.4$; for pore 8 $r_P = 17.8$ and $l_P = 15.5$; for pore 9 $r_P = 14.7$ and $l_P = 18.0$; for pore 10 $r_P = 13.6$ and $l_P = 14.0$; for pore 11 $r_P = 16.0$ and $l_P = 12.0$; for pore 12 $r_P = 14.5$ and $l_P = 10.0$; for pore 13 $r_P = 15.7$ and $l_P = 12.0$; for pore 14 $r_P = 15.0$ and $l_P = 10.9$; for pore 15 $r_P = 21.3$ and $l_P = 19.7$; for pore 16 $r_P = 16.4$ and $l_P = 30.0$; for pore 17 $r_P = 18.5$ and $l_P = 30.0$; for pore 18 $r_P = 18.0$ and $l_P = 30.0$; for pore 19 $r_P = 16.5$ and $l_P = 30.0$; for pore 20 $r_P = 15.5$ and $l_P = 30.0$; and for pore 21 $r_P = 15.5$ and $l_P = 6.0$.

**Supplementary Table 1.** Values of fitting parameters determined from fitting the convolution model to the empirical distributions of $\Delta I$ values (Fig. 3 and 4 in the main text) as well as the resulting calculations of protein volume, $\Lambda$, and shape parameter, $m$.

| Experiment | $E^{\dagger}$ (MV m$^{-1}$) | $\Delta I_{min}$ (pA) | $\Delta I_{max}$ (pA) | $\sigma$ (pA) | $\mu$ (D) | $R^2$ | $\Lambda^*$ (nm$^3$) | $m^*$ |
|---|---|---|---|---|---|---|---|---|
| IgG$_1$, Pore 1 | -1.5 | 329 | 678 | 58 | 596 | 0.998 | 292 | 0.37 |
| IgG$_1$, Pore 2 | -1.6 | 258 | 1,320 | 65 | 1,911 | 1.000 | 223 | 0.13 |
| Intra-event (Fig. 4) | -1.6 | 281 | 938 | 48 | 302 | 0.998 | 232 | 0.21 |
| IgG$_1$, Pore 3 | -0.6 | 164 | 483 | 21 | 2,020 | 0.997 | 319 | 0.24 |
| IgG$_1$, Pore 8 | -1.4 | 266 | 1132 | 64 | 1,493 | 0.999 | 217 | 0.16 |
| GPI-AChE, Pore 3 | -1.0 | 280 | 375 | 14 | 3,530 | 0.999 | 278 *or* **306** | 0.64 *or* **1.8** |
| GPI-AChE, Pore 5 | -1.3 | 279 | 451 | 40 | 1,712 | 0.999 | 222 *or* **259** | 0.50 *or* **3.1** |
| Fab, Pore 6 | -2.1 | 178 | 231 | 11 | 972 | 1.000 | 71 *or* **77** | 0.67 *or* **1.6** |
| β-PE, Pore 6 | -0.8 | 181 | 302 | 31 | 2,125 | 0.999 | **192** *or* 227 | **0.48** *or* 3.5 |
| G6PDH, Pore 7 | -1.0 | 178 | 264 | 12 | 3,590 | 0.999 | 193 *or* **220** | 0.56 *or* **2.3** |
| G6PDH, Pore 15 | -1.1 | 169 | 254 | 58 | 2,822 | 0.997 | 181 *or* **207** | 0.55 *or* **2.4** |
| L-LDH, Pore 8 | -0.8 | 195 | 296 | 16 | 2,802 | 0.999 | **267** *or* 307 | **0.54** *or* 2.5 |
| BSA, Pore 7 | -1.9 | 165 | 258 | 17 | 1,263 | 0.998 | **91** *or* 105 | **0.52** *or* 2.7 |
| BSA, Pore 9 | -1.7 | 165 | 276 | 13 | 2,925 | 0.998 | **110** *or* 130 | **0.48** *or* 3.5 |
| α-Amylase, Pore 10 | -1.6 | 157 | 196 | 5 | 1,243 | 1.000 | 92 *or* 99 | 0.71 *or* **1.5** |
| BChE, Pore 11 | -1.7 | 150 | 364 | 18 | 1,007 | 1.000 | 82 | 0.30 |

$^{\dagger}$ The electric field intensity was calculated according to the following equation: $E = V_A * R_p / (R_{total} * l_p)$, where $R_p$ is the resistance of the pore, $R_{total}$ is the total resistance of the circuit, and $l_p$ is the length of the pore. * Values of $\Lambda$ and $m$ shown in bold are those corresponding to the correct shape (i.e. the shape that matches the crystal structure).

**Supplementary Table 2. Estimated hydration shell thickness of proteins detected in this work.**
Based on the difference between the volume that we measured and the volume determined from crystal structures, we estimated the thickness of the hydration shell and the average number of water molecules required in this ordered water layer. The average hydration shell thickness is 0.34 ± 0.14 nm, which closely matches reported values that range from 0.3 to 0.5 nm[81-84]. An alternative explanation for the difference between the measured and crystal structure volumes is that counter ions, rather than water molecules, bind tightly to the proteins[85].

| Protein | Volume, Λ (nm$^3$) | | Hydration Shell Thickness | |
|---|---|---|---|---|
| | Measured | Crystal Structure | (nm) | $N_{H2O}$[†] |
| IgG$_1$ | 278 | 174 | 0.37 | 1.3 |
| GPI-Acetylcholinesterase | 283 | 145 | 0.53 | 1.9 |
| Fab Fragment | 77 | 56 | 0.21 | 0.8 |
| β-Phycoerythrin | 192 | 139 | 0.29 | 1.0 |
| Glucose-6-Phosphate Dehydrogenase | 220 | 135 | 0.40 | 1.4 |
| L-Lactate Dehydrogenase | 267 | 160 | 0.46 | 1.6 |
| Bovine Serum Albumin | 101 | 78 | 0.19 | 0.7 |
| α-Amylase | 99 | 65 | 0.29 | 1.0 |
| Butyrylcholinesterase | 82 | 69 | 0.12 | 0.4 |
| Streptavidin | 110 | 61 | 0.53 | 1.9 |

[†] The number of water molecules was calculated by dividing the thickness by the diameter of a water molecule (0.28 nm)[86].

**Supplementary Table 3. Estimated volume of proteins detected in this work from dynamic light scattering (DLS) measurements.** The volume of each protein was estimated from the hydrodynamic radius that was obtained *via* DLS; the estimate of volume required assuming that the protein was a perfect sphere (i.e. spherical) or was spheroidal (i.e. ellipsoidal).

| Protein | Hydrodynamic Radius from DLS, $r_H$ (nm) | Spherical Volume from DLS (nm$^3$) | Spheroidal Volume from DLS (nm$^3$)[a] | | Measured Volume from Nanopore Experiments (nm$^3$) |
|---|---|---|---|---|---|
| | | | Measured $m$ | Reference $m$ | |
| IgG$_1$ | 5.29 | 620 | 391 | 339–548 | 278 |
| GPI-AChE | 4.59 | 405 | 330 | 300 | 283 |
| Fab | 3.29 | 149 | 141 | 136–138 | 77 |
| β-PE | 3.83 | 235 | 205 | 179 | 192 |
| G6PDH | 3.95 | 257 | 214 | 206 | 220 |
| L-LDH | 4.07 | 282 | 256 | 261 | 267 |
| BSA | 3.38 | 162 | 143 | 150 | 101 |
| α-Amylase | 2.90 | 102 | 97 | 93 | 99 |
| Streptavidin | 2.82 | 94 | N/A | 94 | 110 |

[a] To calculate the volume of a spheroid particle that would return the hydrodynamic radius measured in DLS experiments, we set the value of $m$ to those determined in nanopore experiments (measured $m$) or to those determined from crystal structures of the proteins (reference $m$).

**Supplementary Table 4.** Average volumes, length-to-diameter ratios, $m$ = A/B, most probable dipole moments, rotational diffusion coefficients, and charges of proteins determined by analysis of resistive pulses and other methods.

| Protein | Volume, $\Lambda^a$ (nm$^3$) Meas. | Ref. | Length-to-Diameter Ratio, $m^a$ Meas. | Ref. | Rotational Diffusion Coef., $D_R^b$ (rad$^2$ s$^{-1}$) Meas. | Ref. Tethered | Bulk[c] | Dipole Moment, $\mu^b$ (D) Meas. | Ref. | Charge, $z$ Meas. | Ref. |
|---|---|---|---|---|---|---|---|---|---|---|---|
| IgG$_1$ | 263 ± 51 | DLS[d]: 391<br>Theor.[e]: 266<br>Lit.: 347 ± 15[87] | 0.23 ± 0.11 | 0.2–0.5[88,89] | 4,586 ± 4,287 | -- | 8.96E5 | 816 ± 88, pH 7.4 | 840[f,g] | -4.2[h], pH 7.4 | -4.6[h,i] |
| GPI-AChE | 283 ± 33 | DLS: 330<br>Theory: 195 | 2.4 ± 0.9 | 2.9[j] | 10,598 ± 3,687 | 10,000[44] | 1.99E6 | 790 ± 95, pH 6.5 | 730[k] | -4.5, pH 6.5 | -- |
| Fab | 77 | DLS: 141<br>Theor.: 97<br>Lit.: 140[90]<br>170 ± 31[2] | 1.6 | 1.7[j]<br>1.8[91] | 22,505 | -- | 5.37E6 | 630 ± 58, pH 7.4 | 630[f]<br>550[k] | -2.9[h], pH 7.4 | -5.3[h,i] |
| β-PE | 192 | DLS: 205<br>Theor.: 194 | 0.48 | 0.35[j] | 7,986 ± 862 | -- | 1.67E6 | 390, pH 7.4 | 395[l] | 6.8, pH 5.1 | 10.5[i,m,n] |
| G6PDH | 220 ± 9 | DLS: 214<br>Theor.: 222 | 2.3 ± 0.1 | 2.5[j] | -- | -- | -- | 188, pH 6.1 | 203[l] | 9.6, pH 6.1 | 15.0[i,m,n] |
| L-LDH | 267 | DLS: 256<br>Theory: 220 | 0.54 | 0.58[j] | -- | -- | -- | 267, pH 7.4 | 206[l] | -5.5, pH 6.1 | -11.7[i,m,n] |
| BSA | 101 ± 13 | DLS: 143<br>Theor.: 111<br>Lit: 109[92]<br>123[93] | 0.50 ± 0.03 | 0.57[j] | -- | -- | -- | 522 ± 78, pH 5.2 | 410[f] | -6.4, pH 5.2 | -3.4[74,i,n] |
| α-Amylase | 99 | DLS: 97<br>Theor.: 89 | 1.5 | 1.8[j] | 32,643 | -- | 5.44E6 | 375, pH 7.4 | 484[l] | -5.3, pH 7.4 | -10.6[i,m,n] |
| BChE | 82 | Theor.: 103 | 0.30 | 0.47[j] | 20,653 | -- | 5.32E6 | 992, pH 7.4 | 1,420[k] | -3.5, pH 7.4 | -- |
| SA[o] | 110 ± 25[p] | DLS: 94<br>Theor.: 88<br>Lit.: 94 ± 18[2]<br>105 ± 3[94] | 1[k] | 1.1[j] | -- | -- | -- | -- | -- | -0.8, pH 7.4 | -2.8[i,m] |

[a] Determined from fitting distributions of maximum $\Delta I$ values; see Supplementary Note 2 for details. [b] Most probable values determined from intra-event fitting; see Supplementary Note 6 for details. [c] An estimate of the rotational diffusion coefficient in bulk solution determined from the crystal structure of the protein using the software HydroPRO. [d] Calculated from the hydrodynamic radius measured *via* DLS; see Supplementary Note 2 for details. [e] An estimate of the volume

of the hydrated protein determined from the crystal structure of the protein using the software HydroPRO. [f] Measured *via* dielectric impedance spectroscopy. [g] This value should be used as a loose approximation due to the low signal-to-noise ratio of the measurement. [h] Results from Yusko *et al.*[2]. [i] Values were reduced by 1 to account for the charge of the lipid anchor. [j] Estimated from the crystal structure of the protein. [k] Calculated from the crystal structure of the protein using the software HydroPRO. [l] Calculated from the crystal structure of the protein using the Weizmann server (http://bioinfo.weizmann.ac.il/dipol/). [m] Estimated using the PROPKA web interface (http://propka.ki.ku.dk/)[75-78]. [n] Values were reduced by 0.93 to account for the reaction of a primary amine on the protein surface with an NHS ester on the crosslinker molecule to form an amide bond[79]. All estimates were done in the absence of ligands except for G6PDH. [o] Parameters determined using pore 4. [p] Since the distribution of Δ$I$ values due to streptavidin translocations was unimodal and Normal, we assumed that streptavidin had a spherical shape, and therefore $m = 1$; to calculate the excluded volume of streptavidin, we solved equation (1) with γ set to a value of 1.5.

## Supplementary References

1 Uram, J. D., Ke, K. & Mayer, M. Noise and bandwidth of current recordings from submicrometer pores and nanopores. *ACS Nano* **2**, 857-872 (2008).
2 Yusko, E. C. *et al.* Controlling protein translocation through nanopores with bio-inspired fluid walls. *Nat. Nanotechnol.* **6**, 253-260 (2011).
3 Bermudez, O. & Forciniti, D. Aggregation and denaturation of antibodies: A capillary electrophoresis, dynamic light scattering, and aqueous two-phase partitioning study. *J. Chromatogr. B* **807**, 17-24 (2004).
4 Jossang, T., Feder, J. & Rosenqvist, E. Photon-correlation spectroscopy of human-IgG. *J. Protein Chem.* **7**, 165-171 (1988).
5 Goodsell, D. *Acetylcholinesterase. June 2004 molecule of the month.*, <http://www.rcsb.org/pdb/101/motm.do?momID=54> (2004).
6 Sigma-Aldrich. *Product information: Acetylcholinesterase from human erythrocytes (c0663)*, <http://www.sigmaaldrich.com/catalog/product/sigma/c0663> (2012).
7 Tan, R. C., Truong, T. N., McCammon, J. A. & Sussman, J. L. Acetylcholinesterase: Electrostatic steering increases the rate of ligand binding. *Biochemistry* **32**, 401-403 (1993).
8 Antosiewicz, J., Wlodek, S. T. & McCammon, J. A. Acetylcholinesterase: Role of the enzyme's charge distribution in steering charged ligands toward the active site. *Biopolymers* **39**, 85-94 (1996).
9 Luk, W. K. W., Chen, V. P., Choi, R. C. Y. & Tsim, K. W. K. N-linked glycosylation of dimeric acetylcholinesterase in erythrocytes is essential for enzyme maturation and membrane targeting. *Febs J.* **279**, 3229-3239 (2012).
10 Porschke, D. *et al.* Electrooptical measurements demonstrate a large permanent dipole moment associated with acetylcholinesterase. *Biophys. J.* **70**, 1603-1608 (1996).
11 Parihar, M. S. & Hemnani, T. Alzheimer's disease pathogenesis and therapeutic interventions. *J. Clin. Neurosci.* **11**, 456-467 (2004).
12 Rath, A. & Deber, C. M. Correction factors for membrane protein molecular weight readouts on sodium dodecyl sulfate-polyacrylamide gel electrophoresis. *Anal. Biochem.* **434**, 67-72 (2013).
13 Skinner, G. M., van den Hout, M., Broekmans, O., Dekker, C. & Dekker, N. H. Distinguishing single- and double-stranded nucleic acid molecules using solid-state nanopores. *Nano. Lett.* **9**, 2953-2960 (2009).
14 Maxwell, J. C. *A treatise on electricity and magnetism*. 3rd edn, 435-441 (Clarendon Press, 1904).
15 Deblois, R. W. & Bean, C. P. Counting and sizing of submicron particles by resistive pulse technique. *Rev. Sci. Instrum.* **41**, 909-915 (1970).
16 Grover, N. B., Naaman, J., Ben-sasson, S. & Doljansk, F. Electrical sizing of particles in suspensions. I.Theory. *Biophys. J.* **9**, 1398-1414 (1969).
17 Han, A. P. *et al.* Label-free detection of single protein molecules and protein-protein interactions using synthetic nanopores. *Anal. Chem.* **80**, 4651-4658 (2008).
18 Ito, T., Sun, L. & Crooks, R. M. Simultaneous determination of the size and surface charge of individual nanoparticles using a carbon nanotube-based Coulter counter. *Anal. Chem.* **75**, 2399-2406 (2003).
19 Golibersuch, D. C. Observation of aspherical particle rotation in Poiseuille flow via the resistance pulse technique. Part 1. Application to human erythrocytes. *Biophys. J.* **13**, 265-280 (1973).
20 Hurley, J. Sizing particles with a Coulter counter. *Biophys. J.* **10**, 74-79 (1970).
21 Soni, G. V. & Dekker, C. Detection of nucleosomal substructures using solid-state nanopores. *Nano. Lett.* (2012).
22 Raillon, C. *et al.* Nanopore detection of single molecule RNAP-DNA transcription complex. *Nano. Lett.* **12**, 1157-1164 (2012).

23  DeBlois, R. W., Uzgiris, E. E., Cluxton, D. H. & Mazzone, H. M. Comparative measurements of size and polydispersity of several insect viruses. *Anal. Biochem.* **90**, 273-288 (1978).
24  Smythe, W. R. Flow around a spheroid in a circular tube. *Phys. Fluids* **7**, 633-638 (1964).
25  Qin, Z. P., Zhe, J. A. & Wang, G. X. Effects of particle's off-axis position, shape, orientation and entry position on resistance changes of micro Coulter counting devices. *Meas. Sci. Technol.* **22** (2011).
26  Fologea, D., Ledden, B., David, S. M. & Li, J. Electrical characterization of protein molecules by a solid-state nanopore. *Appl. Phys. Lett.* **91**, 053901 (2007).
27  Soni, G. V. & Dekker, C. Detection of nucleosomal substructures using solid-state nanopores. *Nano. Lett.* **12**, 3180-3186 (2012).
28  Wei, R., Gatterdam, V., Wieneke, R., Tampe, R. & Rant, U. Stochastic sensing of proteins with receptor-modified solid-state nanopores. *Nat. Nanotechnol.* **7**, 257-263 (2012).
29  Han, A. *et al.* Sensing protein molecules using nanofabricated pores. *Appl. Phys. Lett.* **88**, 093901 (2006).
30  Sexton, L. T. *et al.* Resistive-pulse studies of proteins and protein/antibody complexes using a conical nanotube sensor. *J. Am. Chem. Soc.* **129**, 13144-13152 (2007).
31  Lan, W.-J., Kubeil, C., Xiong, J.-W., Bund, A. & White, H. S. Effect of surface charge on the resistive pulse waveshape during particle translocation through glass nanopores. *J. Phys. Chem. C.* **118**, 2726-2734 (2014).
32  Fan, R. *et al.* DNA translocation in inorganic nanotubes. *Nano. Lett.* **5**, 1633-1637 (2005).
33  Chang, H. *et al.* DNA-mediated fluctuations in ionic current through silicon oxide nanopore channels. *Nano. Lett.* **4**, 1551-1556 (2004).
34  Das, S., Dubsky, P., van den Berg, A. & Eijkel, J. Concentration polarization in translocation of DNA through nanopores and nanochannels. *Physical Review Letters* **108**, 138101 (2012).
35  Japrung, D. *et al.* Single-molecule studies of intrinsically disordered proteins using solid-state nanopores. *Anal. Chem.* **85**, 2449-2456 (2013).
36  Fricke, H. The electric permittivity of a dilute suspension of membrane-covered ellipsoids. *J. Appl. Phys.* **24**, 644-646 (1953).
37  Fricke, H. A mathematical treatment of the electrical conductivity of colloids and cell suspensions. *Journal of General Physiology* **6**, 375-384 (1924).
38  Velick, S. & Gorin, M. The electrical conductance of suspensions of ellipsoids and its relation to the study of avian erythrocytes. *Journal of General Physiology* **23**, 753-771 (1940).
39  Golibersuch, D. C. Observation of aspherical particle rotation in Poiseuille flow via the resistance pulse technique. Part 2. Application to fused sphere dumbbells. *J. Appl. Phys.* **44**, 2580-2584 (1973).
40  Deblois, R. W. & Wesley, R. K. A. Viral sizes, concentrations, and electrophoretic mobilities by nanopar analyzer. *Biophys. J.* **16**, A178-A178 (1976).
41  Osborn, J. A. Demagnetizing factors of the general ellipsoid. *Physical Review* **67**, 351-357 (1945).
42  Vogel, S. *Life in moving fluids: The physical biology of flow*. (Princeton University Press, 1994).
43  Axelrod, D., Koppel, D. E., Schlessinger, J., Elson, E. & Webb, W. W. Mobility measurement by analysis of fluorescence photobleaching recovery kinetics. *Biophys. J.* **16**, 1055-1069 (1976).
44  Yuan, Y. & Axelrod, D. Subnanosecond polarized fluorescence photobleaching - rotational diffusion of acetylcholine-receptors on developing muscle-cells. *Biophys. J.* **69**, 690-700 (1995).
45  Timbs, M. M. & Thompson, N. L. Slow rotational mobilities of antibodies and lipids associated with substrate-supported phospholipid monolayers as measured by polarized fluorescence photobleaching recovery. *Biophys. J.* **58**, 413-428 (1990).
46  Liu, S., Yuzvinsky, T. D. & Schmidt, H. Effect of fabrication-dependent shape and composition of solid-state nanopores on single nanoparticle detection. *ACS Nano* **7**, 5621-5627 (2013).
47  Jr., I. T., Sauer, K., Wang, J. C. & Puglisi, J. D. *Physical chemistry: Principles and applications in biological sciences.*, (Pearson Education, 2002).

48  Freedman, K. J., Haq, S. R., Edel, J. B., Jemth, P. & Kim, M. J. Single molecule unfolding and stretching of protein domains inside a solid-state nanopore by electric field. *Sci. Rep.* **3** (2013).
49  Oukhaled, A. *et al.* Dynamics of completely unfolded and native proteins through solid-state nanopores as a function of electric driving force. *ACS Nano* **5**, 3628-3638 (2011).
50  Howard, J. *Mechanics of motor proteins and the cytoskeleton*.  (Sinauer Associates Inc. , 2001).
51  Ishijima, A. *et al.* Multiple- and single-molecule analysis of the actomyosin motor by nanometer piconewton manipulation with a microneedle: Unitary steps and forces. *Biophys. J.* **70**, 383-400 (1996).
52  Lewalle, A., Steffen, W., Stevenson, O., Ouyang, Z. & Sleep, J. Single-molecule measurement of the stiffness of the rigor myosin head. *Biophys. J.* **94**, 2160-2169 (2008).
53  Spiering, A., Getfert, S., Sischka, A., Reimann, P. & Anselmetti, D. Nanopore translocation dynamics of a single DNA-bound protein. *Nano. Lett.* **11**, 2978-2982 (2011).
54  Mathé, J., Aksimentiev, A., Nelson, D. R., Schulten, K. & Meller, A. Orientation discrimination of single-stranded DNA inside the alpha-hemolysin membrane channel. *Proc. Natl. Acad. Sci. U. S. A.* **102**, 12377-12382 (2005).
55  Di Fiori, N. *et al.* Optoelectronic control of surface charge and translocation dynamics in solid-state nanopores. *Nat. Nanotechnol.* **8**, 946-951 (2013).
56  Menon, M. K. & Zydney, A. L. Effect of ion binding on protein transport through ultrafiltration membranes. *Biotechnol. Bioeng.* **63**, 298-307 (1999).
57  Tao, T. Time-dependent fluorescence depolarization and brownian rotational diffusion coefficients of macromolecules. *Biopolymers* **8**, 609-632 (1969).
58  García de la Torre, J., Huertas, M. L. & Carrasco, B. Hydronmr: Prediction of NMR relaxation of globular proteins from atomic-level structures and hydrodynamic calculations. *Journal of Magnetic Resonance* **147**, 138-146 (2000).
59  Gauthier, M. G. & Slater, G. W. Exactly solvable ogston model of gel electrophoresis. Ix. Generalizing the lattice model to treat high field intensities. *The Journal of Chemical Physics* **117**, 6745-6756 (2002).
60  Davenport, M. *et al.* The role of pore geometry in single nanoparticle detection. *ACS Nano* (2012).
61  Ai, Y. & Qian, S. Direct numerical simulation of electrokinetic translocation of a cylindrical particle through a nanopore using a poisson-boltzmann approach. *Electrophoresis* **32**, 996-1005 (2011).
62  Lines, R. W. in *Particle size analysis*   (eds N. G. Stanley-Wood & R. W. Lines)  352 (The Royal Society of Chemistry, 1992).
63  Atkins, P. & Paula, J. d. *Elements of physical chemistry*.  357 (Oxford University Press, 2012).
64  Chari, R., Singh, S. N., Yadav, S., Brems, D. N. & Kalonia, D. S. Determination of the dipole moments of rnase sa wild type and a basic mutant. *Proteins: Structure, Function, and Bioinformatics* **80**, 1041-1052 (2012).
65  Cheng, Z., Chaikin, P. M. & Mason, T. G. Light streak tracking of optically trapped thin microdisks. *Physical Review Letters* **89** (2002).
66  Sakmann, B. & Neher, E.     (Springer, 2009).
67  van Lengerich, B., Rawle, R. J. & Boxer, S. G. Covalent attachment of lipid vesicles to a fluid-supported bilayer allows observation of DNA-mediated vesicle interactions. *Langmuir* **26**, 8666-8672 (2010).
68  Solomentsev, Y. & Anderson, J. L. Electrophoresis of slender particles. *J. Fluid Mech.* **279**, 197-215 (1994).
69  Rousseeuw, P. J. & Kaufman, L. *Finding groups in data: An introduction to cluster analysis*. (John Wiley & Sons, Inc., 1990).
70  Yusko, E. C. *et al.* Single-particle characterization of Ab oligomers in solution. *ACS Nano* **6**, 5909-5919 (2012).

71  Kerr, M. K. & Churchill, G. A. Bootstrapping cluster analysis: Assessing the reliability of conclusions from microarray experiments. *Proc. Natl. Acad. Sci.* **98**, 8961-8965 (2001).
72  Lawrence, A.-M., Besir, H. & seyin. Staining of proteins in gels with coomassie g-250 without organic solvent and acetic acid. *J Vis Exp*, e1350 (2009).
73  Daniel, Y. L. & Xinsheng Sean, L. On the distribution of DNA translocation times in solid-state nanopores: An analysis using schrödinger's first-passage-time theory. *Journal of Physics: Condensed Matter* **25**, 375102 (2013).
74  Yadav, S., Shire, S. & Kalonia, D. Viscosity analysis of high concentration bovine serum albumin aqueous solutions. *Pharm. Res.* **28**, 1973-1983 (2011).
75  Li, H., Robertson, A. D. & Jensen, J. H. Very fast empirical prediction and rationalization of protein pka values. *Proteins: Structure, Function, and Bioinformatics* **61**, 704-721 (2005).
76  Bas, D. C., Rogers, D. M. & Jensen, J. H. Very fast prediction and rationalization of pka values for protein–ligand complexes. *Proteins: Structure, Function, and Bioinformatics* **73**, 765-783 (2008).
77  Olsson, M. H. M., Søndergaard, C. R., Rostkowski, M. & Jensen, J. H. Propka3: Consistent treatment of internal and surface residues in empirical pka predictions. *Journal of Chemical Theory and Computation* **7**, 525-537 (2011).
78  Søndergaard, C. R., Olsson, M. H. M., Rostkowski, M. & Jensen, J. H. Improved treatment of ligands and coupling effects in empirical calculation and rationalization of pka values. *Journal of Chemical Theory and Computation* **7**, 2284-2295 (2011).
79  Gitlin, I., Mayer, M. & Whitesides, G. M. Significance of charge regulation in the analysis of protein charge ladders. *J. Phys. Chem. B.* **107**, 1466-1472 (2003).
80  Li, J. *et al.* Ion-beam sculpting at nanometre length scales. *Nature* **412**, 166-169 (2001).
81  Fogarty, A. C., Duboue-Dijon, E., Sterpone, F., Hynes, J. T. & Laage, D. Biomolecular hydration dynamics: A jump model perspective. *Chem. Soc. Rev.* **42**, 5672-5683 (2013).
82  Merzel, F. & Smith, J. C. Is the first hydration shell of lysozyme of higher density than bulk water? *Proc. Natl. Acad. Sci.* **99**, 5378-5383 (2002).
83  Austin, R. H., Beeson, K. W., Eisenstein, L., Frauenfelder, H. & Gunsalus, I. C. Dynamics of ligand binding to myoglobin. *Biochemistry* **14**, 5355-5373 (1975).
84  Steinhoff, H. J., Kramm, B., Hess, G., Owerdieck, C. & Redhardt, A. Rotational and translational water diffusion in the hemoglobin hydration shell: Dielectric and proton nuclear relaxation measurements. *Biophys. J.* **65**, 1486-1495 (1993).
85  Balijepalli, A., Robertson, J. W. F., Reiner, J. E., Kasianowicz, J. J. & Pastor, R. W. Theory of polymer–nanopore interactions refined using molecular dynamics simulations. *J. Am. Chem. Soc.* **135**, 7064-7072 (2013).
86  Arrigo, J. S. Screening of membrane surface charges by divalent cations: An atomic representation. *American Journal of Physiology - Cell Physiology* **235**, C109-C117 (1978).
87  Schneider, S. W., Larmer, J., Henderson, R. M. & Oberleithner, H. Molecular weights of individual proteins correlate with molecular volumes measured by atomic force microscopy. *Pflugers Arch.* **435**, 362-367 (1998).
88  Ozinskas, A. J. in *Topics in fluorescence spectroscopy* Vol. 4  (ed Joseph R. Lakowicz)  487 (Kluwer Academic Publishers, 1994).
89  Carrasco, B. *et al.* Crystallohydrodynamics for solving the hydration problem for multi-domain proteins: Open physiological conformations for human IgG. *Biophysical Chemistry* **93**, 181-196 (2001).
90  Janeway, C. A. *Immunobiology: The immune system in health and disease*. 5th edn,  (Garland Publishing, 2001).
91  de la Torre, J. G. & Carrasco, B. Hydrodynamic properties of rigid macromolecules composed of ellipsoidal and cylindrical subunits. *Biopolymers* **63**, 163-167 (2002).

92  Singh, M., Chand, H. & Gupta, K. C. The studies of density, apparent molar volume, and viscosity of bovine serum albumin, egg albumin, and lysozyme in aqueous and rbi, csi, and dtab aqueous solutions at 303.15 k. *Chemistry & Biodiversity* **2**, 809-824 (2005).
93  El Kadi, N. *et al.* Unfolding and refolding of bovine serum albumin at acid ph: Ultrasound and structural studies. *Biophys. J.* **91**, 3397-3404 (2006).
94  Neish, C. S., Martin, I. L., Henderson, R. M. & Edwardson, J. M. Direct visualization of ligand-protein interactions using atomic force microscopy. *Br. J. Pharmacol.* **135**, 1943-1950 (2002).